\documentclass[english,aps,pre,letter,superscriptaddress,preprint,nofootinbib]{revtex4}
\usepackage[T1]{fontenc}
\usepackage[latin9]{inputenc}
\usepackage{xcolor}
\usepackage{pdfcolmk}
\usepackage{babel}
\usepackage{float}
\usepackage{amsmath}
\usepackage{amssymb}
\usepackage{graphicx}
\PassOptionsToPackage{normalem}{ulem}
\usepackage{ulem}
\usepackage[unicode=true,pdfusetitle,
 bookmarks=true,bookmarksnumbered=false,bookmarksopen=false,
 breaklinks=false,pdfborder={0 0 1},backref=false,colorlinks=false]
 {hyperref}

\makeatletter

\providecommand{\tabularnewline}{\\}
\floatstyle{ruled}
\newfloat{algorithm}{tbp}{loa}
\providecommand{\algorithmname}{Algorithm}
\floatname{algorithm}{\protect\algorithmname}
\providecolor{lyxadded}{rgb}{1,0,0}
\providecolor{lyxdeleted}{rgb}{0,0,1}

\DeclareRobustCommand{\lyxsout}[1]{\ifx\\#1\else\sout{#1}\fi}

\@ifundefined{textcolor}{}
{%
 \definecolor{BLACK}{gray}{0}
 \definecolor{WHITE}{gray}{1}
 \definecolor{RED}{rgb}{1,0,0}
 \definecolor{GREEN}{rgb}{0,1,0}
 \definecolor{BLUE}{rgb}{0,0,1}
 \definecolor{CYAN}{cmyk}{1,0,0,0}
 \definecolor{MAGENTA}{cmyk}{0,1,0,0}
 \definecolor{YELLOW}{cmyk}{0,0,1,0}
}

\usepackage{ae,aecompl}

\makeatother

\begin{document}
\global\long\def\V#1{\boldsymbol{#1}}
\global\long\def\M#1{\boldsymbol{#1}}
\global\long\def\Set#1{\mathbb{#1}}

\global\long\def\D#1{\Delta#1}
\global\long\def\d#1{\delta#1}

\global\long\def\norm#1{\left\Vert #1\right\Vert }
\global\long\def\abs#1{\left|#1\right|}

\global\long\def\grad{\M{\nabla}}
\global\long\def\avv#1{\langle#1\rangle}
\global\long\def\av#1{\left\langle #1\right\rangle }

\global\long\def\Mob{\M M}
\global\long\def\J{\M B}
\global\long\def\S{\M B^{\star}}
\global\long\def\L{\M C}

\global\long\def\shalf{\sfrac{1}{2}}
\global\long\def\sthreehalves{\sfrac{3}{2}}

\global\long\def\myhalf{\frac{1}{2}}
\global\long\def\mythreehalves{\frac{3}{2}}

\global\long\def\sM#1{\M{\mathcal{#1}}}
\global\long\def\dprime{\prime\prime}
\global\long\def\Mob{\sM M}
\global\long\def\J{\sM J}
\global\long\def\S{\sM S}
\global\long\def\L{\sM L}
\global\long\def\R{\M{\mathcal{R}}}

\title{Hydrodynamic fluctuations in quasi-two dimensional diffusion}

\author{Raúl P. Peláez}

\affiliation{Departamento de Física Teórica de la Materia Condensada, Universidad
Autónoma de Madrid, Madrid, Spain}

\author{Florencio Balboa Usabiaga}

\affiliation{Courant Institute of Mathematical Sciences, New York University,
New York, NY, USA}

\affiliation{Center for Computational Biology, Flatiron Institute, Simons Foundation,
New York, NY, USA}

\author{Sergio Panzuela}

\affiliation{Departamento de Física Teórica de la Materia Condensada, and Institute
for Condensed Matter Physics, IFIMAC, Universidad Autónoma de Madrid,
Madrid, Spain}

\author{Qiyu Xiao}

\affiliation{Courant Institute of Mathematical Sciences, New York University,
New York, NY, USA}

\author{Rafael Delgado-Buscalioni}

\affiliation{Departamento de Física Teórica de la Materia Condensada, Universidad
Autónoma de Madrid, Madrid, Spain}

\author{Aleksandar Donev}
\email{donev@courant.nyu.edu}

\affiliation{Courant Institute of Mathematical Sciences, New York University,
New York, NY, USA}
\begin{abstract}
We study diffusion of colloids on a fluid-fluid interface using particle
simulations and fluctuating hydrodynamics. Diffusion on a two-dimensional
interface with three-dimensional hydrodynamics is known to be anomalous,
with the collective diffusion coefficient diverging like the inverse
of the wavenumber. This unusual collective effect arises because of
the compressibility of the fluid flow in the plane of the interface,
and leads to a nonlinear nonlocal convolution term in the diffusion
equation for the ensemble-averaged concentration. We extend the previous
hydrodynamic theory to account for a species/color labeling of the
particles, as necessary to model experiments based on fluorescent
techniques. We study the magnitude and dynamics of density and color
density fluctuations using a novel Brownian dynamics algorithm, as
well as fluctuating hydrodynamics theory and simulation. We find that
hydrodynamic coupling between a single tagged particle and collective
density fluctuations leads to a reduction of the long-time self-diffusion
coefficient, even for an ideal gas of non-interacting particles. This
unexpected finding demonstrates that density functional theories that
do not account for thermal fluctuations are incomplete even for ideal
systems. Using linearized fluctuating hydrodynamics theory, we show
that for diffusion on a fluid-fluid interface, nonequilibrium fluctuations
of the total density are small compared to the equilibrium fluctuations,
but fluctuations of color density are giant and exhibit a spectrum
that decays as the inverse cubed power of the wavenumber. We confirm
these predictions through Brownian dynamics simulations of diffusive
mixing with two indistinguishable species. We also examine nonequilibrium
fluctuations in systems with two-dimensional hydrodynamics, such as
thin smectic films in vacuum. We find that nonequilibrium fluctuations
are colossal and comparable in magnitude to the mean, and can be accurately
modeled using numerical solvers for the nonlinear equations of fluctuating
hydrodynamics.
\end{abstract}
\maketitle

\section{Introduction}

Diffusion of colloidal particles confined to two-dimensional surfaces
is a key transport mechanism in several contexts of technological
and biological significance. Colloidal particles can spontaneously
absorb on fluid-fluid interfaces and stabilize Pickering emulsions
\cite{ComplexFluidFluid_Interfaces}. The transverse diffusion of
proteins embedded in lipid bilayers controls their biological function
\cite{MembraneDiffusion_Review}. In man-made colloidal suspensions,
colloidal particles can be confined to primarily diffuse in a plane
by walls \cite{Diffusion2D_Experiments_Rice} or electrostatic forces
\cite{ChargedColloidsInterface_Chaikin}. While much is understood
about complex fluid-fluid interfaces \cite{ComplexFluidFluid_Interfaces},
fundamental questions about diffusive transport at interfaces remain
unanswered \cite{FluidInterfaces_Microrheology}.

Bulk diffusion of particles in liquids is well-known to be controlled
by hydrodynamics, and diffusion on interfaces is no exception. While
the diffusion of colloids and polymers on a fluid-fluid interface
has been studied theoretically since the 1970s \cite{FluidFluidInterface_Polymers,Diffusion2D_TwoViscosities},
\emph{collective diffusion} in a monolayer of colloidal particles
confined to a fluid-fluid interface has only recently been explored
in some detail \cite{ConfinedDiffusion_2D,DDFT_Diffusion_2D,Diffusion2D_IdealGas}.
These recent studies have shown that collective diffusion on interfaces
is anomalous, with the short-time collective diffusion coefficient
diverging as the inverse of the wavenumber. This unexpected finding
has prompted re-examination of previous experimental results \cite{Diffusion2D_Experiments_Weitz,SD_TwoWalls,Diffusion2D_Experiments_Rice},
and it is plausible that the effect may have been overlooked in a
number of other prior experimental and theoretical works as well.
In addition to the practical implications, the anomalous character
of collective diffusion on interfaces brings into question the very
applicability of Fick's law even at macroscopic scales, which brings
into question fundamental assumptions in the phenomenological foundation
of nonequilibrium thermodynamics.

The physical origin of the anomalous collective diffusion has already
been elucidated in prior work by others \cite{ConfinedDiffusion_2D,DDFT_Diffusion_2D,Diffusion2D_IdealGas}.
Stiff normal forces are required to confine the colloidal particles
to a plane. These forces will propagate momentum to the plane via
the incompressible solvent. In the plane, the resulting hydrodynamic
drag created by one particle acts like a flow source that induces
an effective repulsive force on the other particles. This results
in strong displacement correlations which are the origin of the anomalous
collective diffusion. A new interpretation of this phenomena is obtained
if one focuses on the in-plane dynamics. In the plane, the field of
diffusive displacements appears to be compressible due to the momentum
source associated with each confined particle. In the limit of strict
confinement, the flow acts as a compressible two-dimensional cut through
an incompressible three-dimensional flow field. Because of this apparent
compressibility, the Brownian motion of the particles creates an osmotic
pressure proportional to $k_{B}T$, which propagates via the fluid
inducing long-ranged repulsive interactions. This long-ranged repulsion
decays slowly with the particle-particle distance, and dramatically
accelerates collective diffusion at large scales. Furthermore, contrary
to what was previously thought \cite{PartiallyConfined_Quasi2D,PartiallyConfinedDiffusion_2D},
these long-ranged correlations also appreciably modify the single-particle
diffusion, as we show in this work.

Previous work on collective diffusion of colloids on fluid-fluid interfaces
\cite{ConfinedDiffusion_2D,DDFT_Diffusion_2D,Diffusion2D_IdealGas}
has focused on the \emph{ensemble average}. In particular, in \cite{DDFT_Diffusion_2D}
the authors use the ensemble-averaged equations from the Dynamic Density
Functional Theory with Hydrodynamic Interactions (DDFT-HI) developed
in \cite{DDFT_Hydro}. Our focus here is on the \emph{fluctuations}
around the ensemble average, i.e., on the equations for the evolution
of a particular \emph{instance} (trajectory) of diffusive processes
on interfaces. We use the fluctuating DDFT-HI developed in \cite{DDFT_Hydro},
which is closely-related to \emph{fluctuating hydrodynamics} (FHD)
\cite{DiffusionJSTAT}. The (formal) nonlinear equations of FHD are
challenging to interpret \cite{SPDE_Diffusion_DDFT,DDFT_Hydro}, but,
at the same time, they are more physically transparent than the ensemble-averaged
equations, and do not require closures for noninteracting particles
\cite{DDFT_Hydro}. Furthermore, as we will see later, the fluctuating
DDFT-HI equations are a very useful tool for constructing linear-time
Brownian Dynamics (BD) algorithms.

Understanding the magnitude and the dynamics of density fluctuations
is crucial for several reasons. First, in actual experiments one observes
individual instances, not the ensemble average. While in many systems
typical instances are quite similar to the average, i.e., the fluctuations
are small compared to the mean, this is not always the case. It is
well-known that nonequilibrium fluctuations in diffusive processes
are, rather generally, much larger in magnitude than equilibrium fluctuations,
and are also long ranged \cite{FluctHydroNonEq_Book}. Experiments
in microgravity have measured these ``giant fluctuations'' in three
dimensions, and shown that nonequilibrium diffusive fluctuations are
correlated over macroscopic distances \cite{FractalDiffusion_Microgravity,GRADFLEXTransient}.
For diffusion in two-dimensional systems such as thin smectic films
in vacuum \cite{ThinFilms_True2D}, linearized FHD predicts that the
magnitude of the nonequilibrium fluctuations becomes comparable to
the mean \cite{GiantFluctuations_ThinFilms}. The appearance of such
``colossal fluctuations'' implies that the ensemble average is no
longer informative, since each instance looks very different from
the mean. It is therefore essential to understand whether individual
instances of diffusive mixing processes on fluid-fluid interfaces
evolve similarly to the ensemble average. Second, fluctuations can
be measured in experiments and reveal information about the underlying
microscopic mechanism of diffusion. For diffusion in bulk three-dimensional
liquids or in truly two-dimensional liquids (e.g., a hard-disk fluid),
the ensemble average strictly follows the familiar Fick's law \cite{DiffusionJSTAT}.
This means that the ensemble-averaged concentration in a system where
the diffusing particles are strongly correlated by hydrodynamics (e.g.,
two-dimensional thin smectic films \cite{ThinFilms_HIs}), will look
indistinguishable from the ensemble average in a system where the
diffusing particles are uncorrelated (e.g., diffusion in a solid).
But if one examines the magnitude of the nonequilibrium fluctuations,
the hydrodynamic correlations are revealed through an unexpected power-law
dependence of the structure factor on the wavenumber \cite{GiantFluctuations_ThinFilms}.
Third, collective fluctuations can couple bi-directionally to the
motion of each particle and therefore \emph{renormalize} transport
coefficients. For example, the fluctuating hydrodynamic theory developed
in \cite{TracerDiffusion_Demery} shows that collective density fluctuations
reduce the effective diffusion coefficient of a tagged particle in
a dense system of uncorrelated Brownian soft spheres. In this paper
we will show that a similar effect exists even for an ideal gas of
non-interacting but hydrodynamically-correlated particles diffusing
on a fluid-fluid interface.

In this paper we study the ensemble average and fluctuations of the
density of spherical colloidal particles confined to diffuse on a
fluid-fluid interface. We closely mimic the physical setting used
in prior studies \cite{ConfinedDiffusion_2D,DDFT_Diffusion_2D,Diffusion2D_IdealGas},
and make a number of strong simplifying assumptions:
\begin{enumerate}
\item We assume that the interface is perfectly flat and that two fluids
have the same viscosity; the case of unequal viscosity simply amounts
to taking the arithmetic average of the two viscosities \cite{Diffusion2D_TwoViscosities,FluidFluidInterface_Polymers}.
\item In order to isolate the role of hydrodynamics from the role of other
direct interactions such as capillary forces or steric/electrostatic
repulsion, we focus on an \emph{ideal gas} of non-interacting spherical
colloids \cite{Diffusion2D_IdealGas}. While such an idealized system
could not be studied experimentally, it is a natural candidate for
testing existing and developing new theories. Furthermore, light scattering
observational data for colloids at a fluid interface support predictions
made for an ideal gas of particles \cite{Diffusion2D_IdealGas}.
\item We assume that the colloids are strictly confined to the interface
by a strong confining force in the $z$-direction, i.e., they cannot
leave the $x-y$ plane. In reality, the confinement would be partial,
for example, a laser sheet may provide a harmonic confining potential
in the $z$ direction. However, prior work \cite{PartiallyConfined_Quasi2D,PartiallyConfinedDiffusion_2D}
has shown that partial confinement only changes the results for larger
wavenumbers (i.e., for wavelengths smaller than or comparable to the
range of movement in the $z$ direction), and does not affect the
anomalous diffusion in the plane. At the same time, we will show here
that the case of strict confinement can be simulated much more efficiently
using a two-dimensional Brownian dynamics algorithm.
\item We use a minimal far-field description of the hydrodynamics, as used
in prior work by others \cite{ConfinedDiffusion_2D,DDFT_Diffusion_2D,Diffusion2D_IdealGas}.
Such a Rotne-Prager approximation is quantitatively accurate only
for dilute suspensions, and we may question its usefulness in studying
a \emph{collective }effect that is dominant at larger packing densities.
Nevertheless, as already mentioned, the anomalous collective diffusion
arises because of a long-ranged (far-field) repulsion $\sim k_{B}T$.
We therefore believe that short-ranged (near-field) corrections to
the hydrodynamics will not qualitatively change the phenomena studied
here.
\end{enumerate}
While our focus is diffusion on a fluid-fluid interface, which we
will refer to as Quasi2D (abbreviated q2D) diffusion, we will contrast
Quasi2D diffusion to diffusion in truly two-dimensional systems such
as thin films in vacuum \cite{ThinFilms_True2D}, which we will refer
to as True2D (abbreviated t2D). Even though the diffusion is constrained
to a two-dimensional plane in both cases, the hydrodynamics is essentially
three-dimensional in Quasi2D but it is two-dimensional in True2D.
Our computational and analytical tools can easily be extended to other
types of hydrodynamics. For example, it is believed that lateral diffusion
in lipid membranes \cite{MembraneDiffusion_Review} or thin smectic
films in air \cite{ThinFilms_HIs} can be described using a hydrodynamic
model first proposed by Saffman. The (2+1)-dimensional ((2+1)D for
short) Saffman model has already been combined with linearized fluctuating
hydrodynamics in \cite{GiantFluctuations_ThinFilms}, but an experimental
confirmation of the predictions of the theory is still lacking.

We begin by summarizing the relevant aspects of fluctuating DDFT-HI
theory \cite{DDFT_Hydro} in Section \ref{sec:DDFT-HI}, and then
use DDFT-HI to re-examine some observations about the unusual nature
of Quasi2D diffusion made in prior work by others \cite{ConfinedDiffusion_2D,DDFT_Diffusion_2D,Diffusion2D_IdealGas}.
In Section \ref{sec:BD-2D} we develop and validate a novel algorithm
for performing Brownian Dynamics with Hydrodynamic Interactions (BD-HI)
in two dimensions (BD-2D). Our pseudo-spectral algorithm scales linearly
with the number of particles and can efficiently be implemented on
Graphical Processing Units (GPUs). We use the BD-2D algorithm to perform
large scale simulations in Quasi2D and True2D. In particular, in Section
\ref{sec:EquationMean} we study the evolution of the ensemble average
in free diffusion in Quasi2D. Our studies confirm prior results, but
we also perform a new type of numerical experiment by coloring (labeling)
the particles with two colors (species labels). We clarify several
aspects of the anomalous diffusion by studying the evolution of the
ensemble average not just for the total density but also for color
(species) density. In Section \ref{sec:SelfDiffusion} we study the
effect of collective fluctuations on the long-time self-diffusion
coefficient of a tagged (tracer) particle, and quantify how density
affects the long-time self diffusion in Quasi2D systems. In Section
\ref{sec:Fluctuations} we study the magnitude and dynamics of density
and color density fluctuations in True2D and Quasi2D. Finally, we
offer some Conclusions and topics for future work.

\section{\label{sec:DDFT-HI}Fluctuating Dynamic Density Functional Theory
with Hydrodynamic Interactions}

In this section we review some prior results regarding colloidal diffusion
in quasi two-dimensional (Quasi2D) geometry. Our Dynamic Density Functional
Theory with Hydrodynamic Information (DDFT-HI) formulation is based
on several prior works, and in particular \cite{DDFT_Diffusion_2D}.
At the same time, we account for fluctuations using the approach proposed
in \cite{DDFT_Hydro}, and our presentation differs in several important
aspects from prior work even though the results we derive in this
section are already known.

\subsection{From Partial to Strict Confinement}

As a preamble to this section we consider a physical setup where colloids
immersed in an unbounded three-dimensional fluid are confined to remain
near the plane $z=0$ by a strong confining potential, as could be
experimentally done using using a laser sheet trap. Due to the fluid
incompressibility, part of the momentum introduced by a confinement
force applied to one colloid (in the normal direction) will spread
\emph{over the plane} according to the Oseen tensor. As we show next,
in the $z=0$ plane, the resulting hydrodynamic drag acts like a flow-source
which tends to expel other particles around it because the generated
velocity field has a positive divergence.

In this paper we consider the limit of infinitely strong confining
forces. In this limit, it is not difficult to prove \cite{PartiallyConfined_Quasi2D}
that the effective hydrodynamic drag is proportional to the divergence
of the hydrodynamic mobility \emph{evaluated in the plane}. To see
this, let us consider non-interacting particles of radius $a$ confined
to the $x-y$ plane by a quadratic potential $U(z)=(k_{s}/2)z^{2}$.
At a given temperature $T$ the typical displacement of the particles
around the confining plane $z=0$ is $\delta=\left(k_{B}T/k_{s}\right)^{1/2}$.
The Ito equations of Brownian Dynamics with Hydrodynamic Interactions
(BD-HI) for the (correlated) positions $\V Q\left(t\right)=\left\{ \V q_{1}\left(t\right),\dots,\V q_{N}\left(t\right)\right\} $
of $N$ spherical colloidal particles, where $\V q_{i}=\left(x_{i},y_{i},z_{i}\right)\in\Set R^{3}$,
have the general form
\begin{equation}
d\V Q=\M M\V Fdt+\left(2k_{B}T\,\M M\right)^{\frac{1}{2}}d\V{\mathcal{B}}+k_{B}T\left(\partial_{\V Q}\cdot\M M\right)dt,\label{eq:BD_M}
\end{equation}
where $\V{\mathcal{B}}(t)$ is a vector of Brownian motions, and $\V F=-\partial_{\V Q}U$
are the forces arising from the confining potential; other forces
(such as excluded volume interactions) could be added in this discussion
without loss of generality. The symmetric positive semidefinite (SPD)
\emph{mobility}\textbf{\emph{ }}\emph{matrix} $\M M\left(\V Q\right)\succeq\M 0$
encodes all of the information about hydrodynamic interactions (correlations)
\textendash{} we will discuss its form in more detail later. Within
the pairwise Rotne-Prager approximation for $\M M$, we have $\partial_{\V Q}\cdot\M M=0$
because of the incompressibility of the (three-dimensional) flow.

Prior work \cite{ConfinedDiffusion_2D,DDFT_Diffusion_2D,Diffusion2D_IdealGas}
focusing on strict confinement has simply started from the equations
(\ref{eq:BD_M}) with the $z$ component ignored, but these equations
can be derived precisely by taking the limit $k_{s}\rightarrow\infty$,
as we explain next. Two distinct prior works \cite{PartiallyConfined_Quasi2D,PartiallyConfinedDiffusion_2D}
have examined the transition from partial to strict confinement. The
results of these prior studies have shown that partial confinement
only changes the results for wavelengths smaller than or comparable
to the range of movement in the $z$ direction. For larger wavelengths
the system behaves as if the confinement is strict.

For infinitely strong confinement $k_{s}\rightarrow\infty$ and $\delta\rightarrow0$,
particles move strictly in the plane, with $z=0$. It is well known
how to take a limit of overdamped Langevin equations in the presence
of a stiff confining potential. The general theory for this is quite
complicated \cite{ConstrainedBD,ConstrainedStochasticDiffusion,StiffPotential_Langevin,ModelReduction_GENERIC}
because it involves metrics/curvatures of the constraint manifold
and projected mobility matrices, but in our case taking this limit
is trivial because of the following two generic properties. The first
important property is that for particles strictly confined to the
$(x,y)$ plane the mobility matrix
\[
\M M=\left[\begin{array}{cc}
\M M^{\parallel} & \M M^{\parallel,\perp}\\
\left(\M M^{\parallel,\perp}\right)^{T} & \M M^{\perp}
\end{array}\right]=\left[\begin{array}{cc}
\M M^{\parallel}\\
 & \M M^{\perp}
\end{array}\right]
\]
has a block-diagonal structure in which the $z$ direction is decoupled
to the $\left(x,y\right)$ directions, where we have ordered the parallel
or $(x,y)$ degrees of freedom before the perpendicular or $z$ degrees
of freedom. This follows from symmetry arguments: Applying a force
parallel to the plane cannot induce motion normal to the plane because
both sides are symmetric; similarly, normal forces cannot induce a
velocity in the plane. Another important property is that there is
no free energy gradient in the plane due to the confinement because
the confining potential is uniform in the $(x,y)$ plane. This implies
that there is no entropic gradient in the plane coming from non-uniform
confinement to the plane. 

Combining these two properties leads to the trivial conclusion that
one can simply ignore the $z$ component of the original equations
and just take the $x-y$ component of the deterministic terms in the
equation. Specifically, the limiting equation is identical to (\ref{eq:BD_M})
but with $\V q_{i}=\left(x_{i},y_{i}\right)\in\Set R^{2}$ and $\V F_{i}$
now being the $\left(x,y\right)$ components of the particle positions
and applied forces, and $\M M$ replaced by the $(x,y)$ diagonal
block $\M M^{\parallel}$ of the three-dimensional mobility matrix,
\begin{equation}
d\V Q^{\parallel}=\M M^{\parallel}\V F^{\parallel}dt+\left(2k_{B}T\,\M M^{\parallel}\right)^{\frac{1}{2}}d\V{\mathcal{B}}^{\parallel}+k_{B}T\left(\partial_{\V Q^{\parallel}}\cdot\M M^{\parallel}\right)dt,\label{eq:BD_parallel}
\end{equation}
where $\V Q_{\parallel}$ denotes the positions in the plane.

It is instructive to explain how the nonzero stochastic drift term
$k_{B}T\left(\partial_{\V Q_{\parallel}}\cdot\M M^{\parallel}\right)$,
which is necessary for time reversibility (detailed balance) of the
limiting dynamics (\ref{eq:BD_parallel}), arises in the strict confinement
limit $\delta/a\rightarrow0$. For strong confinement, the particles'
height fluctuates rapidly around the $z=0$ plane and thus reach a
quasi-steady equilibrium state in the $z$ direction. This means that
the probability distribution for observing the particles in a given
configuration factorizes in the form
\[
P\left(\V Q,t\right)\approx P_{\parallel}\left(\V Q_{\parallel},t\right)\prod_{i=1}^{N}P_{\perp}\left(z_{i}\right),
\]
 where $P_{\perp}\left(z\right)\sim\exp\left(-\beta U\left(z\right)\right)$
is the Gibbs-Boltzmann distribution in the perpendicular direction.
The parallel velocity of one particle at $z_{1}=0$ due to the drag
created by a second particle rapidly fluctuating in the $z$ direction,
time-averaged over the fast $z$ motion of the particle at $z_{2}$,
is $\V u_{1}^{\parallel}=\int dz_{2}\;P_{\perp}(z_{2})\M M_{12}^{\parallel,\perp}\,F_{2}^{\perp}$,
where $F_{2}^{\perp}=-\partial_{z_{2}}U(z_{2})$ and the component
$\M M_{12}^{\parallel,\perp}$ of the mutual mobility measures the
hydrodynamic coupling between perpendicular forces and parallel flows.
Since $F^{\perp}(z_{2})P_{\perp}(z_{2})=\left(k_{B}T\right)\partial_{z_{2}}P_{\perp}(z_{2})$,
an integration by parts reveals that
\[
\V u_{1}=\int dz_{2}\;\M M_{12}^{\parallel,\perp}\,\partial_{z_{2}}P_{\perp}\left(z_{2}\right)=-\left(k_{B}T\right)\int dz_{2}\;P_{\perp}\left(z_{2}\right)\,\partial_{z_{2}}\M M_{12}^{\parallel,\perp}.
\]
In the limit of strict confinement $P_{\perp}(z)\rightarrow\delta\left(z\right)$
becomes a Dirac $\delta$ function enforcing $z_{1}=z_{2}=0$. Therefore
one concludes that
\[
\V u_{1}^{\parallel}\approx-\left(k_{B}T\right)\left(\partial_{z_{2}}\M M_{12}^{\parallel,\perp}\right)_{z_{1}=z_{2}=0}=\left(k_{B}T\right)\left(\partial_{\V q_{2}^{\parallel}}\cdot\M M_{12}^{\parallel}\right)_{z_{1}=z_{2}=0},
\]
where we have used the incompressibility of the three-dimensional
flow to conclude that the pairwise mobility satisfies $\left(\partial_{\V q_{2}^{\parallel}}\cdot\M M_{12}^{\parallel}\right)+\partial_{z_{2}}\M M_{12}^{\parallel,\perp}=\V 0$.
This demonstrates that the strong confinement leads to an additional
stochastic drift velocity in the plane of the form 
\[
\V u^{\parallel}=\left(k_{B}T\right)\partial_{\V Q_{\parallel}}\cdot\M M^{\parallel},
\]
which is the cause of the unusual collective effects on diffusion
in the plane. For simplicity of notation, henceforth we will use the
same generic (\ref{eq:BD_M}) for both motion in three and two dimensions,
and distinguish only when necessary.

\subsection{DDFT-HI Theory}

We begin by summarizing some key results of DDFT-HI, previously obtained
in \cite{DDFT_Hydro}, by starting from the BD-HI equation (\ref{eq:BD_M}).
We assume here that the limit of strict confinement has already been
taken if one is considering a suspension confined to a plane, so that
particles diffuse in the plane only. In this paper we focus on an
\emph{ideal gas} of hydrodynamically-interacting particles, i.e.,
we take $\V F=0$. In this work we consider periodic boundary conditions
only; however, the formalism is rather general. For simplicity in
this section we assume the system is unbounded (i.e., in free space).
Periodicity can be handled by using the free-space formulation and
summing over periodic images; this is most easily done numerically
in Fourier space and is transparently handled in our algorithm by
using the Fast Fourier Transform (FFT).

To leading order in the far-field, $\M M\left(\V Q\right)$ is given
by the pairwise approximation \cite{DDFT_Hydro}

\begin{equation}
\forall\left(i,j\right):\quad\M M_{ij}\left(\V q_{i},\V q_{j}\right)=\R\left(\V q_{i},\V q_{j}\right),\label{eq:M_R}
\end{equation}
where the tensor $\R\left(\V r,\V r^{\prime}\right)$ is a symmetric
positive-semidefinite (SPD) \emph{hydrodynamic kernel} that depends
on the geometry and boundary conditions \footnote{Note that we have adjusted slightly the notation in \cite{DDFT_Hydro}
to extract $\left(k_{B}T\right)$ outside of $\R$, so that $\R$
only contains hydrodynamic information and involves the viscosity
but not the temperature.}. For particles very far apart $\R$ approaches the Green's function
(also called Oseen tensor) for Stokes flow in the specified geometry
with the specified boundary conditions. A commonly-used model of the
hydrodynamic kernel, suitable in an unbounded three-dimensional system,
is the Rotne-Prager-Yamakawa (RPY) kernel \cite{RotnePrager}. Because
we assume that \eqref{eq:M_R} holds even if $i=j$ (which is true
for the RPY tensor), the (short-time) self-diffusion tensor of a particle
with position $\V r$ is
\[
\M{\chi}\left(\V r\right)=\left(k_{B}T\right)\R\left(\V r,\V r\right).
\]

We consider here a translationally-invariant and isotropic system,
for which symmetry dictates the form
\begin{equation}
\R\left(\V q_{i},\V q_{j}\right)=\R\left(\V r=\V q_{i}-\V q_{j}\right)=f\left(r\right)\M I+g\left(r\right)\frac{\V r\otimes\V r}{r^{2}},\label{eq:R_real_space}
\end{equation}
where $g(0)=0$ and $\otimes$ denotes a diadic product, $\V r\otimes\V r\equiv\V r\V r^{T}$
where superscript $T$ denotes a transpose. The majority of our discussion
is focused on particles on a flat two-dimensional fluid-fluid interface,
in an otherwise unbounded three-dimensional fluid. We refer to this
setup as \emph{quasi two-dimensional} (Quasi2D or q2D) diffusion ($\V r\in\Set R^{2}$),
to be contrasted with \emph{true two-dimensional} (True2D or t2D)
diffusion ($\V r\in\Set R^{2}$) in a thin liquid film suspended in
vacuum, or \emph{true three-dimensional} (True3D or t3D) diffusion
in a bulk liquid ($\V r\in\Set R^{3}$). For True3D systems, the (short-time)
self diffusion of the particles obeys the Stokes-Einstein relationship
\[
\M{\chi}\left(\V r\right)=\left(k_{B}T\right)\R\left(\V 0\right)=\chi_{0}\M I=\frac{k_{B}T}{\left(6\pi\eta a\right)}\M I,
\]
where $\eta$ is the fluid viscosity \footnote{Recall that we assumed equal viscosities of the two fluids on either
side of the plane of diffusion.} and $a$ is the hydrodynamic radius of the colloidal particles. Furthermore,
for $r\gg a$ the hydrodynamic kernel approaches the three-dimensional
Oseen tensor,
\begin{equation}
f\left(r\gg a\right)\approx g\left(r\gg a\right)\approx\frac{1}{8\pi\eta r}.\label{eq:f_g_q2D}
\end{equation}
The subscript zero in $\chi_{0}$ emphasizes that this is the \emph{short-time}
self diffusion coefficient, which can in principle be different from
the \emph{long-time} self diffusion coefficient of the particles because
fluctuations can modify (renormalize) the Stokes-Einstein relationship
\cite{StokesEinstein,TracerDiffusion_Demery}. We return to this difference
in Section \ref{sec:SelfDiffusion}, and for now we use a generic
notation $\chi\approx\chi_{0}$ for the diffusion coefficient.

As explained in detail in \cite{DDFT_Hydro}, we start from (\ref{eq:BD_M})
and define a concentration field from the positions of the particles
via
\begin{equation}
c\left(\V r,t\right)=\sum_{i=1}^{N}\delta\left(\V q_{i}\left(t\right)-\V r\right).\label{eq:c_def}
\end{equation}
For an ideal gas, Ito's rule formally gives a \emph{closed} but \emph{nonlinear
}Ito stochastic advection-diffusion equation for the concentration
(c.f. Eq. (15) in \cite{DDFT_Hydro}) \footnote{We have removed here a term $\V b(\V r,\V r)c(\V r,t)$ from Eq. (15)
in \cite{DDFT_Hydro} since this must vanish for translationally-invariant
systems, and deleted all terms coming from pairwise interactions.},
\begin{equation}
\begin{aligned}\partial_{t}c(\V r,t) & =-\grad\cdot\left(\V w\left(\V r,t\right)c(\V r,t)\right)+\grad\cdot\left(\chi\grad c(\V r,t)\right)\\
 & +\left(k_{B}T\right)\grad\cdot\left(c(\V r,t)\int\R(\V r-\V r^{\prime})\grad^{\prime}c(\V r^{\prime},t)\,d\V r^{\prime}\right).
\end{aligned}
\label{eq:c_fluct_general}
\end{equation}
Here $\V w\left(\V r,t\right)$ is a random velocity field that is
white in time and has a spatial covariance,
\begin{align}
\av{\V w\left(\V r,t\right)\otimes\V w\left(\V r^{\prime},t^{\prime}\right)} & =\left(2k_{B}T\right)\R\left(\V r-\V r^{\prime}\right)\delta\left(t-t^{\prime}\right),\label{eq:C_w}
\end{align}
and has a clear physical interpretation in \emph{fluctuating hydrodynamics}
(FHD) as an overdamped representation of the fluid velocity \cite{DiffusionJSTAT}. 

The nonlinear term on the second line of (\ref{eq:c_fluct_general})
comes from the last term in (\ref{eq:BD_M}) involving the divergence
of the mobility matrix. Integrating by parts, we can rewrite the convolution
as $-\left(k_{B}T\right)\grad\cdot\left(c(\V r,t)\int\left(\grad^{\prime}\cdot\R(\V r-\V r^{\prime})\right)c(\V r^{\prime},t)\,d\V r^{\prime}\right)$,
which mimics the action of a pairwise force $\sim\left(k_{B}T\right)\grad\cdot\R$.
For the True3D or True2D Oseen and RPY tensors, the hydrodynamic kernel
is divergence free, $\grad\cdot\R(\V r)=0,$ and so the nonlinear
term in (\ref{eq:c_fluct_general}) disappears and we obtain a \emph{linear}
fluctuating advection-diffusion equation that can be solved numerically
\cite{DiffusionJSTAT}.

In the case of Quasi2D diffusion, however, the divergence of the mobility
in the plane of confinement is nonzero and we must keep the second
line of (\ref{eq:c_fluct_general}). In particular, in the far field
(\ref{eq:f_g_q2D}) gives
\[
\grad\cdot\R(\V r)\approx\frac{1}{8\pi\eta}\cdot\frac{\V r}{r^{3}}\text{ for }r\gg a.
\]
More precisely, in the equation for particle $i$ the last term in
(\ref{eq:BD_M}) looks like a Coulomb repulsion term \footnote{We have used here Eqs. (A7) and (A10) in \cite{DDFT_Hydro}, as we
explain in more detail in Section \ref{sec:BD-2D}.},
\begin{equation}
\frac{d\V q_{i}\left(t\right)}{dt}\approx\V w\left(\V q_{i},t\right)+\sum_{j\neq i}\frac{k_{B}T}{8\pi\eta}\cdot\frac{\V q_{i}-\V q_{j}}{\norm{\V q_{i}-\V q_{j}}^{3}}.\label{eq:dqi_dt}
\end{equation}
The last term in this equation has the same form as would have arisen
in the absence of hydrodynamic interactions, $M_{ij}=\delta_{ij}/\left(6\pi a\eta\right)$,
had the particles been interacting with a repulsive ``electrostatic''
potential
\begin{equation}
U_{\text{eff}}(r)=\frac{3a}{4r}k_{B}T.\label{eq:effective_repulsion}
\end{equation}
Although this picture of the particles repelling each other with a
long-ranged $r^{-1}$ potential is compelling and intuitive, it is
also misleading because this ``repulsion'' is thermal in origin
and comes from the rapid momentum transport \emph{perpendicular} to
the plane to which the particles are confined by a stiff potential.
As such, the Quasi2D particle system is thermodynamically an ideal
gas, and \emph{not} a gas of point charges. The two terms on the right
hand side of (\ref{eq:dqi_dt}) are in fluctuation-dissipation balance
with each other, and have a common origin in the thermal fluctuations
of the fluid velocity.

As a nonlinear SPDE, (\ref{eq:c_fluct_general}) does not really make
sense and it can be thought of simply as a rewriting of the equations
of BD-HI; interestingly this kind of rewriting is particularly suited
for constructing numerical methods to solve (\ref{eq:BD_M}). In deriving
(\ref{eq:c_fluct_general}), we have performed no coarse-graining
other than forgetting the numbering of the particles, so the concentration
is still a sum of delta functions \cite{DDFT_Hydro}. However, for
non-interacting particles, and for the case of interest to us where
$\R$ is a long-ranged kernel, there is hope that (\ref{eq:c_fluct_general})
may nonetheless be useful in practice, after a suitable renormalization
of the transport coefficients. The more traditional route \cite{DDFT_Hydro_Lowen},
followed in \cite{DDFT_Diffusion_2D}, has been to write deterministic
equations for the ensemble average by averaging over realizations
of $\V w$ (the same as averaging over the noise for the BD-HI equations).
This, however, does \emph{not} give a closed equation since one must
then introduce the two-point correlation function, which is not known,
as we discuss in great detail in Section \ref{sec:EquationMean}.
We therefore keep both the random advection and the nonlinear term
in (\ref{eq:c_fluct_general}), as this is free of approximations
and gives information not just about the ensemble average (which is
hard to measure in experiments), but also about fluctuations present
in each \emph{instance} of the diffusive mixing process (which is
what experiments measure). In this paper we study various nontrivial
consequences of (\ref{eq:c_fluct_general}) and assess its usefulness
by comparing results from DDFT-HI/FHD to results from BD-HI.

\subsection{Linearized Equations}

Let us first consider the case of a spatially uniform system with
concentration $c(\V r,t)=c_{0}+\d c(\V r,t)$, where in some sense
$\d c\ll c_{0}$. Let us for a moment ignore the random advection,
i.e., fluctuations, and simply consider the temporal \emph{relaxation}
of a small perturbation; the more complete derivation of the dynamic
structure factor based on linearized FHD is given in Section \ref{subsec:GiantDensity}.
If we linearize (\ref{eq:c_fluct_general}) around the uniform state
we get the linear non-local diffusion equation
\begin{equation}
\begin{aligned}\partial_{t}\d c(\V r,t) & =\chi\grad^{2}\d c(\V r,t)+\left(k_{B}T\right)\grad\cdot\left(c_{0}\int\R(\V r-\V r^{\prime})\grad^{\prime}\d c(\V r^{\prime},t)\,d\V r^{\prime}\right).\end{aligned}
\label{eq:c_linearized_det}
\end{equation}
This equation can trivially be solved in Fourier space to obtain
\begin{equation}
\hat{\d c}_{\V k}(t)=\hat{\d c}_{\V k}(0)\;\exp\left(-k^{2}\chi_{c}(k)t\right),\label{eq:c_fluct_linearized}
\end{equation}
where hat denotes a Fourier transform \footnote{The scaling convention for the Fourier transforms that we use is given
in \cite{LLNS_S_k}, and ensures that the Fourier spectrum of white
noise is unity.} and the subscript $\V k$ denotes the wavenumber \cite{LLNS_S_k}.
Here the short-time\emph{ collective diffusion coefficient} is
\begin{equation}
\chi_{c}(k)=\chi+\left(k_{B}T\right)c_{0}k^{-2}\left(\V k\cdot\hat{\R}_{\V k}\cdot\V k\right).\label{eq:D_c_general}
\end{equation}
In this work we will not need to distinguish between short-time and
long-time \emph{collective} diffusion (see numerical results in Section
\ref{subsec:S_kt_density}), and we therefore use $\chi_{c}$ to denote
generically a collective diffusion coefficient.

Since $\R$ and thus $\hat{\R}_{\V k}$ are symmetric positive semidefinite
kernels, $\V k\cdot\hat{\R}_{\V k}\cdot\V k\geq0$ and in general
as the density $c_{0}$ increases the collective diffusion is potentially
\emph{accelerated} over the case without hydrodynamics, $\chi_{c}\equiv\chi$.
This somewhat unexpected \emph{collective} effect does not exist in
either True2D or True3D because of the incompressibility of the fluid
flow, $\V k\cdot\hat{\R}_{\V k}\cdot\V k=0$. The collective enhancement
of $\chi_{c}$ becomes important for Quasi2D diffusion and has been
studied in a number of prior works \cite{ConfinedDiffusion_2D,DDFT_Diffusion_2D,Diffusion2D_IdealGas,PartiallyConfinedDiffusion_2D}.

Note that our derivation exposes that the only approximation made
for an ideal gas of particles is the linearization. Although this
linearization seems natural, we will see in Section \ref{sec:SelfDiffusion}
that even for an ideal gas this is only an approximation, and that
fluctuations renormalize the diffusion coefficient $\chi$ by a measurable
amount through a nonlinear (mode) coupling. The derivation in \cite{DDFT_Diffusion_2D}
is in our opinion unnecessarily complicated as it invokes unnecessary
closures to derive the same result (\ref{eq:D_c_general}). This is
because the authors of \cite{DDFT_Diffusion_2D} start from the DDFT-HI
equations for the ensemble average \cite{DDFT_Hydro_Lowen}, which
are not closed, so they need a number of approximations to arrive
at the same equation. Of course, for interacting particles one must
invoke closures, perhaps in the form of density functionals \cite{DDFT_Hydro_Lowen},
and make additional approximations that are very difficult to justify
mathematically.

\subsection{Oseen Approximation in Quasi2D}

To leading order in the far field, i.e., for small $ka\ll1$, we can
compute $\hat{\R}_{\V k}$ in Quasi2D by simply integrating the well-known
Fourier transform of the 3D Oseen tensor along the $z$ axes,
\begin{equation}
\hat{\R}_{\V k}=\frac{1}{2\pi\eta}\int_{k_{z}=-\infty}^{\infty}\frac{dk_{z}}{\left(k^{\prime}\right)^{2}}\left(\M I-\frac{\V k^{\prime}\otimes\V k^{\prime}}{\left(k^{\prime}\right)^{2}}\right),\label{eq:project_2D}
\end{equation}
where $\V k=\left(k_{x},k_{y}\right)$ and $\V k^{\prime}=\left(k_{x},k_{y},k_{z}\right)$.
The integral can be done analytically \cite{ConfinedDiffusion_2D,DDFT_Diffusion_2D,Diffusion2D_IdealGas},
to give the spectral decomposition
\begin{equation}
\hat{\R}_{\V k}=\frac{1}{\eta k^{3}}\left(\frac{1}{2}\V k_{\perp}\otimes\V k_{\perp}+\frac{1}{4}\V k\otimes\V k\right),\label{eq:q2D_small_k}
\end{equation}
where $\V k_{\perp}=\left(k_{y},-k_{x}\right)$ is a vector perpendicular
to $\V k$, i.e., $\V k_{\perp}=\V k\times\hat{z}$. 

This shows that the spectrum of the random velocity field $\V w\left(\V r,t\right)$
decays like $1/k$ and that the field is compressible. Specifically,
for the divergence we get $\V k\cdot\hat{\R}_{\V k}\cdot\V k=k/4\eta,$
which allows us to write (\ref{eq:D_c_general}) in the form
\begin{equation}
\chi_{c}\left(k\right)=\chi\left(1+\frac{1}{kL_{h}}\right),\label{eq:chi_small_k}
\end{equation}
where $L_{h}=4\eta\chi/\left(k_{B}Tc_{0}\right)=2/\left(3\pi ac_{0}\right)$
is a hydrodynamic correlation length. Remarkably, for q2D diffusion
$\chi_{c}$ diverges like $1/k$ \cite{DDFT_Diffusion_2D,ConfinedDiffusion_2D,Diffusion2D_IdealGas}.
For planar packing densities $\phi=\pi c_{0}a^{2}\sim1$, which is
the most interesting regime for collective diffusion, we have $L_{h}\sim a$
and therefore collective diffusion effects manifest themselves strongly
at all length scales of interest.

Note that the real space equivalent of (\ref{eq:q2D_small_k}) shows
a power law $1/r^{3}$ tail rather than a Gaussian tail as it does
for ordinary diffusion \cite{DDFT_Diffusion_2D,ConfinedDiffusion_2D}.
This is yet another dramatic consequence of the nonzero divergence
of the flow in the plane of confinement that we explore further in
Section \ref{sec:EquationMean}.

\section{\label{sec:BD-2D}Brownian dynamics in quasi two-dimensions}

In this section we develop a novel algorithm to perform BD-HI in (quasi-)two-dimensional
systems (BD-2D) in \emph{linear time} in the number of particles.
This improves dramatically the efficiency over prior methods for strict
confinement \cite{Diffusion2D_IdealGas} or partial confinement \cite{PartiallyConfined_Quasi2D,PartiallyConfinedDiffusion_2D}.
The algorithm is based on the DDFT-HI equation (\ref{eq:c_fluct_general}),
viewed from the perspective of fluctuating hydrodynamics \cite{DiffusionJSTAT}.
A closely-related numerical method implemented in the GPU code \emph{fluam}
\cite{ISIBM,BrownianBlobs} has previously been used by some of us
with partial confinement \cite{PartiallyConfined_Quasi2D}. The key
new step here is to assume perfect confinement and eliminate completely
the third dimension, thus significantly reducing the computational
complexity at the expense of introducing a nonzero divergence of the
flow. Our algorithm combines the Fluctuating Force Coupling Method
(F-FCM) \cite{ForceCoupling_Fluctuations,FluctuatingFCM_DC} with
ideas used in the Fluctuating Immersed Boundary method (FIB) \cite{BrownianBlobs}
to construct a simple yet efficient algorithm specifically tailored
to periodic two-dimensional systems, with either True2D and Quasi2D
hydrodynamics.

A rather general form of the hydrodynamic kernel is given by the double
convolution
\begin{equation}
\M{\mathcal{R}}\left(\V r_{1},\V r_{2}\right)=\M{\mathcal{R}}\left(\V r_{1}-\V r_{2}\right)=\int\delta_{a}\left(\V r_{1}-\V r^{\prime}\right)\M{\Set G}\left(\V r^{\prime}-\V r^{\prime\prime}\right)\delta_{a}\left(\V r_{2}-\V r^{\prime\prime}\right)d\V r^{\prime}d\V r^{\prime\prime},\label{eq:R_double_conv}
\end{equation}
where $\M{\Set G}$ is a Green's function for the particular kind
of flow (True2D, Quasi2D, or True3D, or other), and $\delta_{a}$
are regularizing kernels (i.e., smeared delta functions) of width
$a$. The form (\ref{eq:R_double_conv}) is consistent consistent
with both the F-FCM and FIB methods, as well as recent BD-HI algorithms
based on the RPY kernel \cite{SpectralRPY}. For RPY, the delta kernels
are surface delta functions on the surface of a sphere of radius $a$,
and for FCM the delta kernels are Gaussian kernels with standard deviation
$\sigma=a/\sqrt{\pi}$ \cite{ForceCoupling_Monopole}. It is important
to note that for Quasi2D the kernels and the intergrals in (\ref{eq:R_double_conv})
are three dimensional even though the target $\V r_{1}$ and source
$\V r_{2}$ lie in the plane of the interface. By contrast, for True2D
diffusion there is no third dimension and the kernels and intergrals
in (\ref{eq:R_double_conv}) are both two dimensional.

In this work we employ Gaussian kernels for $\delta_{a}$ as in the
FCM because this: (1) allows for spectral accuracy of the numerical
method, without the near-field corrections required for the RPY kernel
\cite{SpectralRPY}; (2) avoids small-scale grid artifacts present
for compactly-supported kernels as used in the FIB method; and (3)
simplifies analytical computations. At the level of accuracy of the
Rotne-Prager pairwise approximation, there is no strong reason to
prefer the RPY kernel over the FCM kernel, as they both capture the
far-field hydrodynamics to a similar accuracy \cite{ForceCoupling_Monopole},
and regularize the near-field hydrodynamics in a similar manner. 

Using the translational invariance and isotropy of (\ref{eq:R_double_conv})
to simplify Eqs. (A7) and (A10) in \cite{DDFT_Hydro}, and using (\ref{eq:R_double_conv}),
and it is not hard to show that the BD-HI equations (\ref{eq:BD_M})
can be written in the form
\begin{equation}
\frac{d\V q_{i}}{dt}=\V w\left(\V q_{i},t\right)+\int\delta_{a}\left(\V q_{i}-\V r^{\prime}\right)\sum_{j}\M{\Set G}\left(\V r^{\prime},\V r^{\prime\prime}\right)\left[\V F_{j}\delta_{a}\left(\V q_{j}-\V r^{\prime\prime}\right)+\left(k_{B}T\right)\left(\V{\partial}\delta_{a}\right)\left(\V q_{j}-\V r^{\prime\prime}\right)\right]d\V r^{\prime}d\V r^{\prime\prime}.\label{eq:dq_dt}
\end{equation}
Here $\partial_{\alpha}\delta_{a}\left(\V r\right)=\partial\delta_{a}\left(\V r\right)/\partial r_{\alpha}$
denotes the gradient of the Gaussian kernel $\delta_{a}\left(\V r\right)$.
It is important to emphasize that this gradient is taken in the ambient
space in which the diffusion happens, i.e., in the plane of confinement
for the case of True2D and Quasi2D diffusion, and in three-dimensional
space for True3D diffusion. Here we have made the sum include $i=j$
because the self term disappears by the fact the kernel and its gradient
are rotationally isotropic. Note that one can perform an integration
by parts and move the gradient from $\delta_{a}$ to the Green's function
as a divergence to show that the last term can be omitted when the
Green's function is divergence free in the space in which the diffusion
happens (e.g., for True2D and True3D).

The particles-only equation (\ref{eq:dq_dt}) is the overdamped and
incompressible limit of coupled fluctuating fluid-particle equations
that include inertia \cite{ISIBM}. We can gain physical intuition
about the term $\left(k_{B}T\right)\left(\V{\partial}\delta_{a}\right)\left(\V q_{j}-\V r^{\prime\prime}\right)$
in (\ref{eq:dq_dt}), which is the cause of the collective diffusion
enhancement, by observing that it arises due to the osmotic pressure
associated with the colloidal particle itself \cite{SELM}. This osmotic
pressure contribution has been justified from a coarse-graining perspective
in \cite{CoarseBlob}.

\subsection{Force Coupling Method in Quasi2D}

One can solve (\ref{eq:dqi_dt}) very efficiently pseudo-spectrally
\cite{DiffusionJSTAT} once an explicit form for the Fourier transform
of the hydrodynamic kernel is available. This Fourier transform must
have the isotropic form
\begin{equation}
\hat{\R}_{\V k}=\frac{1}{\eta k^{3}}\left(c_{2}\left(ka\right)\,\V k_{\perp}\otimes\V k_{\perp}+c_{1}\left(ka\right)\,\V k\otimes\V k\right),\label{eq:R_Fourier}
\end{equation}
where we singled out a $1/k^{3}$ prefactor mimicking (\ref{eq:q2D_small_k}).
The Fourier transform of the Quasi2D FCM kernel can easily be computed
by multiplying the integrand in (\ref{eq:project_2D}) by the square
of the Fourier transform of the Gaussian kernel in agreement with
(\ref{eq:R_double_conv}),
\begin{eqnarray*}
\hat{\R}_{\V k} & = & \frac{1}{2\pi\eta}\int_{k_{z}=-\infty}^{\infty}\frac{dk_{z}}{\left(k^{2}+k_{z}^{2}\right)}\exp\left(-\frac{a^{2}\left(k^{2}+k_{z}^{2}\right)}{\pi}\right)\left(\M I-\frac{\left(\V k,k_{z}\right)\otimes\left(\V k,k_{z}\right)}{\left(k^{2}+k_{z}^{2}\right)}\right).
\end{eqnarray*}

Performing the integrals gives
\begin{eqnarray}
c_{1}\left(K\right) & = & \frac{1}{2\pi}\left(-K\,\exp\left(-\frac{K^{2}}{\pi}\right)-\left(K^{2}+\frac{\pi}{2}\right)\left({\rm erf}\left(\frac{K}{\sqrt{\pi}}\right)-1\right)\right)\label{eq:c12_q2D}\\
c_{2}\left(K\right) & = & \frac{1}{2}\,\left(1-{\rm erf}\left(\frac{K}{\sqrt{\pi}}\right)\right)\nonumber 
\end{eqnarray}
in Quasi2D. An important point is that both expression decay exponentially
like $\exp\left(-a^{2}k^{2}\right)$ in Fourier space, which is crucial
for pseudospectral methods to obtain spectral accuracy. For an unbounded
domain, the FCM correlation tensor $\R\left(\V r\right)$ can be computed
in real space from the expressions given in \cite{ForceCoupling_Monopole}
and has the form (\ref{eq:R_real_space}) with the functions
\begin{eqnarray*}
f(r) & = & \frac{1}{8\pi\eta\,r}\left(\left(1+2\,\frac{a^{2}}{\pi\,r^{2}}\right){\rm erf}\left(\frac{r\sqrt{\pi}}{2a}\right)-2\frac{a}{\pi\,r}\exp\left(-\frac{\pi\,r^{2}}{4a^{2}}\right)\right),\\
g(r) & = & \frac{1}{8\pi\eta\,r}\left(\left(1-6\,\frac{a^{2}}{\pi\,r^{2}}\right){\rm erf}\left(\frac{r\sqrt{\pi}}{2a}\right)+6\frac{a}{\pi\,r}\exp\left(-\frac{\pi\,r^{2}}{4a^{2}}\right)\right).
\end{eqnarray*}
In agreement with (\ref{eq:q2D_small_k}), for small wavenumbers we
have
\begin{eqnarray*}
c_{1}\left(K=ka\ll1\right) & \approx & \frac{1}{4}\\
c_{2}\left(K=ka\ll1\right) & \approx & \frac{1}{2}.
\end{eqnarray*}

We can unify True2D and Quasi2D by defining
\begin{eqnarray}
c_{1}\left(K\right) & = & 0\label{eq:c12_t2D}\\
c_{2}\left(K\right) & = & \frac{a}{K}\,\exp\left(-\frac{K^{2}}{\pi}\right),\nonumber 
\end{eqnarray}
for True2D. We have chosen to keep the standard deviation of the Gaussian
kernel $\delta_{a}$ the same as in True3D, since this empirically
reproduces the diffusion coefficient of a no-slip disk of radius $a$,
see (\ref{eq:f_def}). Note that in True2D we have $c_{2}\left(ka\ll1\right)\approx1/k$.
We can also generically write the short-time self-diffusion coefficient
as
\begin{equation}
\chi_{0}=\left(k_{B}T\right)f(r=0)=\beta\,\frac{k_{B}T}{\eta},\label{eq:chi_self_short_generic}
\end{equation}
where the factor $\beta$ is \cite{StokesEinstein}
\begin{eqnarray}
\beta & = & \frac{1}{6\pi a}\cdot\frac{1}{1+4.41a/L}\approx\frac{1}{6\pi a}\quad\mbox{for Quasi2D, and}\label{eq:f_def}\\
\beta & = & \frac{1}{4\pi}\ln\left(\frac{L}{3.71a}\right)\quad\mbox{for True2D,}\nonumber 
\end{eqnarray}
and $L$ is the size of the square periodic unit cell. We have included
for Quasi2D the leading-order finite-size correction, where we estimated
the coefficient $4.41$ numerically. Observe that in True2D the diffusion
coefficient diverges logarithmically with system size, and the empirical
coefficient $3.71$ matches that for a no-slip disk of radius $a$
\cite{Mobility2D_Hasimoto}, although we have not been able to confirm
this analytically.

From (\ref{eq:R_Fourier}) we get for the flow divergence
\[
c_{0}\left(k_{B}T\right)\V k\cdot\hat{\R}_{\V k}\cdot\V k=c_{0}\left(\frac{k_{B}T}{\eta}\right)kc_{1}(ak)=6\pi\chi ac_{0}kc_{1}(ak)=\left(\chi k^{2}\right)\frac{4c_{1}\left(ka\right)}{kL_{h}},
\]
giving the collective diffusion coefficient 
\begin{equation}
\chi_{c}(k)=\chi\left(1+\frac{4c_{1}\left(ka\right)}{kL_{h}}\right).\label{eq:D_c_full}
\end{equation}
This formula extends (\ref{eq:chi_small_k}) to all wavenumbers for
the FCM kernel.

\subsection{\label{subsec:BD-2D}Efficient Two-Dimensional Brownian Dynamics}

Equation (\ref{eq:dq_dt}) is the basis of our efficient BD-2D algorithm,
summarized in Algorithm \ref{alg:BD-2D}. We now explain the key ideas
behind the steps of the algorithm. This algorithm is now part of the
public-domain GPU code \emph{fluam}, available freely at \url{https://github.com/fbusabiaga/fluam}.

The second term in (\ref{eq:dq_dt}) can be captured by first spreading
to the grid the force density $\V F_{j}\delta_{a}+\left(k_{B}T\right)\V{\partial}\delta_{a}$
localized around each particle $j$, and then solving the fluid equations
in Fourier space by performing the convolution with $\M{\Set G}$
in (\ref{eq:dq_dt}) using FFTs. Convolution by $\M{\Set G}$ amounts
to multiplication by $\hat{\M{\Set G}}_{\V k}$ in Fourier space.
In Fourier space, the FCM representation $\M{\Set G}$ is given by
(\ref{eq:R_Fourier}) but now without the Gaussian ``delta'' functions
in $x$ and $y$ since these are added by the explicit spatial convolution
in (\ref{eq:dq_dt}). Specifically, in Quasi2D
\begin{eqnarray}
\hat{\M{\Set G}}_{\V k} & = & \frac{1}{2\pi\eta}\int_{k_{z}=-\infty}^{\infty}\frac{dk_{z}}{\left(k^{2}+k_{z}^{2}\right)}\exp\left(-\frac{a^{2}k_{z}^{2}}{\pi}\right)\left(\M I-\frac{\left(\V k,k_{z}\right)\otimes\left(\V k,k_{z}\right)}{\left(k^{2}+k_{z}^{2}\right)}\right)=\nonumber \\
 & = & \hat{\R}_{\V k}\exp\left(\frac{a^{2}k^{2}}{\pi}\right)=\frac{1}{\eta}\left[g_{k}\left(ka\right)\,\V k_{\perp}\otimes\V k_{\perp}+f_{k}\left(ka\right)\,\V k\otimes\V k\right],\label{eq:G_hat_k}
\end{eqnarray}
where the explicit form of $f_{k}$ and $g_{k}$ can be read from
(\ref{eq:R_Fourier}). It is important to observe that the convolution
by a Gaussian in the $z$ direction is included in the integral above
since our FFTs and grid are purely two dimensional. In True2D, the
Green's function is the familiar Green's function for 2D Stokes flow,
given by $f_{k}=0$ and $g_{k}=1/\left(\eta k^{2}\right)$.

The first term in (\ref{eq:dq_dt}) represents advection by a random
velocity. This velocity can be generated in Fourier space using (\ref{eq:w_hat}).
From (\ref{eq:R_Fourier}) we get the Fourier representation required
to generate the random velocity efficiently,
\begin{equation}
\widehat{\V w}_{\V k}=\sqrt{\frac{2k_{B}T}{\eta k^{3}}}\left(\sqrt{c_{2}\left(ka\right)}\,\V k_{\perp}\mathcal{Z}_{\V k}^{(2)}+\sqrt{c_{1}\left(ka\right)}\,\V k\mathcal{Z}_{\V k}^{(1)}\right),\label{eq:w_hat}
\end{equation}
where $\mathcal{Z}_{\V k}^{(1/2)}\left(t\right)$ are scalar white
noise processes, independent of each other and independent between
different wavenumbers. The random Gaussian variables $\mathcal{Z}_{\V k}^{(1/2)}$
should obey the symmetry properties required to ensure that the velocity
field is real-valued in real space; see detailed discussion in Section
3.3 in \cite{FBIM}. Note that the same formula (\ref{eq:w_hat})
works for True2D and Quasi2D using the appropriate definitions of
$c_{1}$ and $c_{2}$. In our algorithm the Gaussian factor $\exp\left(-a^{2}k^{2}/\pi\right)$
is included when interpolating the random velocity at the particle
positions, so in (\ref{eq:v_stoch}) in the algorithm the random velocity
is generated using the factors $f_{k}/g_{k}$ and not the factors
$c_{1}/c_{2}$.

\begin{algorithm}
\begin{enumerate}
\item Evaluate particle forces $\V F^{n}=\V F\left(\V Q^{n}\right)$.
\item Compute in real space on a grid the fluid forcing
\[
\V f\left(\V r\right)=\sum_{i}\V F_{i}\delta_{a}\left(\V q_{i}-\V r\right)+\left(k_{B}T\right)\sum_{i}\left(\V{\partial}\delta_{a}\right)\left(\V q_{i}-\V r\right)+\V f_{\text{ext}}\left(\V r\right),
\]
and use the FFT to convert $\V f$ to Fourier space, i.e., to compute
$\hat{\V f}_{\V k}$. For True2D we omit the second term since the
hydrodynamic tensor is divergence free. Here $\V f_{\text{ext}}\left(\V r\right)$
is an external force on the fluid (e.g., gravity), which can be used
to perform non-equilibrium measurements.
\item Compute the fluid velocity resulting from the fluid forcing $\V f$
in Fourier space as a convolution with the Green's function (\ref{eq:G_hat_k}),
\[
\hat{\V v}_{k}^{\text{det}}=\hat{\M{\Set G}}_{\V k}\hat{\V f}_{\V k}.
\]
\item Generate a random fluid velocity with covariance $\left(2k_{B}T/\D t\right)\hat{\M{\Set G}}_{\V k}$
in Fourier space,
\begin{equation}
\hat{\V v}_{k}^{\text{stoch}}=\sqrt{\frac{2k_{B}T}{\eta\D t}}\left(\sqrt{g_{k}\left(k\right)}\,\V k_{\perp}W_{\V k}^{(2)}+\sqrt{f_{k}\left(k\right)}\,\V kW_{\V k}^{(1)}\right),\label{eq:v_stoch}
\end{equation}
where $W_{\V k}^{(1/2)}$ are standard random Gaussian variables with
the symmetry properties required to make $\V v^{\text{stoch}}\left(\V r\right)$
real valued.
\item Compute the total fluid velocity
\[
\hat{\V v}_{k}=\hat{\V v}_{k}^{\text{det}}+\hat{\V v}_{k}^{\text{stoch}},
\]
and use the FFT to convert it to real space and obtain $\V v\left(\V r\right)$.
\item Convolve $\V v\left(\V r\right)$ with a Gaussian in real space to
compute particle velocities,
\[
\V u_{i}=\int\delta_{a}\left(\V q_{i}-\V r\right)\V v\left(\V r\right)d\V r.
\]
\item Advance the particles' positions,
\[
\V q_{i}^{n+1}=\V q_{i}^{n}+\V u_{i}\D t.
\]
\end{enumerate}
\caption{\label{alg:BD-2D}Outline of the $n$-th time step in an efficient
algorithm for performing BD-HI in two dimensions (BD-2D). The time
step size is denoted with $\protect\D t$. Although we employ continuum
notation for simplicity, the actual implementation uses a grid to
discretize the fields, and uses the trapezoidal rule to discretize
the spatial integrals. This algorithm works for \emph{any} translationally
invariant isotropic hydrodynamic interaction kernel.}
\end{algorithm}

The grid used to perform the FFTs in Algorithm \ref{alg:BD-2D} must
be sufficiently fine to accurately resolve the convolutions with the
Gaussian kernel $\exp\left(-a^{2}k^{2}/\pi\right)$ and to capture
all of the active modes in the fluid velocity. We have numerically
estimated that the grid spacing $h$ should satisfy $h\lesssim0.8a$
to ensure a relative error smaller than $10^{-4}$ in the short-time
self diffusion coefficient; the precise value of $h$ should be chosen
to ensure that the number of grid cells factorizes favorably to speed
up the FFT. The Gaussian kernels need to be cut off in real space
after a distance $r_{c}$ to ensure that the kernel $\delta_{a}$
is compactly supported \cite{NUFFT}. We have numerically estimated
that $r_{c}\geq2.5a$ is appropriate for a relative error tolerance
of $10^{-4}$. This cutoff can be efficiently implemented by using
a mask of $\left(2P+1\right)\times\left(2P+1\right)$ cells when spreading
the forces to the grid and interpolating the velocity from the grid,
where $P=4$ for $h=0.8a$.

The temporal integration scheme in Algorithm \ref{alg:BD-2D} is similar
to that used in the FIB method \cite{BrownianBlobs}; see in particular
the simple midpoint scheme presented in section IV.A of \cite{BrownianBlobs}.
We have, however, made several simplifications using the fact that
to high numerical accuracy the hydrodynamic kernel is translationally-invariant,
as well as the fact that we can analytically differentiate Gaussian
kernels to compute $\V{\partial}\delta_{a}$ without requiring a so-called
random finite difference \cite{BrownianBlobs}. A delicate point is
choosing the appropriate time step size $\D t$. A demanding test
that can be used to assess the magnitude of the temporal integration
error is to examine the equilibrium pair correlation function $g_{2}\left(r\right)$,
or, equivalently, the equilibrium structure factor $S(k)$. For interacting
particles the time step size is typically limited by the stiffness
of the steric repulsion; however, even for non-interacting particles
(ideal gas) the time step must be small enough to ensure that particles
move only a fraction of their hydrodynamic radius in a time step,
$\chi\D t\ll a^{2}$. This is the only requirement for an ideal gas
in True2D, however, in Quasi2D the particles ``interact'' through
the thermal fluctuations they induce in the fluid via their osmotic
pressure. We empirically observe that in Quasi2D, the higher the density,
and thus the stronger the collective effects, the smaller the time
step must be in order to maintain the accuracy of $g_{2}\left(r\right)$.
The largest error in $g_{2}$ was found to be for $r\rightarrow0^{+}$.

Most of the simulations reported in this work employ one of two setups,
Setup A for planar packing density $\phi=0.5$ or Setup B for $\phi=1$.
The relevant parameters are given in Table \ref{tab:Setups}.

\begin{table}
\begin{centering}
\begin{tabular}{|c|c|c|c|c|c|}
\hline 
Setup &
$\phi$ &
$L$ &
$N$ &
$k_{B}T$, $\eta$ and $a$ &
$N_{\text{FFT}}^{2}$\tabularnewline
\hline 
\hline 
A &
0.5 &
800 &
$10^{5}$ &
1 &
$1152^{2}$\tabularnewline
\hline 
B &
1.0 &
560.5 &
$10^{5}$ &
1 &
$864^{2}$\tabularnewline
\hline 
\end{tabular}
\par\end{centering}
\caption{\label{tab:Setups}Parameters used for the majority of the simulations
reported in this work, unless otherwise indicated, in arbitrary units.
All simulations use a periodic domain of length $L_{x}=L_{y}=L$ with
a total of $N$ particles covered by an FFT grid of size $N_{\text{FFT}}^{2}$
cells. In \textbf{Setup A}, we set $L=800$ giving a packing density
(fraction) $\phi=c_{0}\pi a^{2}\approx0.5$, where $c_{0}=N/(L_{x}L_{y})$
is the equilibrium number density (concentration). In \textbf{Setup
B}, we set $L=560.5$ to get a higher packing density $\phi=1.0$
and thus stronger collective effects. Unless otherwise noted, in Quasi2D,
we use a time step size $\protect\D t=0.2$, and in True2D we use
$\protect\D t=0.05$ because the diffusion coefficient is larger;
these values were chosen to ensure an error less than 10\% in $g_{2}\left(r\rightarrow0^{+}\right)$
at equilibrium, even for the larger density in Setup B.}
\end{table}

\subsection{\label{subsec:LinearResponse}Linear response theory}

One of the key properties of the dynamics (\ref{eq:BD_M}) is that
it is in detailed balance (i.e., it is time reversible) with respect
to an ideal gas equilibrium distribution. The thermodynamic equilibrium
state of the suspension is consistent with that of an ideal gas. We
have verified that our numerical method satisfies this property by
confirming that the pair correlation function $g_{2}\left(r\right)=1$
for sufficiently small time step sizes (not shown). In order to confirm
that our algorithm preserves the time-reversibility of the dynamics
we now test a key prediction of linear response theory (LRT) \cite{Frenkel_Smit_book},
which is based on time reversibility at equilibrium.

\begin{figure}
\centering{}\includegraphics[width=0.75\textwidth]{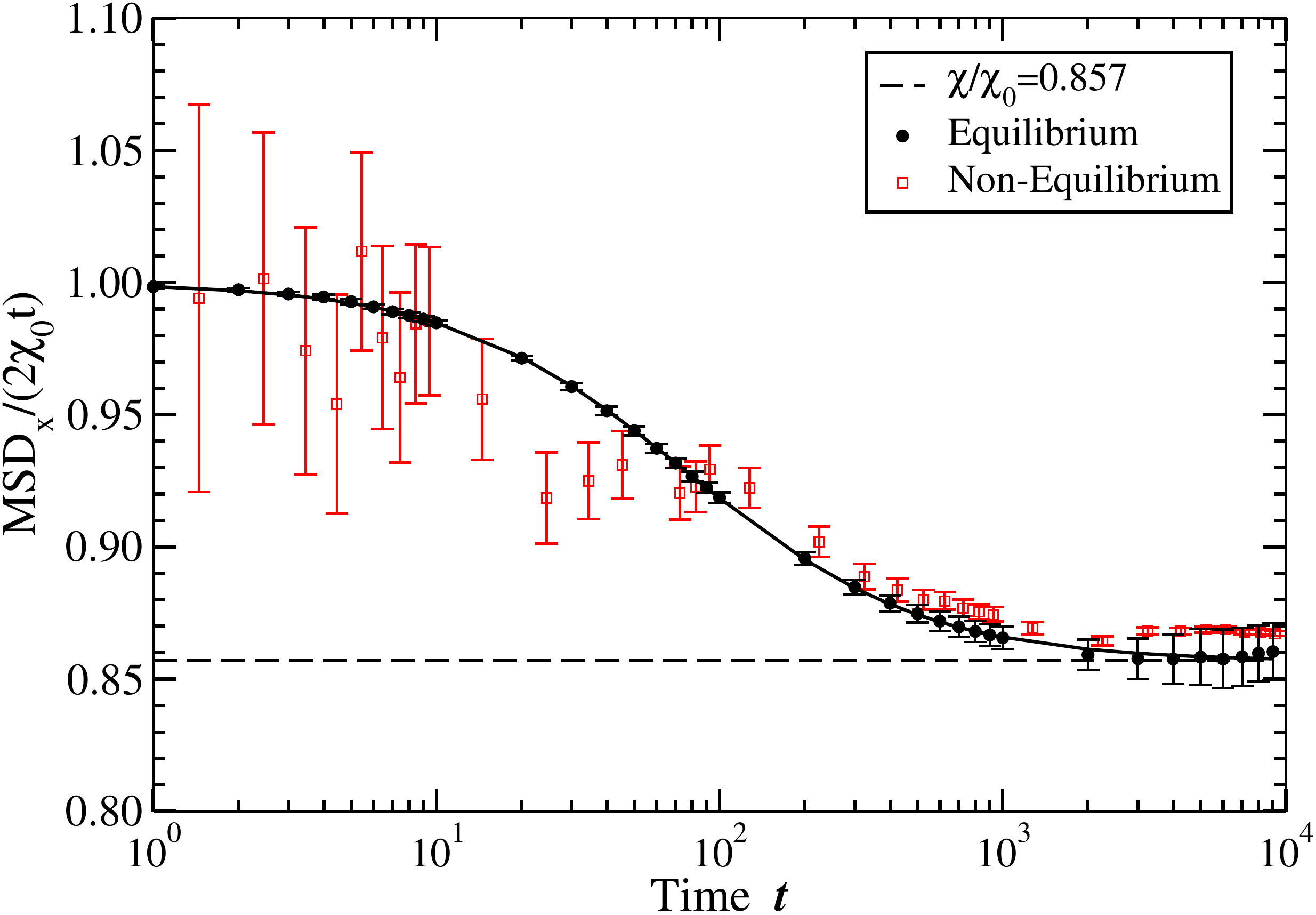}\caption{\label{fig:SelfDiff_LRT}Mean square displacement of a tagged particle
in a Quasi2D ideal gas of density $\phi=1$, computed from the equilibrium
MSD (filled black circles), as well as the nonequilibrium average
displacement of a particle pulled by a force $F=1$ (empty squares).
Error bars indicate one standard deviation confidence intervals. Note
that the statistical errors are quite large for the nonequilibrium
measurements, especially at short times, because only a single particle
is pulled per simulation, whereas the MSD is averaged over all particles.
Parameters are based on Setup B (see Table \ref{tab:Setups}) except
that $\protect\D t=0.1$ to better capture the short time behavior
and reduce the error in $g_{2}\left(r\right)$.}
\end{figure}

Specifically, we analyze the mean square displacement (MSD) of a tagged
particle $p$,
\[
\text{MSD}(t)=\text{MSD}_{x}(t)+\text{MSD}_{y}(t)=\left\langle (x_{p}(t)-x_{p}(0))^{2}\right\rangle +\left\langle (y_{p}(t)-y_{p}(0))^{2}\right\rangle .
\]
This can be computed at equilibrium by averaging the square displacements
over all of the particles. It can also be computed using a non-equilibrium
method based on LRT. If we pull a tagged particle $p$ with a force
$\V F=F\,\hat{\V x}$, LRT predicts that
\begin{equation}
\left\langle x_{p}(t)-x_{p}(0)\right\rangle _{\V F}=-\frac{F}{k_{B}T}\int_{0}^{t}\left\langle x_{p}(0)\dot{x}_{p}(t-t')\right\rangle _{0}\mathrm{d}t'=\frac{F}{2k_{B}T}\left\langle (x_{p}(t)-x_{p}(0))^{2}\right\rangle _{0}.\label{eq:noneq-MSD}
\end{equation}
The average on the left hand side is an average over nonequilibrum
trajectories initialized from the equilibrium distribution, while
the average on the right hand side is an average over equilibrium
trajectories in the absence of the forcing. The formula (\ref{eq:noneq-MSD})
relates the MSD at equilibrium with the mean displacement under a
external force. In the left panel of Fig. \ref{fig:SelfDiff_LRT}
we confirm that this relation is maintained by our Quasi2D BD-HI (BD-q2D
for short) algorithm, confirming that the algorithm preserves detailed
balance (time reversibility).

More interestingly, we observe that the MSD is \emph{not} strictly
linear in time. While the short-time slope of the MSD is consistent
with the short-time self-diffusion coefficient $\chi_{s}^{(s)}=\chi_{0}$
predicted by the Stokes-Einstein relation (\ref{eq:chi_self_short_generic}),
the long-time slope gives a smaller long-time self-diffusion coefficient
$\chi_{s}^{(l)}\approx0.85\,\chi_{0}$ (for this packing density $\phi=1$)
\footnote{It is worth noting that the equality $\chi_{s}^{(s)}\left(\phi\right)=\chi_{0}=\lim_{\phi\rightarrow0}\chi_{s}^{(l)}\left(\phi\right)$
is built into the Rotne-Prager or pairwise approximation of the mobility
we use in this work. When realistic near-field hydrodynamics is accounted
for, $\chi_{s}^{(s)}\left(\phi\right)$ will in general be different
from $\chi_{0}$ for nonvanishing densities.}. We return to this unexpected effect in greater detail in Section
\ref{sec:SelfDiffusion}.

\section{\label{sec:EquationMean}Dynamics of the Ensemble Average}

In this section we study the evolution of the ensemble average or
mean of the total density \cite{DDFT_Diffusion_2D} and the density
of a labeled species (tracer) in Quasi2D diffusion. While for True2D
hydrodynamics the mean strictly follows the familiar Fick law (diffusion
equation), as it would in the absence of hydrodynamics (i.e., for
uncorrelated random walkers), this does \emph{not} carry on to Quasi2D
systems.

\subsection{Ensemble average for density}

As a nonlinear SPDE, (\ref{eq:c_fluct_general}) does not really make
sense mathematically, and it can be thought of simply as a rewriting
of (\ref{eq:BD_M}). To obtain an ensemble average we need to average
over realizations of $\V w$ (i.e., over stochastic realizations),
but this does \emph{not} give a closed equation and we must also introduce
the unknown two-point correlation function $c^{(2)}\left(\V r,\V r^{\prime},t\right)$.
Specifically, ensemble averaging (\ref{eq:c_fluct_general}) gives
\cite{DDFT_Hydro_Lowen,DDFT_Hydro}
\begin{eqnarray}
\partial_{t}c^{(1)}\left(\V r,t\right) & = & \grad\cdot\left(\M{\chi}(\V r)\grad c^{(1)}\left(\V r,t\right)\right)+\left(k_{B}T\right)\grad\cdot\left(\int\R\left(\V r,\V r^{\prime}\right)\grad^{\prime}c^{(2)}\left(\V r,\V r^{\prime},t\right)\,d\V r^{\prime}\right),\label{eq:c_av_general}
\end{eqnarray}
where the single-particle distribution function (mean number density)
is $c^{(1)}\left(\V r,t\right)=\av{c\left(\V r,t\right)}$. For an
ideal gas, the standard closure for the two-particle correlation function
is \cite{DDFT_Lowen,DDFT_Hydro_Lowen}
\[
c^{(2)}\left(\V r,\V r^{\prime},t\right)\approx c^{(1)}\left(\V r,t\right)c^{(1)}\left(\V r^{\prime},t\right).
\]

After making this approximation in (\ref{eq:c_av_general}) we get
the \emph{closed} nonlinear nonlocal diffusion equation
\begin{equation}
\partial_{t}c^{(1)}\left(\V r,t\right)=\grad\cdot\left(\M{\chi}(\V r)\grad c^{(1)}\left(\V r,t\right)+\left(k_{B}T\right)c^{(1)}\left(\V r,t\right)\int\R\left(\V r,\V r^{\prime}\right)\grad^{\prime}c^{(1)}\left(\V r^{\prime},t\right)\,d\V r^{\prime}\right),\label{eq:c_av_closure}
\end{equation}
which can be solved efficiently and accurately using standard pseudospectral
methods. Note that this equation looks exactly like the fluctuating
(\ref{eq:c_fluct_general}) but without the noise term. This suggests
that this approximation can also be thought of as neglecting fluctuations.
It is important to note that in True2D or True3D (\ref{eq:c_av_closure})
reduces to the standard (linear) diffusion equation $\partial_{t}c^{(1)}=\chi\grad^{2}c^{(1)}$
after integration by parts, and is an \emph{exact} consequence of
(\ref{eq:BD_M}); divergence-free hydrodynamic interactions affect
the spectrum of the fluctuations but do not alter the mean \cite{DiffusionJSTAT}.
In Quasi2D, however, (\ref{eq:c_av_closure}) is only an approximation
(i.e., a closure) that must be tested numerically.

\begin{figure}
\begin{centering}
\emph{\includegraphics[width=0.5\textwidth]{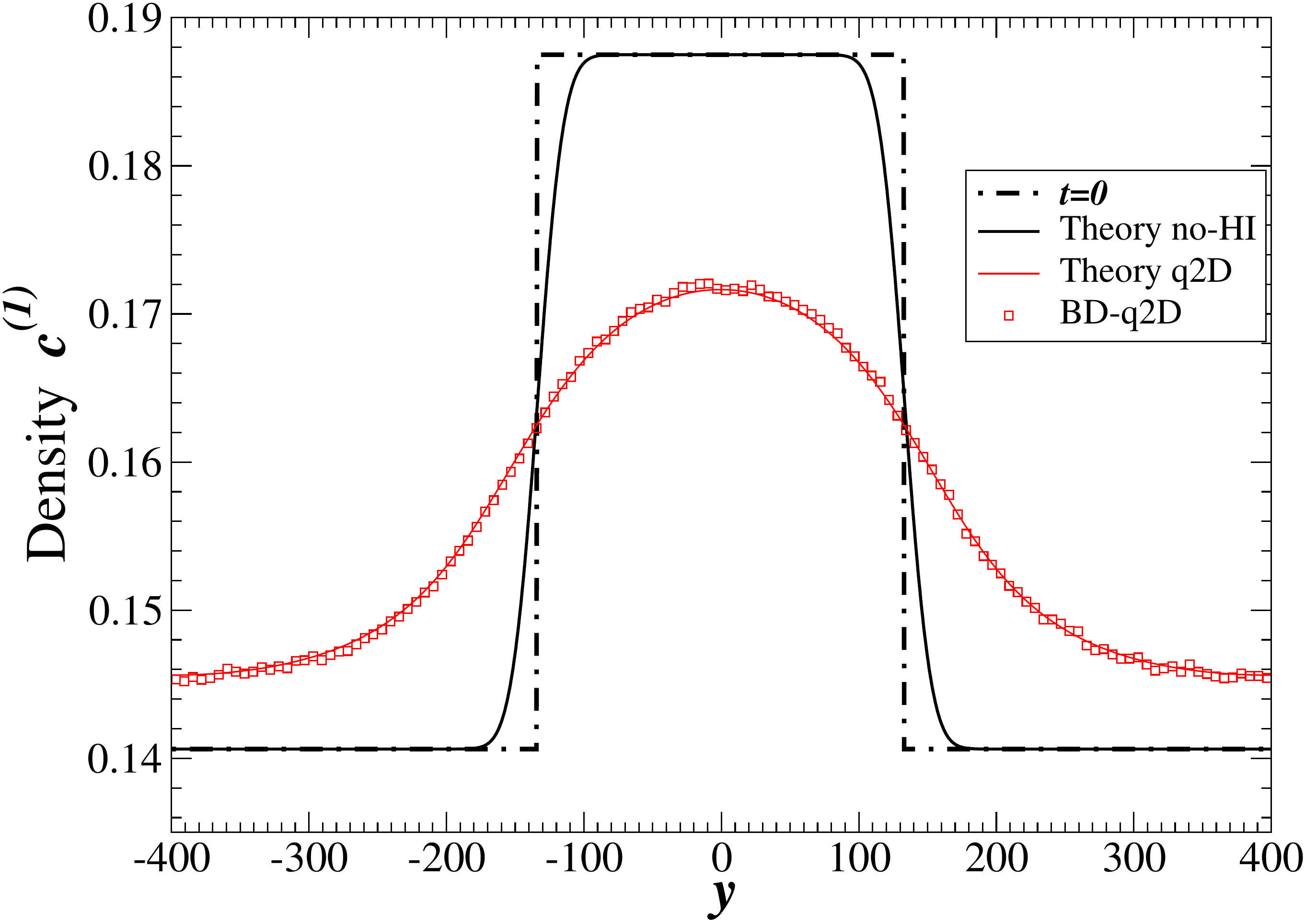}\includegraphics[width=0.5\textwidth]{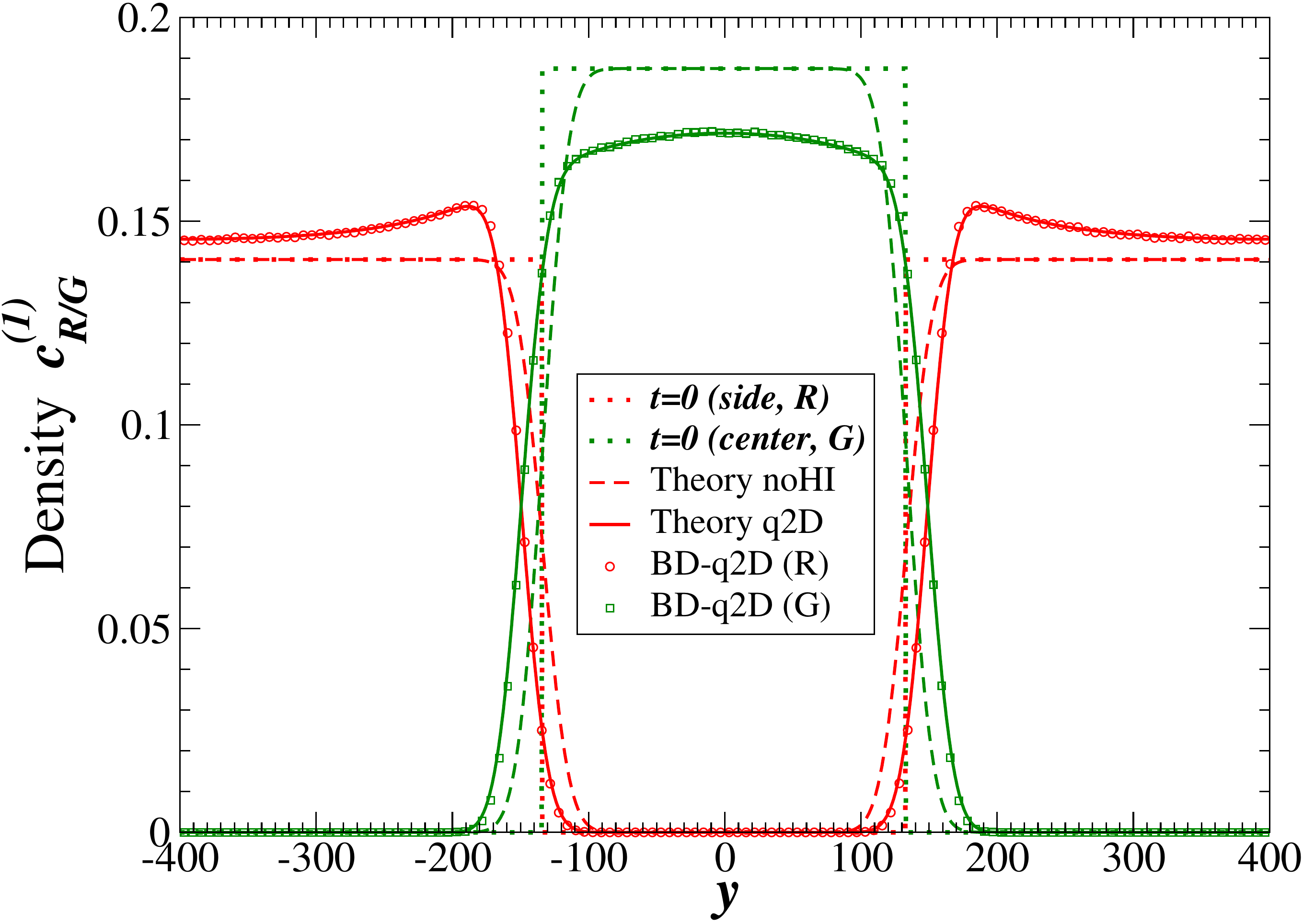}}
\par\end{centering}
\centering{}\caption{\label{fig:MeanDensityNoColor}Ensemble-averaged one-dimensional density
profiles illustrating the diffusion of a density jump in the $y$
coordinate. The density is initially higher in a stripe along the
$x$ axes covering the middle third of the domain. Parameters are
based on Setup A (see Table \ref{tab:Setups}). (Left) The total density
$c^{(1)}(y,t)$ at time $t=0$ (dashed-solid black line) and at time
$t=2100$ for BD-q2D (red squares). Numerical solutions of the mean-field
equation (\ref{eq:c_av_closure}) are shown with a solid red line.
The solution of the standard diffusion equation, which applies without
hydrodynamic interactions, is shown with a solid black line. (Right)
Density of ``red'' particles $c_{R}^{(1)}(y,t)$ (red lines and
red circles) and ``green'' particles $c_{G}^{(1)}(y,t)$ (green
lines and green squares) at time $t=0$ (dotted lines) and at $t=2100$
for BD-q2D (symbols). The numerical solution of (\ref{eq:c_av_RG})
is shown with a solid line of the same color as the corresponding
symbols. For comparison we also show the exact solution without hydrodynamics
with a dashed line of the same color.}
\end{figure}

We compare results from BD-HI simulations with a numerical solution
of (\ref{eq:c_av_closure}) in the left panel of Fig. \ref{fig:MeanDensityNoColor}.
In this numerical experiment, we use parameters from Setup A (see
Table \ref{tab:Setups}), $\phi=0.5$, and follow the dynamics up
to time $T=2100$. We average the numerical results over one thousand
simulations. In the initial configuration, we randomly and uniformly
distribute $0.9\cdot10^{5}$ particles throughout the domain, and
then randomly and uniformly distribute an additional $0.1\cdot10^{5}$
particles in the stripe $-L_{y}/6\leq y\leq L_{y}/6$. This creates
a one-dimensional density gradient along the $y$ direction, see black
dashed line in the left panel of Fig. \ref{fig:MeanDensityNoColor};
see Fig. \ref{fig:InstanceDensity} for a visual representation in
a similar setup. In what follows we study the density profile averaged
along the $x$ axis, i.e., we study the ensemble average $c^{(1)}(y,t)=\av{c(y,t)}$
of $c(y,t)=\int_{x=0}^{L_{x}}c(x,y,t)dx$.

In the left panel of Fig. \ref{fig:MeanDensityNoColor} we compare
results for $c^{(1)}(y,T)$ for a system without hydrodynamic interactions
(black solid line) and a system with Quasi2D hydrodynamics (red squares,
BD-q2D) to the numerical solution of (\ref{eq:c_av_closure}) in Quasi2D
(red solid line). Note that the case without hydrodynamics is \emph{exactly}
described by the standard diffusion equation. The good agreement between
the numerical results for BD-q2D and the theoretical predictions suggests
that the mean-field approximation (\ref{eq:c_av_closure}) is reasonable.
When solving (\ref{eq:c_av_closure}) we used $\chi=\chi_{s}^{(s)}=\chi_{0}$
and not the long-time self diffusion coefficient $\chi_{s}^{(l)}\approx0.88\chi_{0}$
(at this density $\phi=0.5$, see Fig. \ref{fig:SelfDiff_LRT}); however,
the difference between the two values is too small to appreciate on
the scales of Fig. \ref{fig:MeanDensityNoColor} because the nonlinear
term in (\ref{eq:c_av_closure}) dominates.

\emph{}

\subsection{Ensemble average for color}

As already studied in detail in prior work \cite{ConfinedDiffusion_2D,DDFT_Diffusion_2D},
the evolution of density perturbations in Quasi2D is very different
from the familiar diffusive decay, not just in the time domain but
also in the space domain. In particular, density perturbations spread
much further in Quasi2D and develop inverse cubed power-law tails
instead of the familiar Gaussian tails. Such an algebraic spatial
decay of localized density perturbations is quite unexpected and unusual,
and raises some questions. Do some/all/few particles quickly displace
very far, so as to create a power-law tail for $t>0$? Or, do all
particles still displace by small diffusive displacements, but a power-law
tail arises in the average due to the inter-particle correlations?

To further interrogate the individual particle displacements we color
(label) the particles ``red'' and ``green''. This could be done
in experiments by using fluorescent labels. The generalization of
(\ref{eq:c_av_closure}) to account for species labels is straightforward
to write down and takes the form of two coupled nonlinear nonlocal
diffusion equations,
\begin{eqnarray}
\partial_{t}c_{R/G}^{(1)}\left(\V r,t\right) & = & \grad\cdot\left(\chi\grad c_{R/G}^{(1)}\left(\V r,t\right)+\left(k_{B}T\right)c_{R/G}^{(1)}\left(\V r,t\right)\int\R\left(\V r,\V r^{\prime}\right)\grad^{\prime}c^{(1)}\left(\V r^{\prime},t\right)\,d\V r^{\prime}\right),\label{eq:c_av_RG}
\end{eqnarray}
where the subscript R/G labels the species by color and $c^{(1)}=c_{R}^{(1)}+c_{G}^{(1)}$.
If we add these two equations we get back equation (\ref{eq:c_av_closure}),
as we must.

\begin{figure}
\begin{centering}
\includegraphics[width=0.5\textwidth]{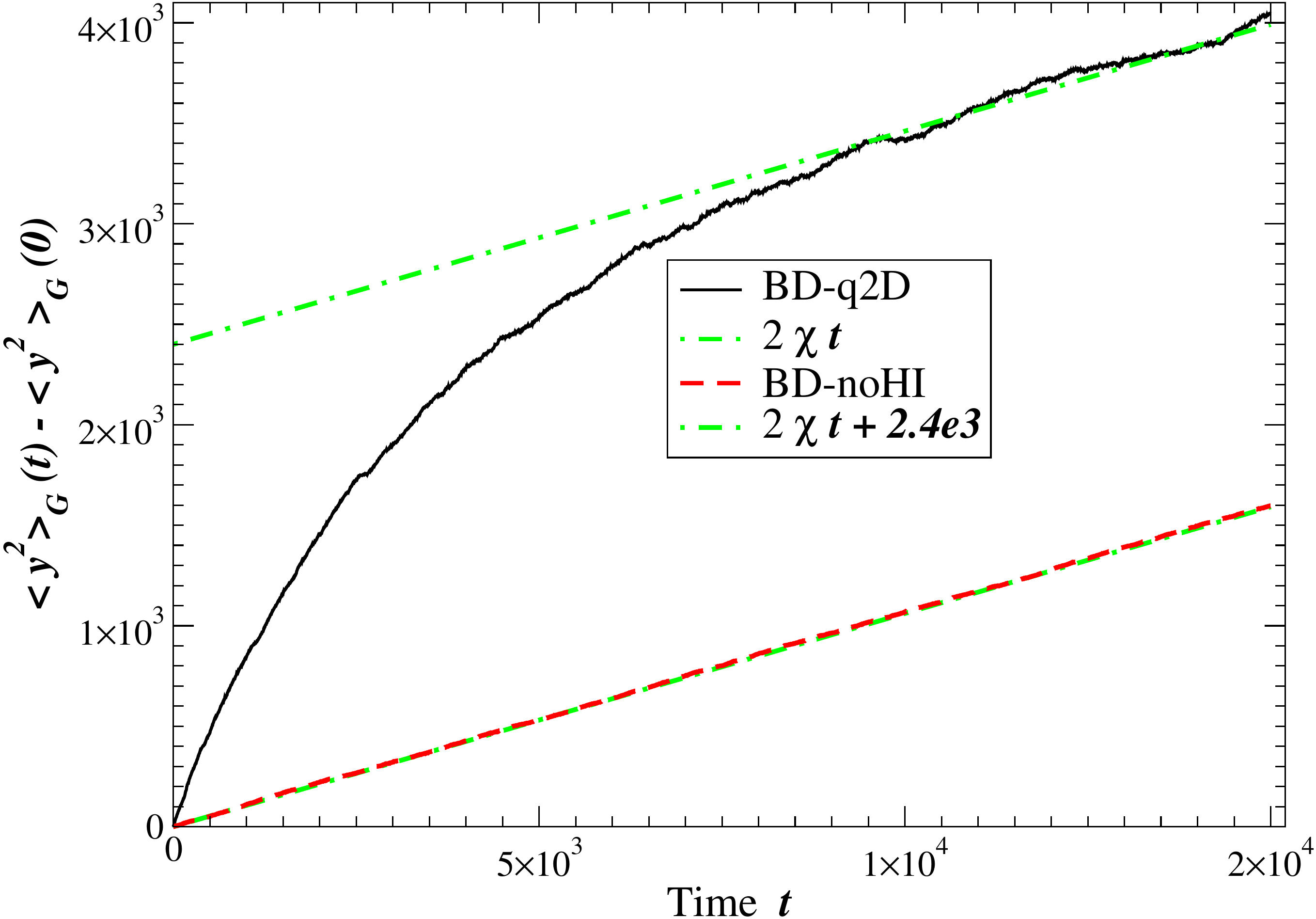}\includegraphics[width=0.5\textwidth]{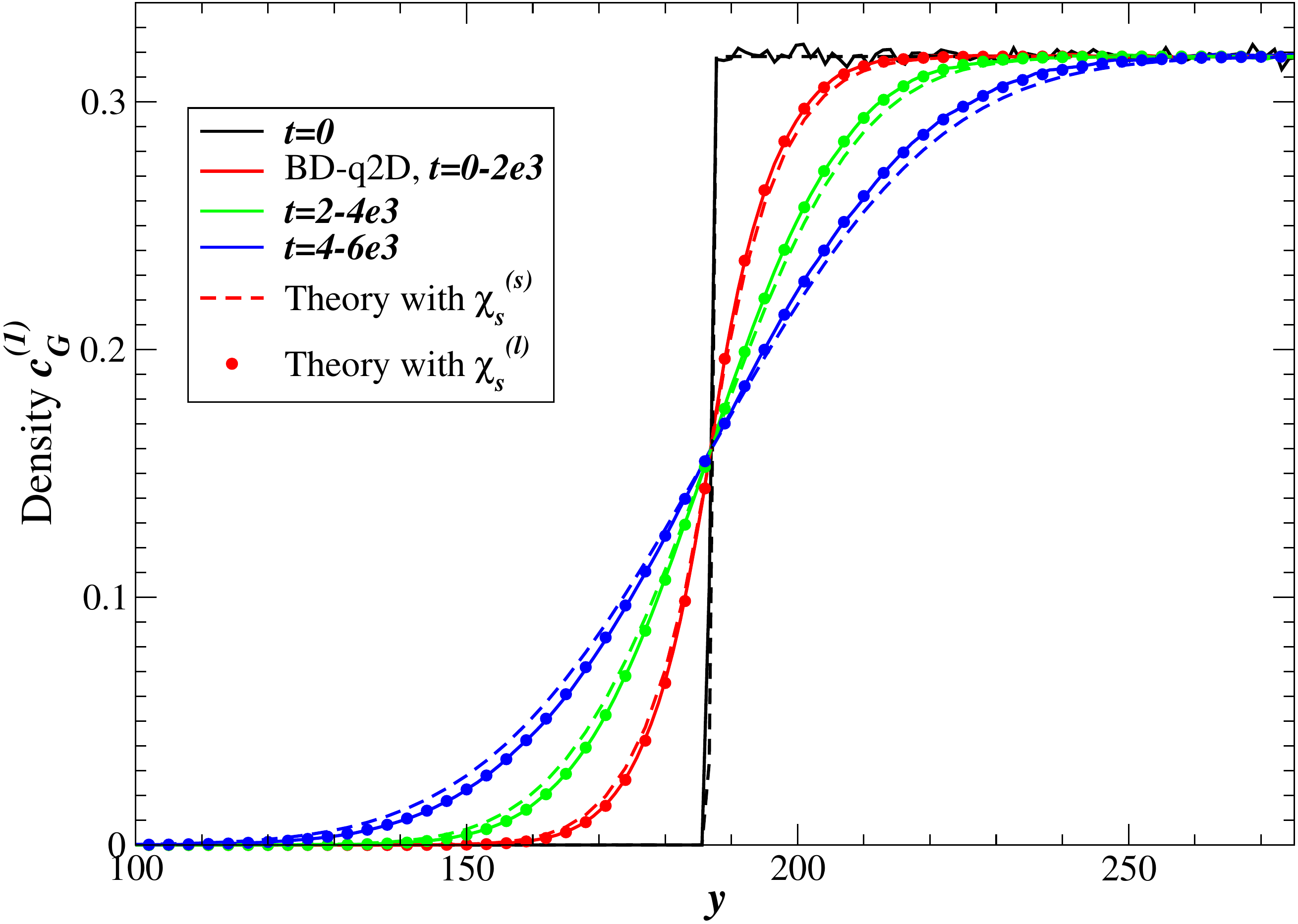}
\par\end{centering}
\centering{}\caption{\label{fig:MeanDensityColor}(Left) Mean squared displacement $\protect\av{y^{2}}_{G}\left(t\right)-\protect\av{y^{2}}_{G}\left(0\right)$
(see (\ref{eq:y2_G})) along the $y$ axis for the ``green'' particles,
i.e., the ones initially at a higher density in the middle third stripe
of the domain, for the same simulations used to produce Fig. \ref{fig:MeanDensityNoColor}.
Without hydrodynamics the MSD follows the familiar diffusive evolution
$2\chi t$ at all times. For Quasi2D the MSD initially grows much
more rapidly due to the collective interactions, but eventually settles
to the expected diffusive growth at long times. (Right) Ensemble-averaged
density of ``green'' particles $c_{G}^{(1)}(y)$ illustrating the
diffusive mixing of red and green particles in Quasi2D, starting from
a uniform density with a perfectly sharp red-green interface at $t=0$
(black lines). Parameters are based on Setup B (see Table \ref{tab:Setups}).
The density is averaged over time intervals $0<t\protect\leq2\cdot10^{3}$
(red lines), $2\cdot10^{3}<t\protect\leq4\cdot10^{3}$ (green lines)
and $4\cdot10^{3}<t\protect\leq6\cdot10^{3}$ (blue lines). Theoretical
results based on the analytical solution of the standard diffusion
equation are shown with dashed lines for $\chi=\chi_{0}$, and with
solid circles for $\chi=\chi_{s}^{(l)}\approx0.85\chi_{0}$.}
\end{figure}

For the setup studied in the left panel of Fig. \ref{fig:MeanDensityNoColor},
we color all of the particles in the central stripe $-L_{y}/6\leq y\leq L_{y}/6$
green, and color the remaining particles red. We compare ensemble-averaged
results from BD-q2D simulations with a numerical solution of (\ref{eq:c_av_RG})
in the right panel of Fig. \ref{fig:MeanDensityNoColor}. Without
hydrodynamics we get familiar diffusive mixing with Gaussian tails,
as illustrated with dashed lines in the figure. For Quasi2D hydrodynamics,
however, we see that the the red particles are pushed away by the
excess of green particles in the middle stripe. We find that it is
$c_{R}^{(1)}$ that shows the fat power-law $y^{-3}$ tails, while
$c_{G}^{(1)}$ has rapidly decaying Gaussian tails (not shown). This
can also be appreciated by computing the second moment
\begin{equation}
\av{y^{2}}_{G}\left(t\right)=\frac{\av{\int y^{2}c_{G}\left(y,t\right)dy}}{\av{\int c_{G}\left(y,t\right)dy}},\label{eq:y2_G}
\end{equation}
which is finite and grows linearly in time for long times both with
and without hydrodynamics, as shown in the left panel of Fig.\textbf{
}\ref{fig:MeanDensityColor}.

Taken together, these results demonstrate that particles still exhibit
only Gaussian diffusive displacements even in Quasi2D. Along the way,
however, they effectively repel nearby particles, thus creating density
waves that exhibit the unusual power-law behavior. Indeed, it has
already been observed in \cite{ConfinedDiffusion_2D,DDFT_Diffusion_2D}
that if one places a localized density perturbation in an otherwise
empty domain, the tails are Gaussian rather than power law, since
there are no particles to ``push'' away and thus create the density
wave. In particular, if we color particles in an otherwise uniform
suspension, $c^{(1)}(y)=c_{0}=\text{const.}$, the nonlinear convolution
term in (\ref{eq:c_av_RG}) disappears and one obtains a pair of \emph{uncoupled}
diffusion equations, despite the fact that red and green particles
still ``repel'' each other with the long-ranged $1/r$ hydrodynamic
interaction (\ref{eq:effective_repulsion}).

We explore this claim further in Fig.\textbf{ }\ref{fig:MeanDensityColor}
by performing numerical experiments using parameters from Setup B
(see Table \ref{tab:Setups}), $\phi=1$, except that we have used
a small time step size $\D t=0.01$ to minimize temporal integration
errors. See Fig. \ref{fig:InstanceColor} for a visual representation
of a single instance of the diffusive mixing. In the right panel of
Fig.\textbf{ }\ref{fig:MeanDensityColor} we show simulation results
for $c_{G}^{(1)}\left(y\right)$, averaged over 100 simulations and
additionally averaged over time intervals of duration $2\cdot10^{3}$
in order to improve the statistics. We compare BD-q2D (solid lines)
with the solution of an uncoupled system of two diffusion equations
(dashed lines), one for each species. For the theory, we have used
for the diffusion coefficient $\chi$ the short-time diffusion coefficient
$\chi_{s}^{(s)}=\chi_{0}$ given by (\ref{eq:chi_self_short_generic}).
The agreement between simulation and theory is significantly improved
if we use instead the long-time self diffusion coefficient (circles),
$\chi_{s}^{(l)}\approx0.85\chi_{0}$ at this density $\phi=1$, see
Fig. \ref{fig:SelfDiff_LRT} and Section \ref{sec:SelfDiffusion}.
This clearly demonstrates that the mean-field equations (\ref{eq:c_av_closure})
and (\ref{eq:c_av_RG}) are \emph{not exact} even for a Quasi2D ideal
gas of particles because collective density fluctuations play a non-negligible
role in the dynamics of the ensemble average as well. Note that incorporating
the time-dependent self diffusion coefficient observed in Fig. \ref{fig:SelfDiff_LRT}
in the DDFT equations would require introducing memory (i.e., non-Markovian
behavior), which is beyond the scope of this work.

\section{\label{sec:SelfDiffusion}Long-time self diffusion}

\begin{figure}
\centering{}\includegraphics[width=0.5\textwidth]{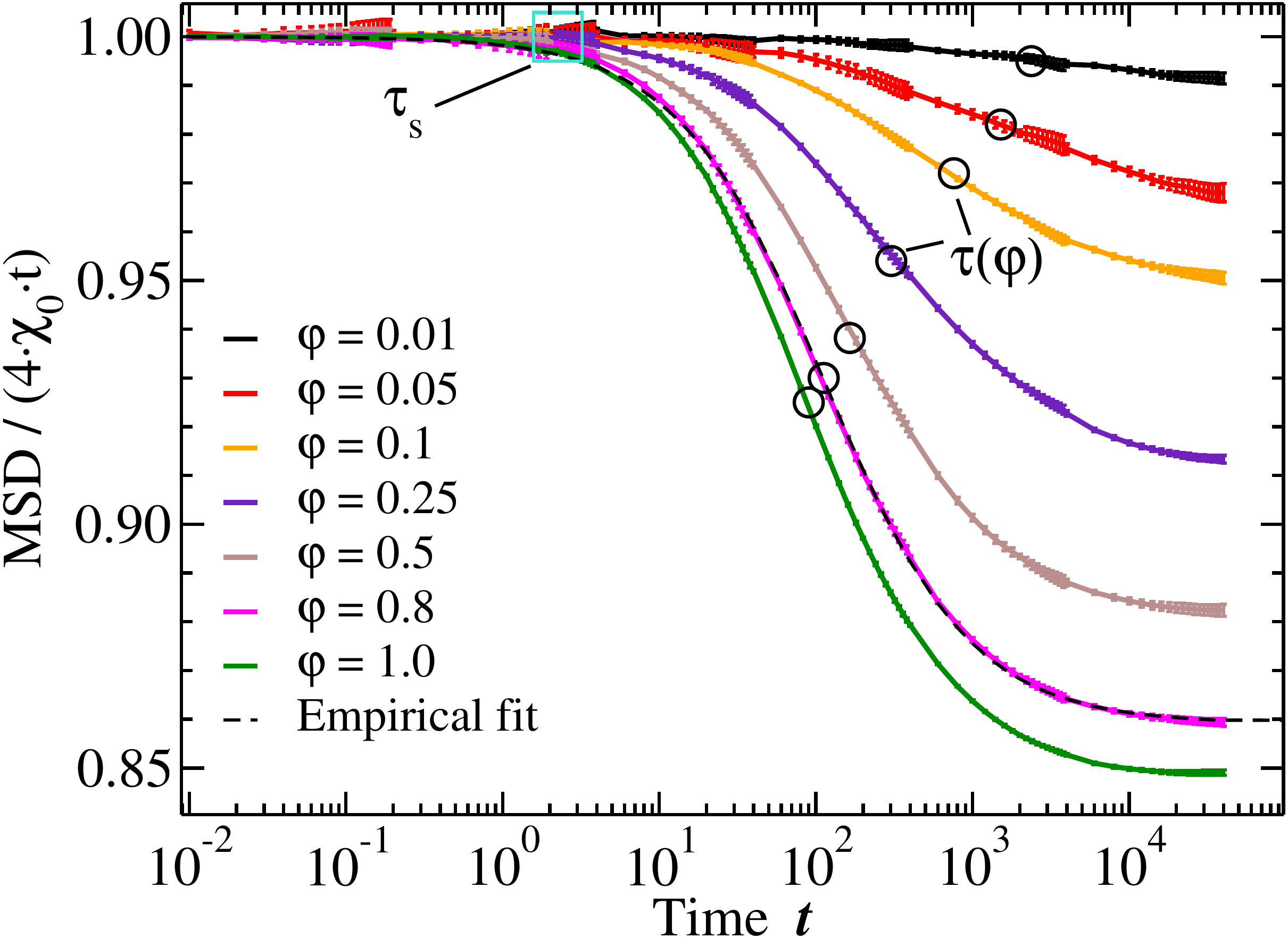}\includegraphics[width=0.5\textwidth]{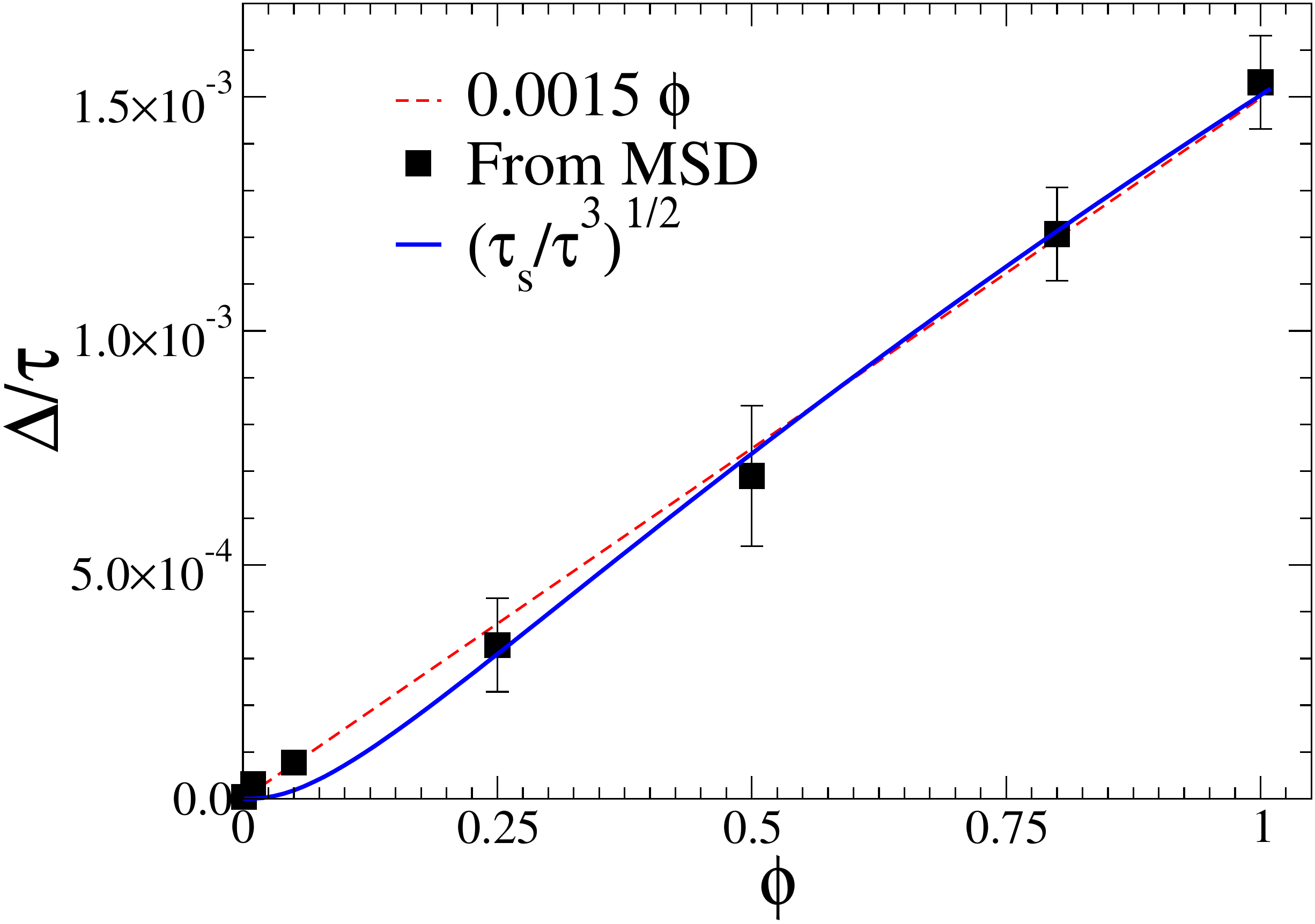}\caption{\label{fig:MSD_t_phi}(Left) Mean square displacement of a tagged
particle in Quasi2D for different packing densities $\phi$ (see legend).
Each data set can be fit rather well by the empirical fit (\ref{eq:MSD_fit}),
as illustrated for $\phi=0.8$ (dashed line). Black circles denote
the cross-over time $\tau(\phi)$, and the beginning of the subdiffusive
regime $\tau_{s}$ is indicated with a cyan square. We have confirmed
that finite size affects are negligible (not shown). (Right) The coefficient
$\Delta/\tau$ appearing in the short-time expansion of the MSD (\ref{eq:MSD_expansion})
obtained in two different ways. The squares are obtained from linear
fits to the short-time decay of the MSD shown in the left panel. The
dashed red line confirms the linear relation with $\phi$ predicted
by (\ref{eq:msd_linear}) with a fit coefficient giving $\alpha=0.088$.
The solid blue line is obtained from the empirical fit (\ref{eq:delta})
and the empirical relation $\Delta=(\tau_{s}/\tau(\phi))^{1/2}$.
Both ways to obtain $\Delta/\tau$ prove to be consistent when the
fit exponent $b\approx1$ (see inset in left panel of Fig. \ref{Fig:MSD_fit_delta}).}
\end{figure}

In Fig. \ref{fig:SelfDiff_LRT} we observed a notable decrease of
the time-dependent diffusion coefficient from the short-time or bare
value $\chi_{s}^{(l)}=\chi_{0}$ to a long-time or renormalized value
$\chi_{s}^{(l)}$. This is due to a collective effect since the difference
between the two values grows with the density, as we show in\textbf{
}the left panel of Fig. \ref{fig:MSD_t_phi}.

It is known that the time evolution of the MSD depends on the gradients
of the mobility matrix $\M M$. In particular, the theoretical series
expansion for the MSD of a given particle $i$ (see (3.24) in \cite{BrownianSuspensions_MSD})
is
\begin{equation}
\text{MSD}_{i}(t)\approx4\chi_{0}t-t^{2}\left(k_{B}T\right)^{2}\av{\frac{\partial M_{il}^{\alpha\beta}}{\partial r_{k}^{\gamma}}\frac{\partial M_{ik}^{\alpha\gamma}}{\partial r_{l}^{\beta}}}+O\left(t^{3}\right),\label{eq:MSD_short_time}
\end{equation}
where we imply summation over all repeated indices except for $i$,
and use Latin indices to denote particles and Greek ones to denote
dimensions. The expression inside the average can be written as the
trace of a positive definite matrix and therefore is always positive,
i.e., for short times the effective diffusion coefficient is \emph{reduced}
from its bare value\emph{.} By examining the form of the second term
in (\ref{eq:MSD_short_time}) we can see that in Quasi2D within the
Oseen approximation (\ref{eq:q2D_small_k}),
\[
\left(k_{B}T\right)^{2}\av{\frac{\partial M_{il}^{\alpha\beta}}{\partial r_{k}^{\gamma}}\frac{\partial M_{ik}^{\alpha\gamma}}{\partial r_{l}^{\beta}}}=4\alpha\left(\frac{k_{B}T}{6\pi\eta}\right)^{2}\av{\sum_{j}\frac{1}{r_{i,j}^{4}}}\approx4\alpha\left(\frac{k_{B}T}{6\pi\eta}\right)^{2}\int_{r=a}^{\infty}\frac{2\pi c_{0}r}{r^{4}}dr=4\alpha c_{0}\chi_{0}^{2}.
\]
The positive non-dimensional coefficient of proportionality $\alpha$
is non-trivial to compute explicitly, but is, in any case, a constant
independent of density. In passing from a discrete sum to the integral
above we could have used the FCM kernel to avoid the cutoff at $r=a$;
however, the integral converges and the cutoff does not affect the
result beyond the value of the unknown coefficient $\alpha$. The
above relation predicts that for short times
\begin{equation}
\text{MSD}(t)\approx4\chi_{0}t-4\alpha c_{0}\chi_{0}^{2}t^{2}=4\chi_{0}t\left(1-\frac{\alpha\phi\chi_{0}}{\pi a^{2}}t\right).\label{eq:msd_linear}
\end{equation}
Linear fits to the MSD at short times give coefficients that are in
good agreement with (\ref{eq:msd_linear}) providing $\alpha\approx0.088$,
see black squares in the right panel of Fig. \ref{fig:MSD_t_phi}.

\begin{figure}
\centering{}\includegraphics[width=0.5\textwidth]{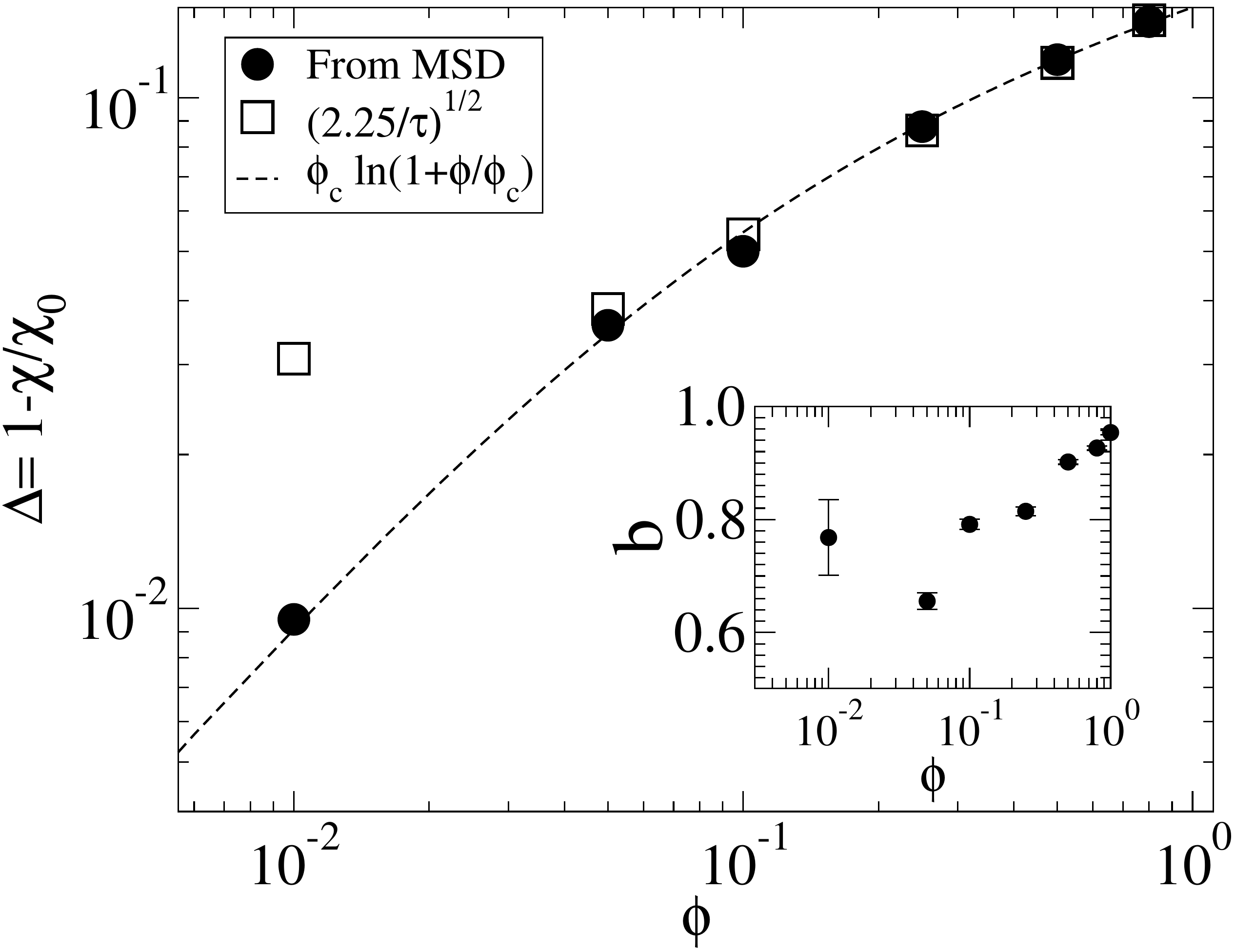}\includegraphics[width=0.5\textwidth]{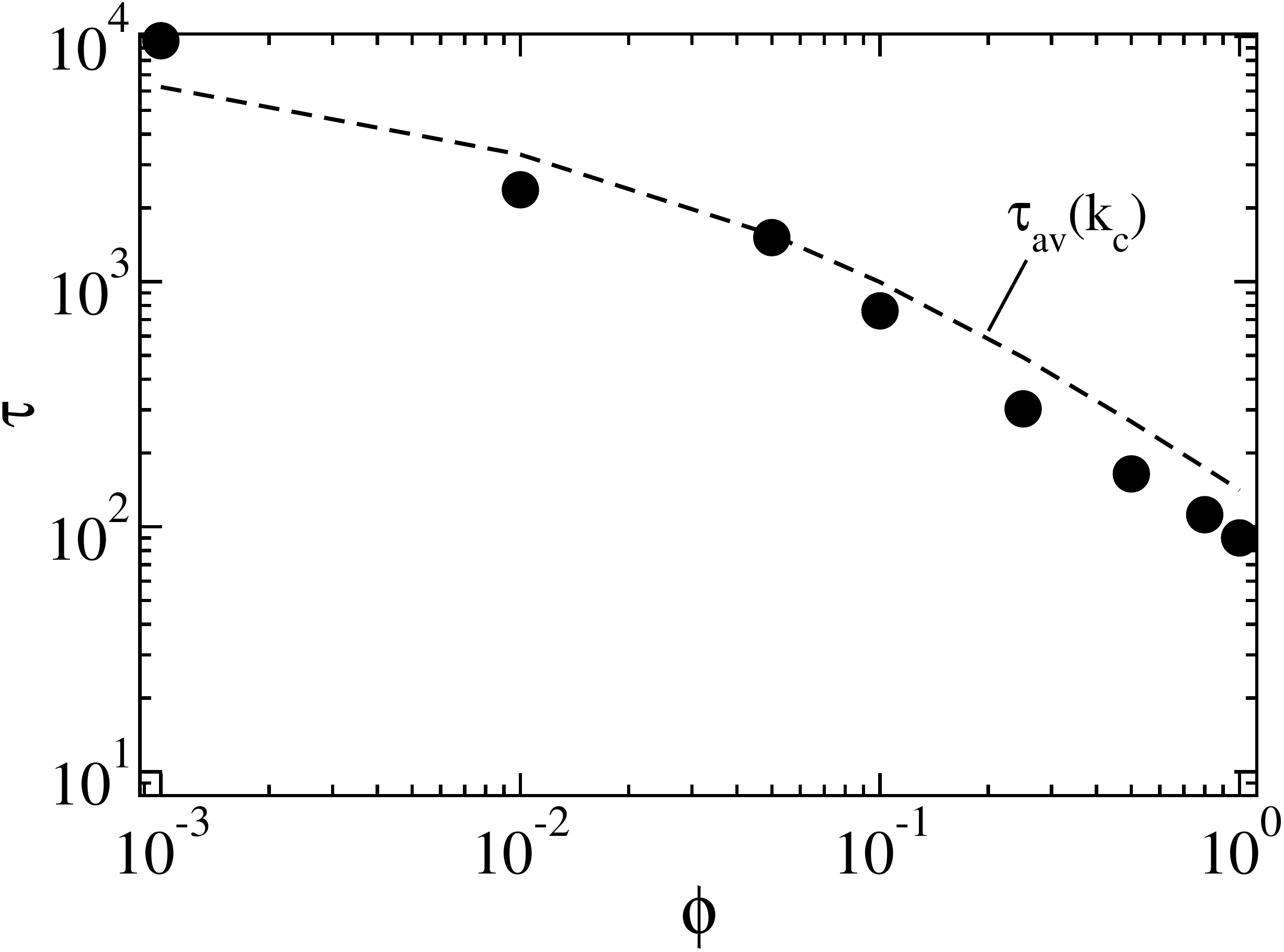}\caption{\label{Fig:MSD_fit_delta}Parameters in the fit (\ref{eq:MSD_fit})
to the MSD as a function of the surface packing fraction $\phi$.
(Left) The relative decrease in the self-diffusion coefficient $\Delta$
(filled circles). The empty squares show the empirical relation $\Delta\approx(\tau_{s}/\tau(\phi))^{1/2}$,
where $\tau_{s}=2.25$. The dashed line is an empirical fit $\Delta=\phi_{c}\ln(1+\phi/\phi_{c})$
with $\phi_{c}=0.0493$. The inset shows the exponent $b(\phi)$.
(Right)\textbf{\label{fig:MSD_fit_tau}} The cross-over time $\tau$
for the subdiffusive regime (circles), compared to the average collective
relaxation time $\tau_{c}(k_{c};\phi)$ defined in (\ref{eq:tav})
(dashed line), where $k_{c}=\left(\alpha/\pi a^{2}\right)^{1/2}=0.17$.}
\end{figure}

At longer times, however, the MSD strongly deviates from the quadratic
approximation (\ref{eq:msd_linear}). We find that
\begin{equation}
\frac{\text{MSD}(t)}{4\chi_{0}t}=1-\frac{\Delta}{\left(1+\left(\tau/t\right)^{b}\right)}\label{eq:MSD_fit}
\end{equation}
accurately fits the numerical results, as shown in the left panel
of Fig. \ref{fig:MSD_t_phi}. The fitting parameters shown in Fig.
\ref{Fig:MSD_fit_delta} include the exponent $b(\phi)$, the duration
of the subdiffusive regime (cross-over time) $\tau(\phi)$, and the
relative jump of the self diffusion coefficient $\Delta(\phi)=1-\chi_{s}^{(l)}/\chi_{0}$.
Interestingly, we find that $\Delta\approx(\tau_{s}/\tau(\phi))^{1/2}$,
as illustrated in the left panel of Fig. \ref{Fig:MSD_fit_delta}.
Here the onset of the sub-diffusive regime $\tau_{s}=2.25$ is a constant
independent of $\phi$, and is indicated with a large open square
in the left panel of Fig. \ref{fig:MSD_t_phi}. On the other hand,
the left panel of Fig. \ref{Fig:MSD_fit_delta} shows that the overall
decrease of the self-diffusion coefficient is in reasonable agreement
with the empirical fit 
\begin{equation}
\Delta\approx\phi_{c}\,\ln\left(1+\phi/\phi_{c}\right),\label{eq:delta}
\end{equation}
where the fitting parameter $\phi_{c}\approx0.0493$ is the surface
fraction at which the collective renormalization of the diffusion
coefficient becomes noticeable. Note that for $\phi<\phi_{c}$ the
collective effects are quite small and our fits are subject to large
statistical errors.

In order to connect the fit in (\ref{eq:MSD_fit}) with the expansion
in (\ref{eq:msd_linear}), we need to Taylor expand (\ref{eq:MSD_fit}).
The inset in the left panel of Fig. \ref{Fig:MSD_fit_delta} shows
that the exponent $b$ increases slowly from $b\approx0.8$ for $\phi=0.05$
to $b\approx1$ for $\phi=1$. If $b=1$, then the formula (\ref{eq:MSD_fit})
is analytic and
\begin{equation}
\text{MSD}(t\ll\tau)\approx4\chi_{0}t\left(1-\frac{\Delta}{\tau}t\right).\label{eq:MSD_expansion}
\end{equation}
Comparing (\ref{eq:msd_linear}) and (\ref{eq:MSD_expansion}), we
conclude that
\begin{equation}
\frac{\Delta}{\tau}\approx\frac{\alpha\chi_{0}}{\pi a^{2}}\,\phi=\chi_{0}k_{c}^{2}\,\phi,\label{eq:delta_over_tau}
\end{equation}
where $\alpha$ is defined by (\ref{eq:msd_linear}). The right panel
of Fig. \ref{fig:MSD_t_phi} shows $\Delta/\tau$ computed by combining
(\ref{eq:delta}) with the relationship $\Delta=(\tau_{s}/\tau)^{1/2}$.
The plot confirms an approximately linear relation between the relative
drop in self-diffusion $\Delta$ and the duration of the diffusive
regime $\tau$, consistent with the value of $\alpha$ estimated from
the short-time behavior of the MSD. The relation (\ref{eq:delta_over_tau})
defines a ``collective'' length scale with wavenumber $k_{c}\equiv\left(\alpha/\pi a^{2}\right)^{1/2}\approx0.17a^{-1}$.

It is quite surprising to find a sub-diffusive regime in the dynamics
of an ideal gas of non-interacting particles. This unexpected behavior
arises because of collective density fluctuations, which lead to collective
drift forces that affect the motion of a tagged particle. This suggests
that $\tau(\phi)$ should be related to the relaxation time for collective
density fluctuations at length scales smaller than or comparable to
$k_{c}^{-1}$. We recall that in Quasi2D, the relaxation time of
a density fluctuation of wavenumber $k$ is $\tau_{c}(k)=\left(\chi_{c}k^{2}\right)^{-1}$,
where $\chi_{c}=\chi_{0}\left(1+\left(kL_{h}\right)^{-1}\right)$
with $L_{h}=2a/3\phi$, see (\ref{eq:q2D_small_k}). This leads us
to an estimate of an average \emph{collective} relaxation time,
\begin{equation}
\tau_{\text{av}}(k_{c})=\frac{\int_{0}^{k_{c}}2\pi k\,\tau_{c}(k)S(k)dk}{\int_{0}^{k_{c}}2\pi kS(k)dk}=\frac{2}{\chi_{0}k_{c}^{2}}\ln\left(1+k_{c}L_{h}\right).\label{eq:tav}
\end{equation}
The right panel of Fig. \ref{fig:MSD_fit_tau} shows that the cross-over
time $\tau$ extracted from the fits of the MSD agrees reasonably
well with $\tau_{\text{av}}(k_{c}=0.17a^{-1})$. This confirms that
the sub-diffusion of individual particles is related to the collective
configurational memory of the system. Empirically we have found that
the self diffusion of a particle is determined by collective density
fluctuations of wavelengths smaller than about $2\pi/k_{c}\approx35\,a$.
A theoretical explanation of the empirical relations we have found
is at present lacking, and requires generalizing the field theory
presented in \cite{TracerDiffusion_Demery} to account for hydrodynamics,
as we discuss in more detail in the Conclusions.

\section{\label{sec:Fluctuations}Fluctuations in Two-Dimensional Systems}

In this section we examine the magnitude and dynamics of density fluctuations
for diffusion confined to a plane, both at thermodynamic equilibrium
and out of equilibrium. In particular, we look at the ensemble-averaged
spectrum of the fluctuations at a given time $t$,
\[
S(\V k;t)=\av{\left(\widehat{\d c}(\V k,t)\right)\left(\widehat{\d c}(\V k,t))\right)^{\star}},
\]
which we call the \emph{static} structure factor. To simplify the
notation, we usually omit the (potential) time dependence and denote
the static structure factor with $S\left(\V k\right)$. The dynamics
at steady state (either equilibrium or non-equilibrium) can be characterized
by the dynamic structure factor
\[
S(\V k,\omega)=\av{\left(\widehat{\d c}(\V k,\omega)\right)\left(\widehat{\d c}(\V k,\omega)\right)^{\star}},\quad\mbox{or, equivalently,}\quad S(\V k,t)=\av{\left(\widehat{\d c}(\V k,t)\right)\left(\widehat{\d c}(\V k,0)\right)^{\star}},
\]
where $\omega$ is the wavefrequency. Here the average is both over
time, and, since our system is ergodic, the steady-state ensemble.
Note that $S(\V k)\equiv S(\V k,t=0)$.

In order to model density fluctuations, we add stochastic fluxes to
(\ref{eq:c_linearized_det}) in the spirit of linearized fluctuating
hydrodynamics (FHD) \cite{FluctHydroNonEq_Book}. We begin with equilibrium
fluctuations and then consider non-equilibrium fluctuations.

\subsection{\label{subsec:LFHD}Equilibrium Fluctuations}

In this section we consider a uniform system with background number
density $c_{0}$ at thermodynamic equilibrium. We first examine the
density fluctuations and then extend the results to account for particle
color (species labels).

\subsubsection{\label{subsec:S_kt_density}Structure Factor for Density}

Obtaining the linearized FHD equations from (\ref{eq:c_linearized_det})
is not trivial (see discussion in Sections 3.2 and 4.1 of \cite{DiffusionJSTAT}),
however, the appropriate equations can essentially be guessed or at
least inferred from fluctuation-dissipation balance. Basically, we
know that at thermodynamic equilibrium (i.e., no macroscopic gradients),
in the absence of direct particle interactions, we have no spatial
correlations, $g_{2}\left(r\right)=1$ for all $r$, and $S\left(\V k\right)=c_{0}$
for all wavenumbers. From this we can infer that the linearized FHD
equation for density at thermodynamic equilibrium is

\begin{equation}
\begin{aligned}\partial_{t}\d c(\V r,t) & =\chi\grad^{2}\d c(\V r,t)+\sqrt{2c_{0}\chi}\,\grad\cdot\M{\mathcal{W}}\\
 & +c_{0}\left(k_{B}T\right)\grad\cdot\left(\int\R(\V r-\V r^{\prime})\grad^{\prime}\d c(\V r^{\prime},t)\,d\V r^{\prime}\right)-c_{0}\left(\grad\cdot\V w\right),
\end{aligned}
\label{eq:c_linearized_FHD}
\end{equation}
where $\M{\mathcal{W}}\left(\V r,t\right)$ is a standard white-noise
Gaussian vector field with uncorrelated components. The first line
in this equation is the standard linearized stochastic diffusion equation
that would apply in the absence of hydrodynamics \cite{SPDE_Diffusion_Dean,DDFT_Hydro},
and it maintains fluctuation-dissipation balance even if it weren't
for the second line. The second term in the second line comes from
the random advection term
\begin{equation}
\grad\cdot\left(\V wc\right)=c\left(\grad\cdot\V w\right)+\V w\cdot\grad c\label{eq:div_wc}
\end{equation}
in (\ref{eq:c_fluct_general}). Upon linearization at equilibrium,
the second term $\V w\cdot\grad c$ disappears; but it is this term
that leads to the giant fluctuations out of equilibrium studied in
Section \ref{subsec:GiantDensity}. It is important to recall that
both (\ref{eq:c_linearized_det}) and (\ref{eq:c_linearized_FHD})
are approximate, not only because we have linearized the fluctuations
but also because in Quasi2D the value of $\chi$ is ambiguous. In
what follows we simply use the short-time self diffusion coefficient,
but at larger length and time scales the somewhat smaller long-time
self diffusion coefficient should be used.

In Fourier space (\ref{eq:c_linearized_FHD}) reads
\begin{equation}
\partial_{t}\left(\widehat{\d c}\right)=-\left(k^{2}+c_{0}\left(k_{B}T\right)\left(\V k\cdot\widehat{\R}\cdot\V k\right)\right)\widehat{\d c}+\sqrt{2c_{0}\chi}\,\left(ik\mathcal{Z}_{\V k}\right)-c_{0}\left(i\V k\cdot\widehat{\V w}\right),\label{eq:LFHD_density}
\end{equation}
where $\mathcal{Z}_{\V k}$ is a standard white-noise process (one
per wavenumber) associated with $\widehat{\M{\mathcal{W}}}_{\V k}$.
Using the fact that $\av{\widehat{\V w}\widehat{\V w}^{\star}}=2\left(k_{B}T\right)\widehat{\R}$
and $\widehat{\V w}$ and $\mathcal{Z}_{\V k}$ are uncorrelated,
we obtain the equilibrium spectrum
\[
S(\V k)=\av{\abs{\widehat{\d c}}^{2}}=\frac{c_{0}\chi k^{2}+c_{0}^{2}\left(k_{B}T\right)\left(\V k\cdot\widehat{\R}\cdot\V k\right)}{\chi k^{2}+c_{0}\left(k_{B}T\right)\left(\V k\cdot\widehat{\R}\cdot\V k\right)}=c_{0},
\]
which is the correct result since the system is an ideal gas at thermodynamic
equilibrium. 

\begin{figure}
\centering{}\includegraphics[width=0.49\textwidth]{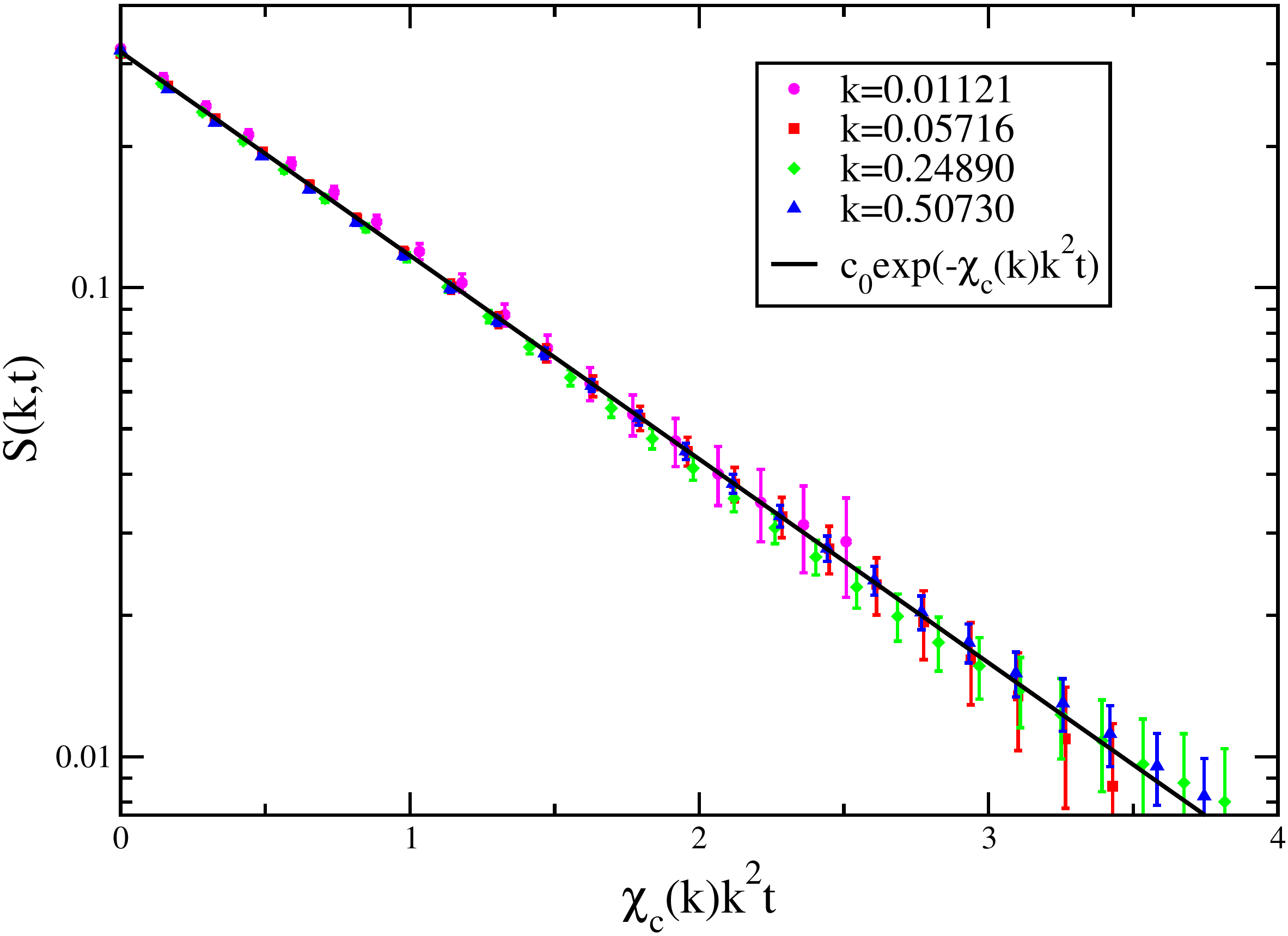}\includegraphics[width=0.49\textwidth]{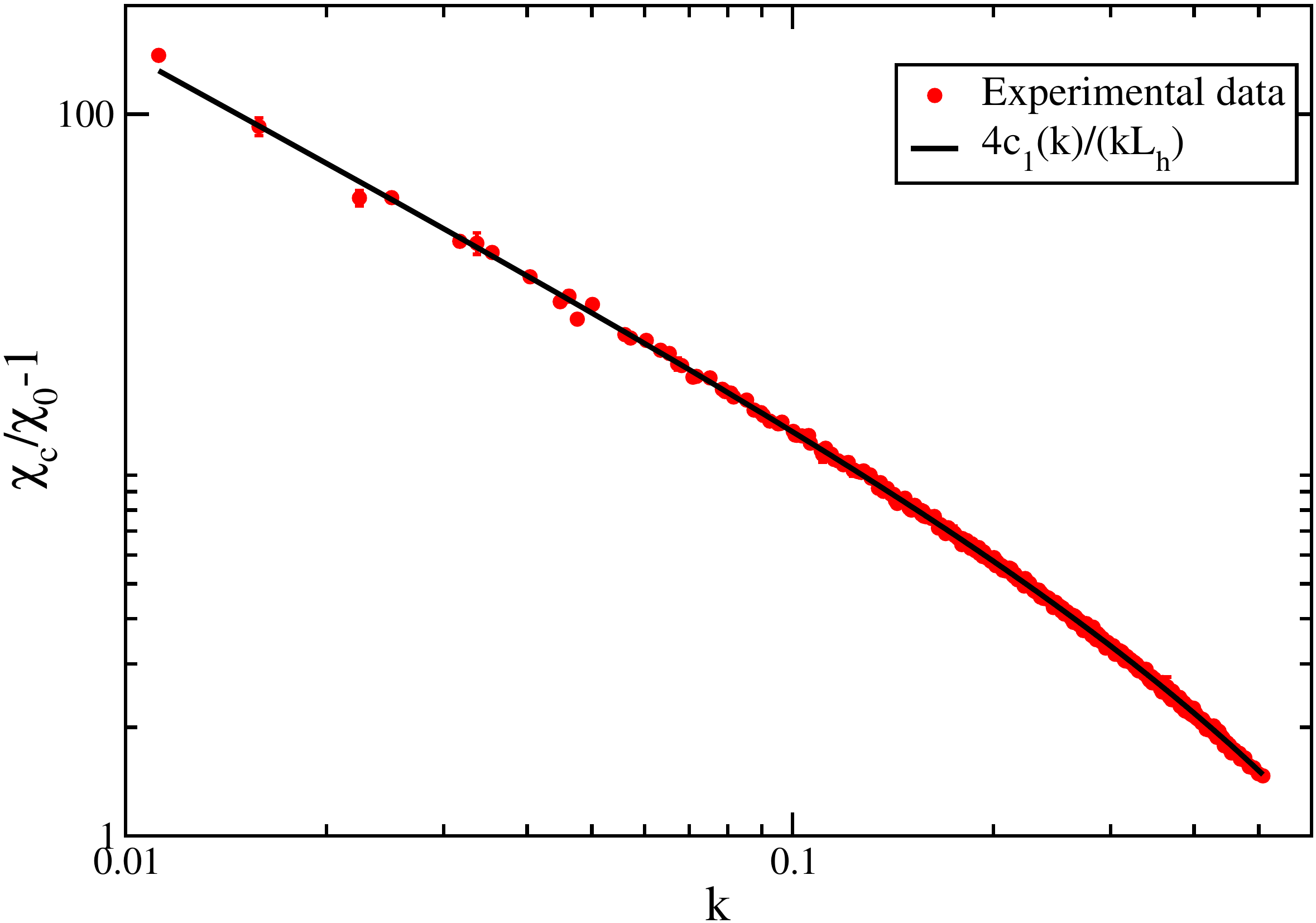}\caption{(Left)\label{fig:S_k_t_density} Numerical results from BD-q2D for
dynamic structure factors for several wavenumbers (see legend), compared
to the theoretical prediction (\ref{eq:c_linearized_FHD}) with $\chi_{c}(k)$
given in (\ref{eq:D_c_full}). The parameters used are taken from
Setup B (see Table \ref{tab:Setups}), $\phi=1$, and the results
are averaged over 1000 simulations, each of length $2\cdot10^{4}$
time steps. (Right)\label{fig:Dc_k_ideal} The collective diffusion
coefficient estimated from the numerical dynamic structure factors
shown in the left panel as a function of $k$ (symbols), compared
to the theoretical prediction (\ref{eq:D_c_full}) (line). To estimate
$\chi_{c}\left(k\right)$ we fitted $S(\protect\V k,t)$ to an exponential
over the time interval during which the value decays by one order
of magnitude.}
\end{figure}

The dynamic structure factor is given from (\ref{eq:LFHD_density})
by
\begin{equation}
S(\V k,t)=c_{0}\,\exp\left(-k^{2}\chi_{c}(k)t\right),\label{eq:S_k_t_rho}
\end{equation}
where the collective diffusion coefficient $\chi_{c}$ is given in
(\ref{eq:D_c_general},\ref{eq:D_c_full}). In the left panel of Fig.
\ref{fig:S_k_t_density} we show $S(\V k,t)$ for several selected
wavenumbers, rescaling the time by the theoretical prediction for
$k^{2}\chi_{c}(k)$ to overlap the data. In the right panel of Fig.
\ref{fig:S_k_t_density} we compare numerical measurements of the
collective diffusion coefficient from the rate of decay of $S(\V k,t)$
with the theoretical prediction; we see an excellent agreement over
all wavenumbers. To within statistical accuracy, in the left panel
of Fig. \ref{fig:S_k_t_density} we see a \emph{mono-exponential}
decay over the accessible time period for our ideal gas system, without
a statistically-significant difference between short and long-time
diffusion coefficients. Observe that because of the large relative
value of the enhancement of the collective diffusion coefficient relative
to the self-diffusion coefficient, it is difficult to see the difference
between the short- and long-time self diffusion coefficients on this
figure. To see this difference more clearly we consider dynamic structure
factors for color fluctuations.

\subsubsection{\label{subsec:S_kt_color}Dynamic Structure Factor for Color}

The linearized FHD equations for the density of red/green particles
can be inferred from (\ref{eq:c_av_RG}) by accounting for stochastic
mass fluxes in a manner that preserves fluctuation-dissipation balance.
At thermodynamic equilibrium,
\begin{align}
\partial_{t}\d c_{R/G}(\V r,t) & =\chi\grad^{2}\d c_{R/G}(\V r,t)+\sqrt{2c_{R/G}\chi}\,\grad\cdot\M{\mathcal{W}}^{(R/G)}-c_{R/G}\left(\grad\cdot\V w\right)\nonumber \\
 & +\left(k_{B}T\right)\grad\cdot\left(c_{R/G}\int\R(\V r-\V r^{\prime})\grad^{\prime}\d c(\V r^{\prime},t)\,d\V r^{\prime}\right),\label{eq:dc_RG_lin}
\end{align}
where for notational simplicity we denoted the the average equilibrium
color density with $c_{0}^{R/G}\equiv c_{R/G}$ without the subscript
zero. The structure factor \emph{matrix}, either static or dynamic,
is defined as the covariance of the fluctuations of density in Fourier
space,
\[
\M S=\left[\begin{array}{cc}
\av{\left(\widehat{\d c}_{R}\right)\left(\widehat{\d c}_{R})\right)^{\star}} & \av{\left(\widehat{\d c}_{R}\right)\left(\widehat{\d c}_{G})\right)^{\star}}\\
\av{\left(\widehat{\d c}_{G}\right)\left(\widehat{\d c}_{R})\right)^{\star}} & \av{\left(\widehat{\d c}_{G}\right)\left(\widehat{\d c}_{G})\right)^{\star}}
\end{array}\right]=\left[\begin{array}{cc}
S_{RR} & S_{RG}\\
S_{GR} & S_{GG}
\end{array}\right].
\]

From (\ref{eq:dc_RG_lin}) it is straightforward to show that at thermodynamic
equilibrium the static structure factor is
\[
\M S\left(\V k\right)=\M S_{0}=\left[\begin{array}{cc}
c_{R} & 0\\
0 & c_{G}
\end{array}\right],
\]
which is consistent with an ideal gas mixture of red and green particles.
To obtain the dynamic structure factors for color, we write (\ref{eq:dc_RG_lin})
in Fourier space in matrix form,
\begin{align}
\partial_{t}\left[\begin{array}{c}
\widehat{\d c_{R}}\\
\widehat{\d c_{G}}
\end{array}\right] & =-k^{2}\left[\begin{array}{cc}
\chi+\D{\chi}_{c}^{R} & \D{\chi}_{c}^{R}\\
\D{\chi}_{c}^{G} & \chi+\D{\chi}_{c}^{G}
\end{array}\right]\left[\begin{array}{c}
\widehat{\d c_{R}}\\
\widehat{\d c_{G}}
\end{array}\right]+\text{stochastic forcing}=\nonumber \\
 & =-k^{2}\M M\left[\begin{array}{c}
\widehat{\d c_{R}}\\
\widehat{\d c_{G}}
\end{array}\right]+\text{stochastic forcing},\label{eq:dc_M}
\end{align}
where we introduced the collective diffusion enhancements for red
and green particles via
\[
\D{\chi}_{c}^{R/G}(k)=\chi\left(\frac{4c_{1}\left(ka\right)}{kL_{h}^{R/G}}\right)\approx\frac{k_{B}T}{4\eta k}c_{R/G},
\]
where $L_{h}^{R/G}=2/\left(3\pi ac_{R/G}\right)$. It is a standard
result from (\ref{eq:dc_M}) that
\[
\M S\left(\V k,t\right)=\exp\left(-\M Mk^{2}t\right)\M S_{0}.
\]
Computing the matrix exponential for an equimolar mixture, $c_{R}=c_{G}=c_{0}/2$,
gives
\begin{eqnarray*}
S_{RR}\left(k,t\right)=S_{GG}\left(k,t\right) & = & \frac{c_{0}}{4}\left(\exp\left(-\chi_{c}k^{2}t\right)+\exp\left(-\chi k^{2}t\right)\right),\\
S_{RG}\left(k,t\right)=S_{GR}\left(k,t\right) & = & \frac{c_{0}}{4}\left(\exp\left(-\chi_{c}k^{2}t\right)-\exp\left(-\chi k^{2}t\right)\right).
\end{eqnarray*}
Observe that the structure factors are sums of two modes, one slow
one related to the self-diffusion coefficient $\chi$, and one faster
one related to the enhanced collective diffusion coefficient $\chi_{c}$.

\begin{figure}
\centering{}\includegraphics[width=0.6\textwidth]{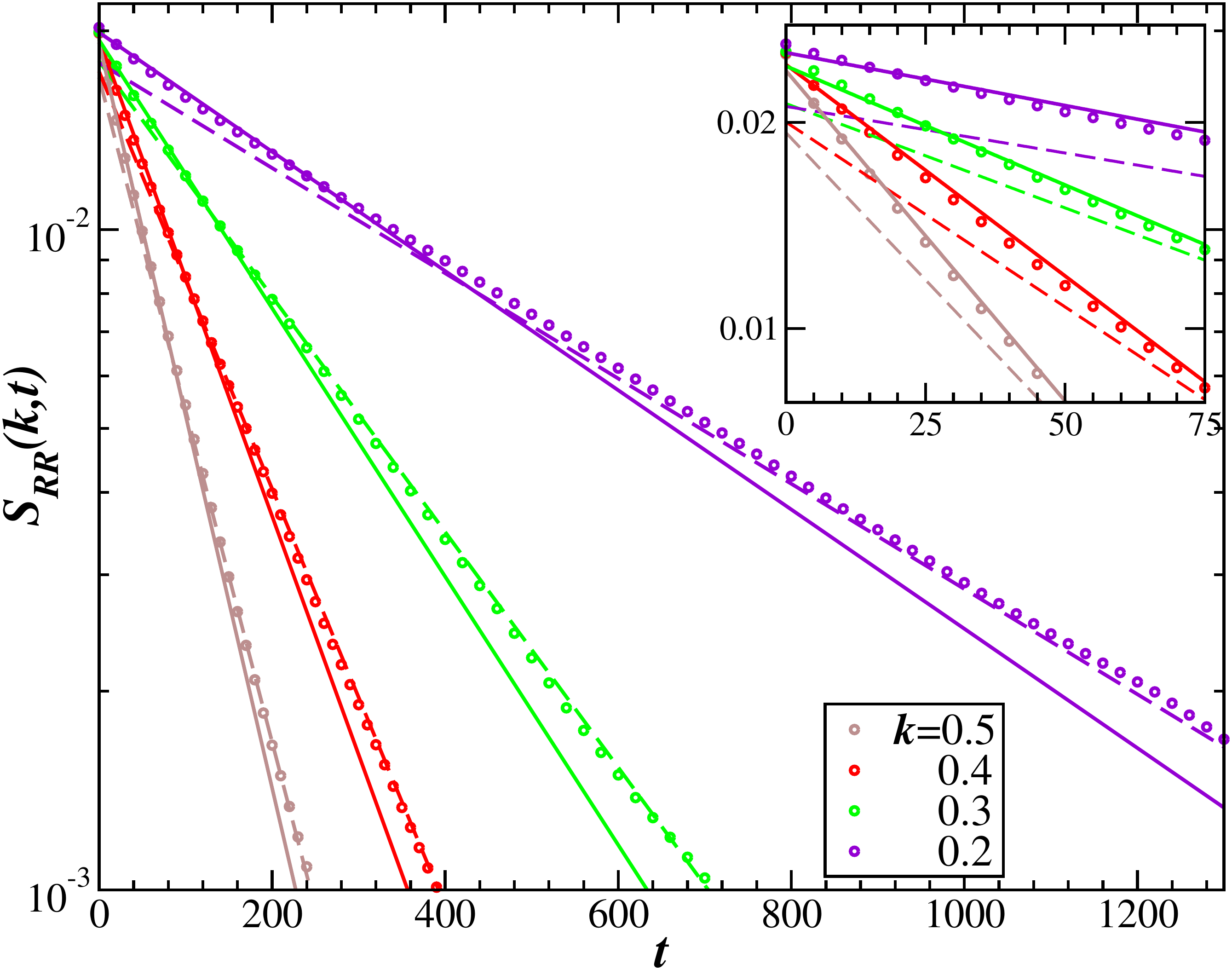}\caption{\label{fig:S_k_t_color}Dynamic structure factor for the density of
red particles when a small fraction $c_{R}/c_{0}=1/16$ are colored
red. The solid lines show the theoretical prediction (\ref{fig:MeanDensityColor})
with $\chi$ taken to be the short-time self diffusion coefficient
$\chi_{0}$, while the dashed lines use the long-time value $\chi_{s}^{(l)}$
(see Section \ref{sec:SelfDiffusion}). The inset focuses on short
times. The parameters are the same as for Fig. \ref{fig:S_k_t_density}
(Setup B, $\phi=1$, see Table \ref{tab:Setups}).}
\end{figure}

If we label with a ``red'' tag only a small fraction of the particles,
$c_{R}\ll c_{G}\approx c_{0}$, and only trace the red particles,
we get
\begin{equation}
S_{RR}\left(\V k,t;\;c_{R}\ll c_{G}\right)\approx c_{R}\exp\left(-\chi k^{2}t\right),\label{eq:S_RR_k_t}
\end{equation}
which allows us to focus on the self-diffusion contribution. In Fig.
\ref{fig:S_k_t_color} we show numerical results from BD-q2D for $S_{RR}\left(\V k,t\right)$
when only $1/16$-th of the particles are colored red. The results
in Fig. \ref{fig:S_k_t_color} show that there is a transition from
exponential decay with rate set by the short-time self-diffusion coefficient
$\chi_{0}$ at short times, to a decay with rate set by the long-time
self diffusion coefficient $\chi_{s}^{(l)}$ at long times. This means
that even at the level of linearized FHD there are collective memory
effects that arise due to renormalization of the transport coefficient
by fluctuations. In particular this means that the equations (\ref{eq:c_linearized_FHD})
and (\ref{eq:dc_RG_lin}) are \emph{not }exact even for an ideal gas
mixture, even at the linearized level, i.e., at the level of a Gaussian
approximation for the fluctuations.

\subsection{Giant Non-equilibrium Fluctuations}

It is well known that nonequilibrium fluctuations in diffusive systems
are long-ranged and strongly enhanced (giant) compared to the short-ranged
equilibrium fluctuations \cite{FluctHydroNonEq_Book}. These giant
fluctuations have been measured experimentally in three dimensions
in microgravity \cite{FractalDiffusion_Microgravity}, but have not
yet been measured in two-dimensional systems to our knowledge. Linearized
fluctuating hydrodynamics predicts that in True2D the fluctuations
are truly giant or ``colossal'' because the magnitude of the fluctuations
is comparable to that of the mean, see Fig. 2 and associated discussion
in \cite{GiantFluctuations_ThinFilms}. The contribution from the
compressibility of the fluid in the plane of confinement has not yet
been evaluated because the theoretical calculations in \cite{GiantFluctuations_ThinFilms}
are based on the Saffman mobility kernel for membranes \cite{MembraneDiffusion_Review},
which \emph{assumes} that the flow is incompressible in the plane
of the membrane.

In this section we study giant fluctuations in the presence of density
and/or color gradients for a Quasi2D colloidal suspension confined
to a planar interface, and compare to a True2D system such as a thin
film in vacuum \cite{GiantFluctuations_ThinFilms,ThinFilms_True2D}.
We begin by considering a density gradient, and then consider a color
gradient in the absence of a density gradient. In linearized FHD the
general case of a density and color gradient is a simple superposition
of the two cases we consider here. 

In general, out of equilibrium the FHD equations should be linearized
around the solution of the ``deterministic'' equations \footnote{The precise statement is that the linearized FHD equations give the
central limit theorem and describe the Gaussian fluctuations from
the law of large numbers for the diffusive process.}. Here we take these unknown deterministic macroscopic equations to
be the same as the approximate equations for the ensemble average
(\ref{eq:c_av_closure},\ref{eq:c_av_RG}). It is difficult to do
better than this because to obtain the macroscopic equations we would
have to perform a very nontrivial coarse graining in space and time
\cite{DiffusionJSTAT}. The best approximation of the macroscopic
equations known to us at present is (\ref{eq:c_av_closure},\ref{eq:c_av_RG})
with $\chi$ being the long-time self-diffusion coefficient.

The linearization of the FHD equations around the (time-dependent)
solution of the deterministic equations can often be performed numerically
by artificially reducing the magnitude of the stochastic forcing terms
\cite{MultiscaleIntegrators}. Doing the same analytically is in general
quite challenging \cite{GiantFluctuations_Cannell}. Here we follow
the traditional theoretical route \cite{FluctHydroNonEq_Book} and
assume that the macroscopic gradient is imposed externally and is
constant in space and time. Furthermore, we assume that the gradient
is sufficiently weak, so that the macroscopic density can be assumed
not to vary in space. While in our numerical simulations the gradient
varies both in space and time, this theoretical approximation enables
simple estimates that give us the basic physical picture. For a more
detailed comparison we numerically solve the FHD equations in Section
\ref{subsec:NFHD}.

\subsubsection{\label{subsec:GiantDensity}Density Gradient}

In this section we consider a weak imposed macroscopic gradient $\grad c_{0}$
of the total number density. In the presence of the weak gradient
the linearized FHD equation (\ref{eq:c_linearized_FHD}) has an extra
term $\V w\cdot\grad c_{0}$ coming from (\ref{eq:div_wc}),
\begin{equation}
\begin{aligned}\partial_{t}\d c(\V r,t) & =\chi\grad^{2}\d c(\V r,t)+\sqrt{2c_{0}\chi}\,\grad\cdot\M{\mathcal{W}}\\
 & +c_{0}\left(k_{B}T\right)\grad\cdot\left(\int\R(\V r-\V r^{\prime})\grad^{\prime}\d c(\V r^{\prime},t)\,d\V r^{\prime}\right)-c_{0}\left(\grad\cdot\V w\right)-\V w\cdot\grad c_{0}.
\end{aligned}
\label{eq:c_LFHD_grad}
\end{equation}
Let us consider a gradient in the $y$ direction, $\V g=\grad c_{0}=g\hat{\V y}$,
and take a wavenumber perpendicular to the gradient, $\V k\perp\V y$.
In this case, in Fourier space (\ref{eq:c_LFHD_grad}) becomes
\begin{eqnarray}
\partial_{t}\left(\widehat{\d c}\right) & = & -\left(\chi k^{2}+c_{0}\left(\frac{k_{B}T}{\eta}\right)kc_{1}(ka)\right)\widehat{\d c}\nonumber \\
 & + & \sqrt{2c_{0}\chi}\,\left(ik\mathcal{Z}_{\V k}\right)-ic_{0}\sqrt{\frac{2\left(k_{B}T\right)kc_{1}\left(ka\right)}{\eta}}\,\mathcal{Z}_{\V k}^{(1)}\nonumber \\
 & - & g\sqrt{\frac{2\left(k_{B}T\right)c_{2}\left(ka\right)}{\eta k}}\,\mathcal{Z}_{\V k}^{(2)},\label{eq:chat_LFHD_grad}
\end{eqnarray}
where $\mathcal{Z}_{\V k}^{(1/2)}$ are independent standard white-noise
processes (one per wavenumber) associated with the vortical/longitudinal
modes of $\V w$, see (\ref{eq:v_stoch}). From this equation we obtain
the static structure factor
\begin{equation}
S(k)=\av{\abs{\widehat{\left(\d c\right)}}^{2}}=S_{0}+\D S=c_{0}+g^{2}\frac{\pi a^{2}c_{2}\left(ka\right)}{k^{2}\left(\pi a^{2}\beta k+c_{1}\left(ka\right)\phi\right)},\label{eq:S_k_neq}
\end{equation}
where $\beta$ is defined in (\ref{eq:f_def}). The second term above
represents the nonequilibrium fluctuations in excess of the equilibrium
spectrum $S_{0}=c_{0}$.

To get an idea of how big the nonequilibrium contribution may be,
let's consider small wavenumbers $ka\ll1$, to get
\[
\frac{\D S}{S_{0}}\left(ka\ll1\right)\approx g^{2}\begin{cases}
\frac{2}{c_{0}^{2}}\cdot\frac{1}{k^{2}} & \mbox{for Quasi2D}\\
\frac{4\pi}{c_{0}\,\ln\left(\frac{L}{3.71a}\right)}\cdot\frac{1}{k^{4}} & \mbox{for True2D}.
\end{cases}
\]
The $1/k^{4}$ divergence for True2D is well-known and leads to ``giant''
nonequilibrium fluctuations \cite{DiffusionJSTAT,GiantFluctuations_ThinFilms}.
For Quasi2D the divergence is only $1/k^{2}$. To get an idea of the
magnitude of the nonequilibrium correction, let us assume that the
gradient is imposed with boundary conditions over a length scale $L$
comparable to the size of the system, $g=c_{0}/L$, and consider a
small wavenumber $k=2\pi/L$. Denoting $L/a=N_{L}$, we estimate
\[
\max\frac{\D S}{S_{0}}\approx\begin{cases}
\frac{1}{2\pi^{2}}\ll1 & \mbox{for Quasi2D}\\
\frac{N_{L}^{2}\phi}{4\pi^{4}\ln\left(N_{L}/3.71\right)}\gg1 & \mbox{for True2D}.
\end{cases}
\]

\begin{figure}[H]
\begin{centering}
\includegraphics[width=1\textwidth]{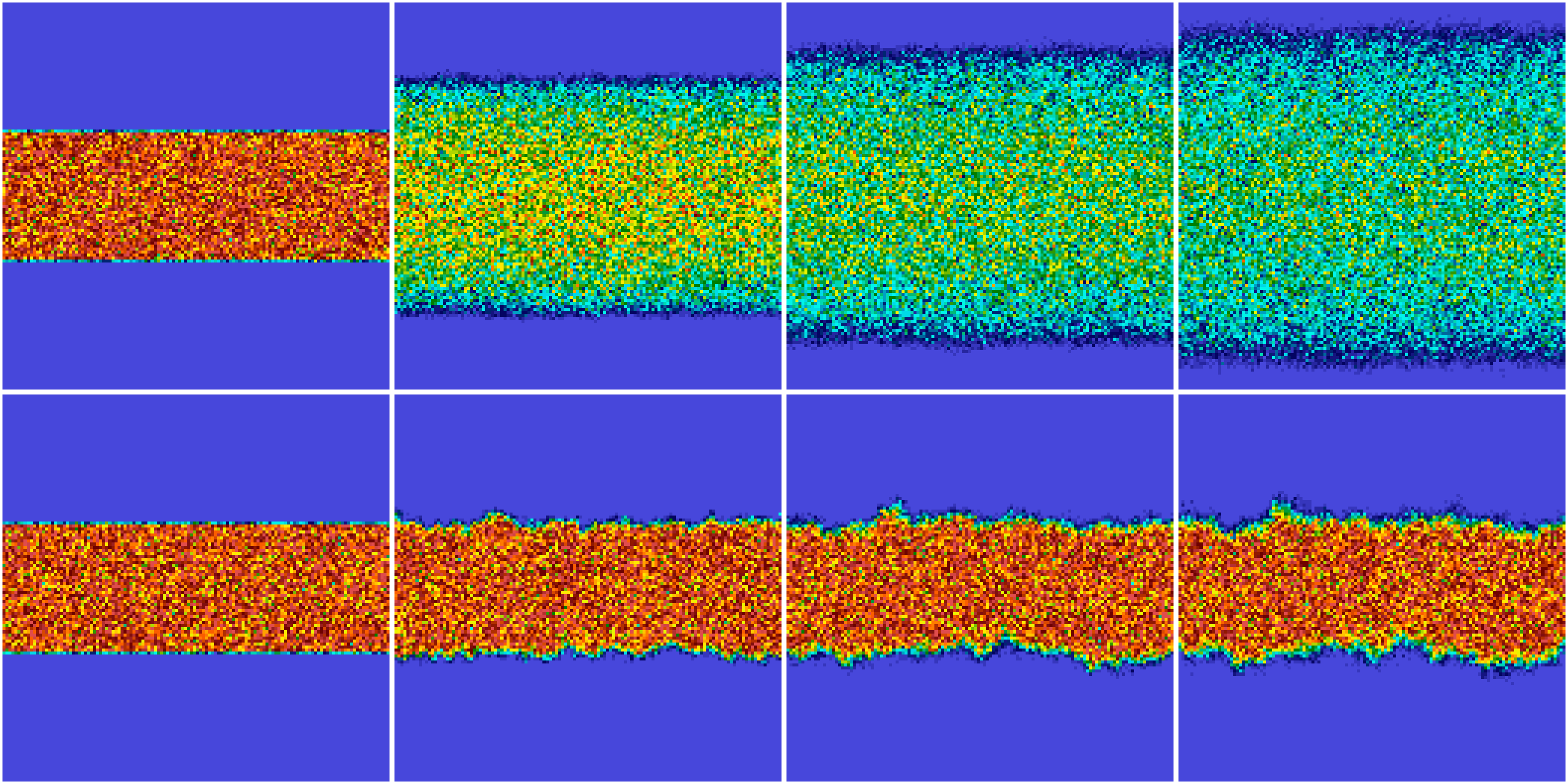}
\par\end{centering}
\centering{}\caption{\label{fig:InstanceDensity}Diffusion of a density perturbation initially
localized in the middle third of the domain. We show snapshots at
several points of equal relative time for BD-q2D (top row) and BD-t2D
(bottom row). Snapshots are shown at times $0$, $T/3$, $2T/3$ and
$T$ (left to right), where the total simulation time $T$ is $T_{q2D}=6775$
and $T_{t2D}=791$. The images show the number density computed by
counting the number of particles in each cell of a $128^{2}$ grid;
the color bar is fixed to range from $0$ (blue) to $0.4$ (red) in
all snapshots.}
\end{figure}

We conclude that in Quasi2D it is not possible to measure the nonequilibrium
fluctuations since they are much smaller than equilibrium fluctuations
for all wavenumbers. This is in agreement with numerical observations
that there are no giant fluctuations in Quasi2D, as illustrated in
the top row of Fig. \ref{fig:InstanceDensity}. In this numerical
experiment, we use parameters from Setup B (see Table \ref{tab:Setups}),
but double the system size in both directions. We place $\sim1.333\cdot10^{5}$
particles only in the middle third stripe of the domain, giving an
initial packing density $\phi=1$ in the middle and zero elsewhere.
For Quasi2D (BD-q2D) we follow the evolution to time $T_{q2D}=6775$,
and for True2D (BD-t2D) we follow the evolution to $T_{t2D}=791$.
This ensures that at the final time the diffusive mixing has progressed
to the same relative time, $\chi_{q2D}T_{q2D}=\chi_{t2D}T_{t2D}$.
In Fig. \ref{fig:InstanceDensity} we show several snapshots of the
density (time increases from left to right) during the diffusive spreading
of the initial density perturbation. Visually we do not see any large-scale
structure of the density fluctuations for Quasi2D. We have examined
the spectrum of the fluctuations averaged along the $y$ directions,
and have been unable to measure $\D S$ compared to $S_{0}$ to within
statistical accuracy. In True2D, the particles have spread much less
because of the absence of the collective ``repulsion'', however,
we can already visually appreciate the diffusive growth of giant fluctuations.
At later times (not shown) we observe fully-developed ``colossal''
fluctuations in True2D, as illustrated in the bottom row in Fig. \ref{fig:InstanceColor}
for a slightly different setup.

Note that in True3D the magnitude of the long-ranged nonequilibrium
fluctuations is much larger than at equilibrium, in particular, in
True3D one gets $\max\left(\D S/S_{0}\right)\sim\phi N_{L}^{2}\gg1$.
Therefore, it is somewhat surprising that the giant fluctuations in
the presence of a density gradient are strongly suppressed in Quasi2D,
where the hydrodynamics is still three dimensional. It appears that
the apparent long-ranged ``repulsion'' between the particles suppresses
the fluctuations. This suggests that in order to see giant fluctuations
we should eliminate density gradients so that the repulsion does not
act, i.e., we should instead impose gradients of color (species label)
only.

\subsubsection{\label{subsec:GiantColor}Color Gradient}

In this section we consider fluctuations of color in the presence
of a macroscopic color gradient without a gradient in density. We
impose the gradient $\grad c_{R}=-\grad c_{G}=\V g$ and perform the
same computation as in the previous section. The linearized FHD equations
consist of (\ref{eq:dc_RG_lin}) with an additional forcing term $\pm\V w\cdot\V g$
on the right hand side. Considering wavenumbers perpendicular to
the gradient, in Fourier space we have
\begin{eqnarray*}
\partial_{t}\left[\begin{array}{c}
\widehat{\d c_{R}}\\
\widehat{\d c_{G}}
\end{array}\right] & = & -\left(\chi k^{2}\M I+\left(\frac{k_{B}T}{\eta}\right)kc_{1}(ak)\left[\begin{array}{cc}
c_{R} & c_{R}\\
c_{G} & c_{G}
\end{array}\right]\right)\left[\begin{array}{c}
\widehat{\d c_{R}}\\
\widehat{\d c_{G}}
\end{array}\right]+ik\left[\begin{array}{c}
\sqrt{2\chi c_{R}}\;\mathcal{Z}_{\V k}^{\left(R\right)}\\
\sqrt{2\chi c_{G}}\;\mathcal{Z}_{\V k}^{\left(G\right)}
\end{array}\right]\\
 & - & i\sqrt{\frac{2\left(k_{B}T\right)kc_{1}\left(ka\right)}{\eta}}\left[\begin{array}{c}
c_{R}\\
c_{G}
\end{array}\right]\mathcal{Z}_{\V k}^{(1)}-\sqrt{\frac{2\left(k_{B}T\right)c_{2}\left(ka\right)}{\eta k}}\left[\begin{array}{c}
g\\
-g
\end{array}\right]\mathcal{Z}_{\V k}^{(2)},
\end{eqnarray*}
where $\mathcal{Z}_{\V k}^{\left(R/G\right)}$ are independent scalar
white noise processes associated with $\M{\mathcal{W}}^{(R/G)}$.
Solving this linear system of Ornstein-Uhlenbeck equations we obtain
\begin{equation}
\M S\left(k\right)=\M S_{0}+\D{\M S}=\left[\begin{array}{cc}
c_{R} & 0\\
0 & c_{G}
\end{array}\right]+g^{2}\frac{c_{2}\left(ka\right)}{\beta\,k^{3}}\left[\begin{array}{cc}
1 & -1\\
-1 & 1
\end{array}\right],\label{eq:S_k_neq_color}
\end{equation}
where $\beta$ is defined in (\ref{eq:f_def}). Note that, if we look
at the total density $\d c=\d c_{R}+\d c_{G}$, we only see equilibrium
fluctuations, as expected,
\[
\av{\left(\widehat{\d c}\right)\left(\widehat{\d c})\right)^{\star}}=S_{RR}+S_{GG}+S_{RG}+S_{GR}=c_{R}+c_{G}=c.
\]

In True2D there is no difference between color gradient or density
gradient since the particles are \emph{passive} non-interacting tracers.
Therefore, we focus now on Quasi2D. Our theory (\ref{eq:S_k_neq_color})
predicts that the nonequilibrium fluctuations of color in Quasi2D
have a spectrum $\sim1/k^{3}$, and might therefore be measurable
in simulations or experiments, unlike the fluctuations of total density
in the presence of a density gradient. Specifically, for small $k$
we predict a nontrivial correlation between the density fluctuations
of red and green particles,
\begin{equation}
\D S_{RG}\left(ka\ll1\right)\approx-g^{2}\frac{3\pi a}{k^{3}}.\label{eq:dS_q2D_small_k}
\end{equation}
To estimate how large the color fluctuations may be, we again take
$c_{R}=c_{G}=c_{0}$, $g=c_{0}/L$, $k=2\pi/L$ and $L=N_{L}a$, to
estimate
\[
\max\frac{-\D S_{RG}}{c_{0}}\approx\frac{3\phi N_{L}}{8\pi^{3}}.
\]
For $N_{L}=800$ and $\phi=0.5$, for example, we estimate $\max\left(\D S/S_{0}\right)\approx5$,
which is much larger than $1$ but not nearly as large as the corresponding
value $\max\left(\D S/S_{0}\right)\sim150$ for True2D.

\begin{figure}[H]
\begin{centering}
\includegraphics[width=1\textwidth]{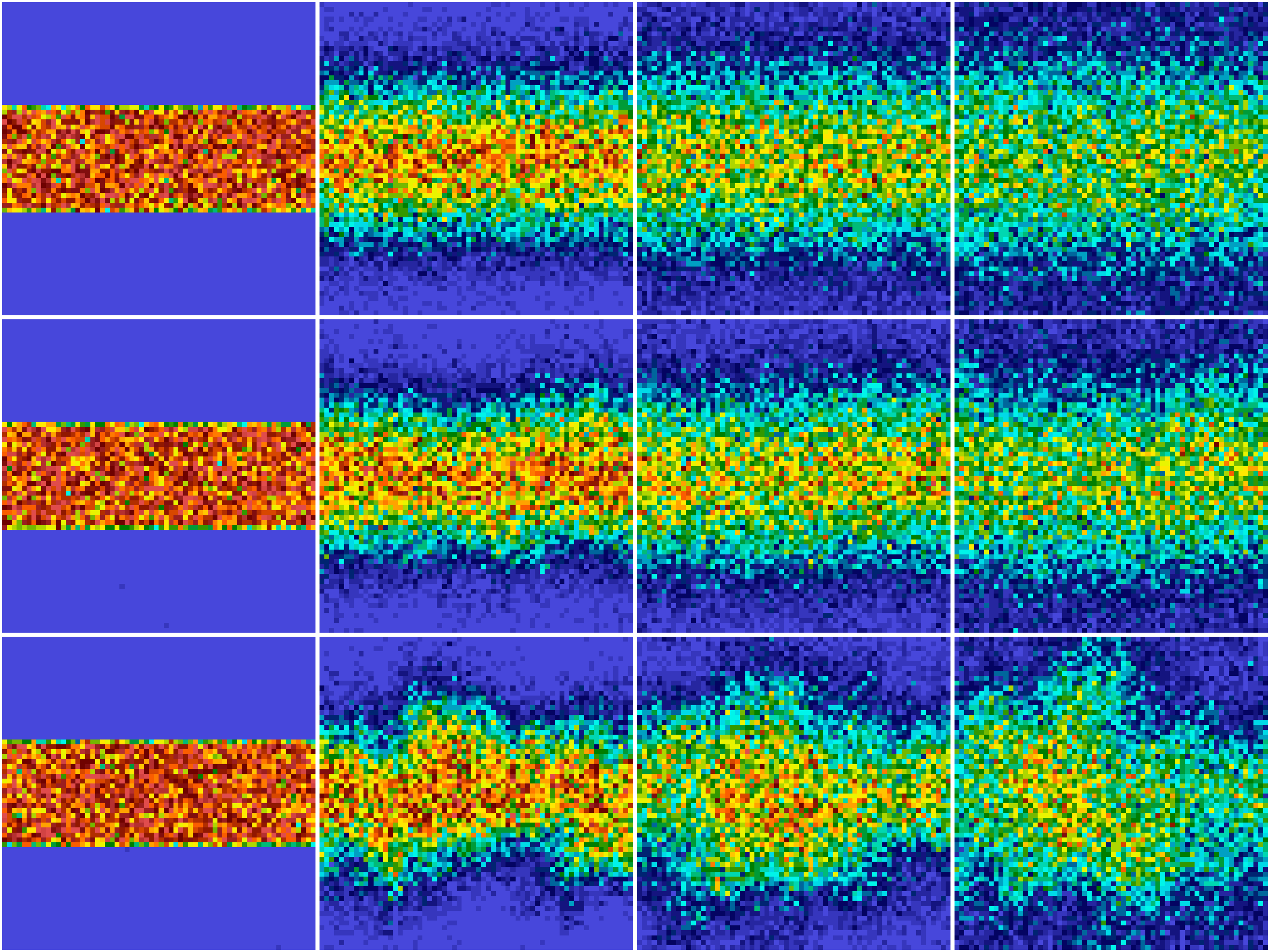}
\par\end{centering}
\centering{}\caption{\label{fig:InstanceColor}Diffusion of a perturbation of color (species)
density in the absence of hydrodynamics (BD-noHI, top row), with Quasi2D
hydrodynamics (BD-q2D, midde row), and with True2D hydrodynamics (BD-t2D,
bottom row). The images show the number density of ``green'' particles
computed by counting the number of green particles in each cell of
a $64^{2}$ grid; the color bar is fixed to range from $0$ (blue)
to $0.4$ (red) in all snapshots. The green particles are initially
localized in the middle third of the domain and the rest of the domain
is filled with red particles at the same packing density $\phi=1$,
giving a uniform total density over the domain. We show several snapshots
at times $0$, $0.3T$, $0.6T$ and $0.9T$ (left to right), where
the total simulation time $T$ is $T_{q2D}=1.505\cdot10^{5}$ and
$T_{t2D}=2\cdot10^{4}$.}
\end{figure}

Unlike the case of a pure density gradient illustrated in Fig. \ref{fig:InstanceDensity},
one can appreciate the giant color fluctuations in Quasi2D even visually.
In Fig. \ref{fig:InstanceColor} we show snapshots of the density
of ``green'' particles (i.e., particles which are initially localized
in the middle third of the domain) comparing uncorrelated walkers
(BD-noHI, top row), Quasi2D (BD-q2D, middle row), and True2D (BD-t2D,
bottom row). In these simulations, we use parameters from Setup B
(see Table \ref{tab:Setups}), $\phi=1$, and start with a uniform
system where we colored the particles in the middle third of the domain
green. For Quasi2D (BD-q2D) we follow the evolution to time $T_{q2D}=1.505\cdot10^{5}$,
and for True2D (BD-t2D) we follow the evolution to $T_{t2D}=2\cdot10^{4}$.
This ensures that at the final time the diffusive mixing has progressed
to the same relative time, $\chi_{q2D}T_{q2D}=\chi_{t2D}T_{t2D}$.
For comparison we have also simulated uncorrelated walkers (BD-noHI)
to the same relative time. A close examination of the figure reveals
visibly enhanced large-scale fluctuations in the presence of Quasi2D
hydrodynamics compared to no hydrodynamics, in agreement with the
$1/k^{3}$ theoretical prediction (\ref{eq:S_k_neq_color}). For True2D
we see the development of ``colossal'' fluctuations, the magnitude
of which is \emph{comparable} to the mean as predicted by linearized
FHD \cite{GiantFluctuations_ThinFilms}. Interestingly, however, the
very fact that fluctuation are not small compared to the mean invalidates
the linearized FHD theory used to predict the colossal fluctuations
in the first place, as we demonstrate numerically in Section \ref{subsec:GiantColor}.

\begin{figure}
\begin{centering}
\includegraphics[width=0.49\textwidth]{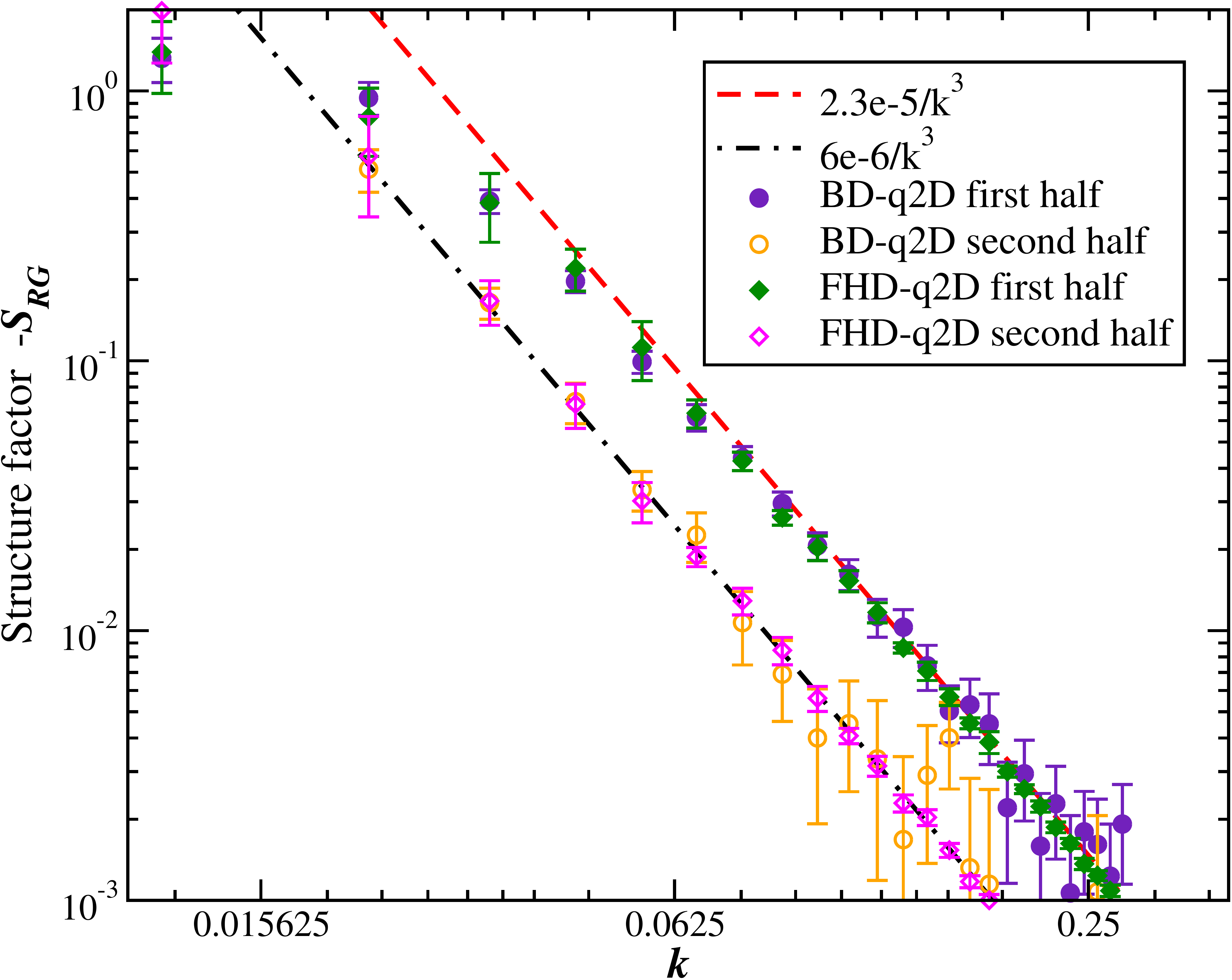}\includegraphics[width=0.49\textwidth]{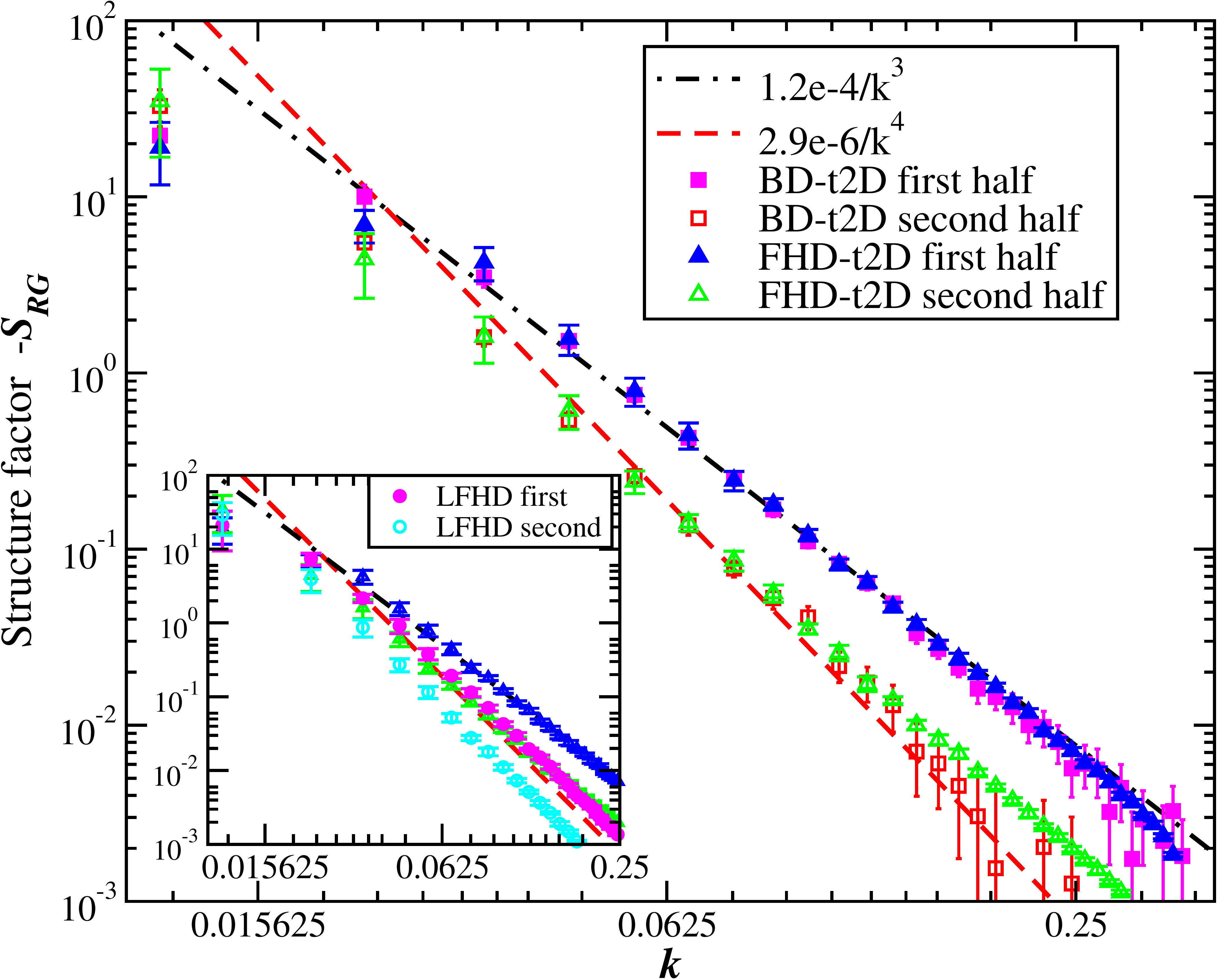}
\par\end{centering}
\centering{}\caption{\label{fig:S_k_q2D}Spectrum of the fluctuations of color density
during the diffusive mixing illustrated in Fig. \ref{fig:InstanceColor},
for Quasi2D (left) and True2D (right) hydrodynamics. The spectrum
is averaged over the two halves of the simulations (filled symbols
for first half and empty symbols for second half). We average the
spectra over 64 simulations and show two standard deviation error
bars. (Left) Results from BD-q2D (circles) for the first (solid circles)
and second half (empty circles), compared to the $k^{-3}$ asymptotic
power-law predicted by (\ref{fig:MSD_t_phi}) (lines). Diamonds show
results from a pseudo-spectral FHD solver for (\ref{eq:lin_FHD_color_q2D}).
(Right) Results from BD-t2D (squares) compared to results obtained
by solving the nonlinear FHD equation (\ref{eq:lin_FHD_color_t2D})
using a pseudo-spectral code (triangles) with an FFT grid of $64^{2}$
cells. The empirical power law $S_{RG}\sim-k^{-3}$ fits the data
well over a broad range of wavenumbers. The inset demonstrates that
solving the linearized FHD equations using the same pseudo-spectral
code (circles) gives $S_{RG}\sim-k^{-4}$ in agreement with the theoretical
prediction (\ref{eq:S_k_neq_color}), but this prediction is in disagreement
with the particle and nonlinear simulations.}
\end{figure}

For a more quantitative study of giant color fluctuations, in Fig.
\ref{fig:S_k_q2D} we show the ensemble- and time-averaged structure
factor $S_{RG}\left(k_{x}=k,k_{y}=0\right)$, i.e., the Fourier spectrum
for wavenumbers perpendicular to the gradient, computed using the
FFT on a grid of $512$ cells. We average the spectrum separately
over the first (time $0<t\leq T/2$) and second (time $T/2<t\leq T$)
halves of the simulation. In the first half of the simulation, the
spectrum is still evolving toward its asymptotic power-law behavior
but the fluctuations are larger because the gradient is largest initially.
In the second half of the simulation, the spectrum has reached its
asymptotic shape but the magnitude decays in time like the square
of the gradient. Note that the fluctuations at the smallest wavenumbers
grow slowest and are subject to strong finite size effects, and are
therefore not expected to obey the prediction of the simple linearized
FHD theory.

\subsubsection{\label{subsec:NFHD}Nonlinear FHD equations}

In order to perform a more quantitative comparison between FHD predictions
and results from the particle simulations shown in Fig. \ref{fig:S_k_q2D},
we need to solve the FHD equations numerically for the time dependent
diffusive mixing under consideration. This is straightforward to do
in the linearized setting in True2D and True3D, and (numerically)
linearized FHD simulations \cite{MultiscaleIntegrators} favorably
compare to experimental measurements for a three-dimensional system
\cite{GRADFLEXTransient}. However, the right panel of Fig. \ref{fig:S_k_q2D}
shows that results from BD-t2D (squares) deviate from the theoretical
$k^{-4}$ decay over the first half of the simulation. This is because
nonlinearities play a role since the fluctuations are comparable to
the mean.

The nonlinear DDFT equations are formal and cannot be given a well-defined
meaning as stochastic partial differential equations because their
solution is a distribution and not a function. Instead, one must spatially
coarse grain the dynamics in order to give nonlinear FHD meaning,
as discussed at length in \cite{DiffusionJSTAT,DDFT_Hydro,FluctDiff_FEM,FluctReactDiff}.
However, we have been able to reproduce the particle results for the
correlation between red and green particles (i.e., for $S_{RG}\left(k\right)$)
by numerically solving (\ref{eq:c_fluct_general}) specialized for
the case of an incompressible velocity field,
\begin{eqnarray}
\partial_{t}c_{R/G} & =-\epsilon^{\frac{1}{2}}\V w\cdot\grad c_{R/G}+\chi\grad^{2}c_{R/G} & ,\label{eq:lin_FHD_color_t2D}
\end{eqnarray}
where we remind the reader that the advective term is to be interpreted
in the Ito sense \footnote{For $\epsilon=1$, equation (\ref{eq:lin_FHD_color_t2D}) can also
be written as $\partial_{t}c_{R/G}=-\V w_{h}\odot\grad c_{R/G},$
where $\odot$ denotes a dot product interpreted in the Stratonovich
sense \cite{DiffusionJSTAT,DDFT_Hydro}.}. Here $\epsilon$ is a parameter that can be used to control the
magnitude of the fluctuations \cite{MultiscaleIntegrators}; we define
the structure factor of interest as $S_{RG}=\epsilon^{-1}\av{\left(\widehat{\d c}_{R}\right)\left(\widehat{\d c}_{G}\right)^{\star}}$.
Setting $\epsilon\ll1$ makes the nonequilibrium fluctuations weaker
and thus \emph{numerically} linearizes the equations \footnote{in the limit $\epsilon\rightarrow0$ one can give a precise meaning
to (\ref{eq:lin_FHD_color_t2D}) in the spirit of linearized fluctuating
hydrodynamics \cite{MultiscaleIntegrators}}. The initial condition we have used is $c_{R/G}(\V r,t=0)=c_{R/G}^{(1)}(\V r,t=0)$,
which is a function and not a distribution like the initial condition
(\ref{eq:c_def}) used in the particle simulations \footnote{Our numerical experiments suggest that even for $\epsilon=O(1)$ the
nonlinear equation (\ref{eq:lin_FHD_color_t2D}) is well-behaved and
its solution remains a function and not a distribution, but it is
difficult to make more precise statements at present.}. Because of this, (\ref{eq:lin_FHD_color_t2D}) does \emph{not} reproduce
the equilibrium component of the fluctuations and instead gives $S_{RR}=S_{GG}=-S_{RG}$,
as predicted by the linearized theory (\ref{eq:S_k_neq_color}) for
the nonequilibrium contribution to the spectrum.

We solve the True2D nonlinear equations (\ref{eq:lin_FHD_color_t2D})
numerically using a standard anti-aliased pseudo-spectral method;
our MATLAB codes are available at \url{https://github.com/stochasticHydroTools/FHDq2D}.
In True2D the particles are passive tracers and there is no difference
between short- and long-time self-diffusion coefficients, so we have
set $\chi=\chi_{0}$. The results for $S_{RG}(k)$ for $\epsilon=1$
are shown with triangles in the right panel of Fig. \ref{fig:S_k_q2D}.
The FHD results are in excellent agreement with the results of the
particle simulations (squares). The inset in the right panel of Fig.
\ref{fig:S_k_q2D} shows that the numerical solutions of the linearized
FHD equations, which we estimated using $\epsilon=0.01$, show a clear
$-k^{-4}$ trend for both halves of the run (circles), and are \emph{not}
in agreement with the results from nonlinear FHD or BD-t2D. This suggests
that in True2D one must numerically solve \emph{nonlinear} FHD equations
in order to quantitatively describe giant fluctuations.

In Quasi2D, it is significantly more challenging to write meaningful
nonlinear FHD equations because of the nonlinearity of the convolution
term arising due to the compressibility of the hydrodynamic kernel,
as we discuss in more detail in the Conclusions. Instead, we have
solved using a pseudospectral method a variant of (\ref{eq:lin_FHD_color_t2D})
adapted to Quasi2D,
\begin{equation}
\partial_{t}c_{R/G}=-\epsilon^{\frac{1}{2}}\V w^{\perp}\cdot\grad c_{R/G}+\chi\grad^{2}c_{R/G},\label{eq:lin_FHD_color_q2D}
\end{equation}
where $\V w^{\perp}$ is the incompressible or vortical component
of $\V w$. More precisely, $\av{\V w^{\perp}\left(\V r,t\right)\otimes\V w^{\perp}\left(\V r^{\prime},t^{\prime}\right)}=\left(2k_{B}T\right)\R^{\perp}\left(\V r-\V r^{\prime}\right)\delta\left(t-t^{\prime}\right),$
where in Fourier space $\hat{\R}_{\V k}^{\perp}=\left(c_{2}\left(ka\right)/\eta k^{3}\right)\,\V k_{\perp}\otimes\V k_{\perp}$
is the incompressible component of the Quasi2D hydrodynamic kernel.

In the left panel of Fig. \ref{fig:S_k_q2D} we show numerical results
for $S_{RG}\left(k\right)$ for $\epsilon=0.01$, but little change
is observed if $\epsilon$ is further reduced or increased to the
natural value $\epsilon=1$. This indicates that the equations are
essentially linear and nonlinearities play little role. In our simulations
we have used the (renormalized) long-time self-diffusion coefficient,
$\chi=\chi_{l}^{(s)}\approx0.86\chi_{0}$, as suggested by the numerical
results for the mean shown in the right panel of Fig. \ref{fig:MeanDensityColor};
the matching is not as good if we use the short-time value $\chi=\chi_{0}$.
We observe an excellent agreement between the results from BD-q2D
and the numerical solution of (\ref{eq:lin_FHD_color_q2D}) in the
left panel of Fig. \ref{fig:S_k_q2D}. This demonstrates that the
\emph{nonequilibrium} fluctuations in Quasi2D are controlled by the
vortical component of the velocity, just as in True2D or True3D. This
can be appreciated from the fact that the nonequilibrium contribution
$\D{\M S}$ in (\ref{eq:S_k_neq_color}) only involves $c_{2}(k)$
and does not depend on the amplitude of the compressible or longitudinal
component $c_{1}(k)$, which only affects the \emph{equilibrium} contribution
$\M S_{0}$.

\section{Conclusions and Discussion}

We developed an efficient algorithm for Brownian dynamics with hydrodynamic
interactions, suitable for modeling diffusion of spherical colloids
of hydrodynamic radius $a$ confined to a two-dimensional plane. We
used this algorithm to perform large-scale particle simulations and
studied collective diffusion on fluid-fluid interfaces (Quasi2D) and
in two-dimensional liquids (True2D). The nonzero compressibility of
the three dimensional flow at the fluid-fluid interface leads to a
nonzero divergence of the mobility matrix. Under the action of hydrodynamic
fluctuations in the fluid, this compressibility acts like a pairwise
repulsive potential of order $k_{B}T\left(a/r\right)$, and changes
the nature of diffusion dramatically.

\subsection{Summary of Findings}

We first examined the evolution of ensemble averages of the particle
number density. Consistent with prior studies, we obtained good agreement
between a simple closure for the equations of dynamic density functional
theory and particle simulations of an ideal gas of hydrodynamically-correlated
Brownian particles. At the same time, we found that the effective
particle-particle repulsion leads to a nontrivial reduction of the
long-time self-diffusion coefficient as the packing density increases,
even for an ideal gas of non-interacting particles. By coloring the
particles with two species labels, we elucidated the mechanism by
which an initially localized density perturbation develops inverse-cubed
power-law tails at later times, even though individual particles still
have Gaussian displacements at long times consistent with ordinary
diffusion. We found that the diffusive mixing of two colors is consistent
with simple diffusion but with a diffusion coefficient equal to the
long-time self-diffusion coefficient, indicating that fluctuations
renormalize the equations for the ensemble average and must be retained
in a quantitatively-accurate description. 

We further examined the magnitude and dynamics of collective density
fluctuations at thermodynamic equilibrium and out of equilibrium.
We found that for an ideal gas mixture, equilibrium dynamic structure
factors show nearly mono-exponential decay, with the difference between
short and long times coming mainly from the difference between the
short- and long-time self-diffusion coefficient. Out of equilibrium,
we found that nonequilibrium density fluctuations are small compared
to equilibrium fluctuations, indicating that in Quasi2D each instance
of the diffusive spreading of density perturbations looks similar
to the ensemble average. By contrast, in True2D nonequilibrium fluctuations
grow to be comparable to the mean and make each instance of the mixing
process qualitatively different from the ensemble average. We found
that nonequilibrium color fluctuations exhibit a power-law spectrum
typical of giant fluctuations seen in other geometries, but are much
smaller in magnitude in Quasi2D than in True2D. Fluctuating hydrodynamics
can accurately model the giant color fluctuations in both Quasi2D
and True2D, with nonlinearities playing a significant role only in
True2D. We found that the compressibility of the hydrodynamic tensor
in Quasi2D does not affect giant color fluctuations in the absence
of a density gradient; the magnitude of the red-green correlations
could be accurately predicted by linearized FHD based on an incompressible
fluid velocity. At the same time, we found that the hydrodynamic ``repulsion''
among the particles strongly suppresses density fluctuations in the
presence of a density gradient. The difference between True2D and
Quasi2D is striking and emphasizes the important role of hydrodynamics
and fluctuations in diffusion in liquids. In True2D the mean is the
same as in the absence of hydrodynamics, but the fluctuations dominate
the mean and exhibit a yet unexplained $k^{-3}$ power law. In Quasi2D,
the mean behavior differs strongly from standard diffusion, but the
fluctuations are small compared to the mean and well-described by
linearized fluctuating hydrodynamics.

In Section \ref{sec:SelfDiffusion} we examined in some detail the
time dependence of the self-diffusion coefficient, i.e., the mean
square displacement. We found that the reduction of the self-diffusion
coefficient with time is a collective effect governed by density fluctuations
at length scales notably larger than the particle size. It remains
a challenge to develop a theoretical understanding of this effect.
In \cite{TracerDiffusion_Demery}, following prior theoretical work
of Demery and Dean \cite{TracerDiffusion_Demery_adiabatic,TracerDiffusion_Demery_PathIntegral},
the authors have developed a theory for the effective diffusion of
a tracer particle in a dense suspension of soft repulsive colloids,
in the absence of hydrodynamics. This theory predicts a reduction
of the diffusion coefficient for repulsive soft spheres, just as we
observe in Quasi2D, consistent with the interpretation of (\ref{eq:effective_repulsion})
as a soft repulsive potential \footnote{The $1/r$ behavior only applies in the far field and needs to be
regularized at short distances. In our formulation based on the FCM
method, in Fourier space the apparent repulsive potential can be written
as $\hat{U}(k)=\left(6\pi a/k\right)c_{1}\left(ka\right)$.}. In \cite{TracerDiffusion_Demery}, the authors use fluctuating hydrodynamics
to describe the collective fluctuations of the fluid excluding the
tracer particle, and couple this bidirectionally to the equation of
motion for the tracer particle itself. Generalizing this approach
to Quasi2D diffusion is nontrivial. Importantly, our particles are
correlated hydrodynamically and are driven not by independent noise
but rather advected by a smooth in space and white in time random
velocity field $\V w$. Furthermore, our system is in detailed balance
(time reversible) with respect to an \emph{ideal gas} equilibrium
distribution, unlike a gas of particles that interact with a direct
repulsive potential.

\subsection{Future Directions}

It remains a challenge for the future to formulate nonlinear FHD equations
in Quasi2D. A natural first guess is to consider a regularized (truncated)
version of the formal fluctuating DDFT-HI equations \cite{DDFT_Hydro},
\begin{eqnarray}
\partial_{t}c_{R/G}\left(\V r,t\right) & = & \grad\cdot\left(-\V w\left(\V r,t\right)c_{R/G}(\V r,t)+\chi\grad c_{R/G}\left(\V r,t\right)+\sqrt{2\chi c_{R/G}(\V r,t)}\,\V{\mathcal{W}}^{(R/G)}(\V r,t)\right)\nonumber \\
 & + & \left(k_{B}T\right)\grad\cdot\left(c_{R/G}\left(\V r,t\right)\int\R\left(\V r,\V r^{\prime}\right)\grad^{\prime}c\left(\V r^{\prime},t\right)\,d\V r^{\prime}\right),\label{eq:nonlin_FHD_color}
\end{eqnarray}
where we have now employed an Ito interpretation for the advective
terms. Our attempts to solve (\ref{eq:nonlin_FHD_color}) using a
pseudospectral numerical method \footnote{In our pseudo-spectral method, we simply truncate the Fourier modes
at $k_{\max}=\pi/h$ where $h$ is the grid spacing, which is analogous
to spatial discretization in finite volume methods.} have failed to produce results in agreement with particle simulations
(BD-q2D). The fluctuation-dissipation balance between the advection
by the compressible (longitudinal) component of $\V w$ (which acts
as a source of fluctuations) and the nonlinear convolution term (which
damps or dissipates fluctuations) is rather delicate and difficult
to preserve in spatially coarse-grained equations.

In this work we focused on an ideal gas of non-interacting particles
in order to emphasize the hydrodynamic effects. In any experimentally-realizable
system, however, the particles will interact with each other with
short-ranged steric forces at a minimum, and most likely with some
additional longer-ranged interaction as well, such as partially-screened
repulsive electrostatic forces \cite{ChargedColloidsInterface_Chaikin}
or attractive capillary forces \cite{Diffusion2D_IdealGas}. Preliminary
investigations for partially-confined Quasi2D suspensions \cite{PartiallyConfined_Quasi2D}
have already shown that the strong collective effects in Quasi2D diffusion
persist in the presence of steric repulsion. At the same time, however,
steric repulsion induces caging and introduces a new cage-breaking
time scale. We therefore expect to see a strong difference between
short and long times in both dynamic structure factors and nonequilibrium
dynamics. Direct interactions can formally be included in the fluctuating
DDFT-HI equations \cite{DDFT_Hydro}, however, the equations are no
longer closed and making further progress requires introducing various
approximations and density functionals inspired by static DFT \cite{SPDE_Diffusion_Formal,DDFT_Pep}.
The errors made because of such approximations would in practice overwhelm
the modest corrections to the self-diffusion coefficient we were able
to isolate in this work by using an ideal gas.

Although we focused here on purely three-dimensional hydrodynamic
interactions in an unbounded system, the same kind of compressibility
effect likely plays a role in other geometries and systems as well.
In particular, membranes immersed in fluid are another system where
diffusion is confined to a surface but the hydrodynamics is three
dimensional. Examples include thin smectic films in air \cite{ThinFilms_True2D}
and lipid bilayers at the surface of vesicles. The Saffman model \cite{SaffmanThinFilm,SaffmanOseen_Membrane}
for membranes gives a modified form for the hydrodynamic correlation
tensor (see Eq. (13) in the review \cite{MembraneDiffusion_Review}),
which is consistent with our general form (\ref{eq:G_hat_k}) with
$f_{k}=0$ and $g_{k}=1/\left(\eta k\left(k+k_{c}\right)\right)$.
A number of modifications of these functions have been computed in
slightly different geometries such as, for example, triply periodic
systems \cite{MembraneDiffusion_Periodic}. Here $k_{c}$ is a roll-over
wavenumber that depends on the ratio of the viscosity of the surrounding
fluid to the two-dimensional viscosity of the membrane/film itself;
for $k_{c}=0$ we get True2D hydrodynamics. The Saffman model assumes
incompressibility in the plane, $f_{k}=0$, and therefore there would
be no anomalous collective diffusion with Saffman hydrodynamic correlations;
this justifies the simplified fluctuating hydrodynamic equations used
in \cite{GiantFluctuations_ThinFilms}. Much of the work on diffusion
in membranes has focused on self-diffusion, and, to our knowledge,
collective diffusion in membranes has not yet been carefully studied
either experimentally or numerically. The True2D Brownian dynamics
algorithm developed here can be used to study collective diffusion
in membranes with a simple change of the function $g_{k}$, at least
under the assumption that the Saffman model applies on the time and
length scales of interest. Future work should probe the transition
from Quasi2D hydrodynamics to Saffman-like hydrodynamics in dense
monolayers of hard spheres on fluid-fluid interfaces, as well as in
lipid bilayers.

A nonzero divergence of the hydrodynamic mobility matrix would also
arise in confined geometries. For example, dense colloids would sediment
close to a bottom wall and then diffusive primarily in the plane of
the wall. Similarly, colloids may be confined to diffuse in a plane
by two walls in a slit channel. The anomalous short-time collective
diffusion we studied here will be important in these types of wall-bounded
systems as well. In fact, a re-analysis of earlier experiments \cite{Diffusion2D_Experiments_Rice}
has shown anomalous behavior of the collective diffusion coefficient
with wavenumber in several experimental realizations of quasi one-
and two-dimensional systems. The confinement by the walls screens
the hydrodynamic interactions and will therefore reduce the enhancement
of the collective diffusion. For particles confined to diffuse above
a single no-slip wall, the hydrodynamic interactions decay like the
cubed of the distance, but, counter-intuitively, the interactions
decay like the square of the distance for particles in a slit channel
\cite{HI_Confined_Decay}. In both cases the compressibility of the
hydrodynamic kernel in the plane of diffusion is nonzero. Future work
should explore the consequences of this for collective diffusion both
with simulations and experiments. The hydrodynamic kernel can easily
be changed in our two-dimensional BD-HI algorithm to take into account
the Green's function for Stokes flow in confined geometries; however,
even for particles diffusing strictly in a plane the method requires
an uncontrolled approximation. Specifically, when confinement breaks
the symmetry between the two half spaces bounded by the plane of diffusion,
forces parallel to the plane can induce motion perpendicular to the
plane; a similar effect appears for diffusion on curved interfaces
as well. This effect can be taken into account but at a significant
increase in complexity. Instead, it may be more appropriate (and physically
realistic) to use three-dimensional Brownian dynamics resolving the
confining potential, at least for the case of particles sedimented
above a bottom wall \cite{MagneticRollers}.
\begin{acknowledgments}
We thank Eric Vanden-Eijnden and Johannes Bleibel for informative
discussions. This work was supported in part by the National Science
Foundation under collaborative award DMS-1418706 and by DMS\textendash 1418672,
and by the U.S. Department of Energy Office of Science, Office of
Advanced Scientific Computing Research, Applied Mathematics program
under award DE-SC0008271. We thank the NVIDIA Academic Partnership
program for providing GPU hardware for performing some of the simulations
reported here. R.D-B and SP acknowledge  the support  of  the Spanish  Ministry  of Science  and Innovation MINECO  (Spain) under grant FIS2013-47350-C5-1-R  and the ``Mar\'{\i}a de Maeztu''  Programme for Units of  Excellence in R\&D (MDM-2014-0377). Part  of the  simulations were done  in Marenostrum under grant FI-2017-2-0023. R.D-B and RP acknowledges support the donors of The American  Chemical Society  Petroleum Research Fund  for partial support of this research via PRF-ACS ND9 grant.
\end{acknowledgments}


\end{document}